\documentclass[11pt,a4paper]{article}
\usepackage[T1]{fontenc}
\usepackage{lmodern}
\usepackage[margin=28mm]{geometry}
\usepackage{amsmath,amssymb,amsthm,mathtools,microtype}
\usepackage{mathrsfs,aliascnt}
\usepackage{xcolor}

\usepackage{graphicx}
\usepackage{tikz}
\usepackage{pgfplots}
\usepackage{caption}
\usepackage{float}

\usepackage{authblk}

\usetikzlibrary{arrows.meta}
\pgfplotsset{compat=1.18}

\usepackage[colorlinks=true,linkcolor=blue!50!black,citecolor=blue!50!black,urlcolor=blue!50!black]{hyperref}
\usepackage[nameinlink,noabbrev]{cleveref}
\usepackage{eucal}
\numberwithin{equation}{section}
\newtheorem{theorem}{Theorem}[section]
\newaliascnt{proposition}{theorem}
\newtheorem{proposition}[proposition]{Proposition}
\aliascntresetthe{proposition}
\newaliascnt{lemma}{theorem}
\newtheorem{lemma}[lemma]{Lemma}
\aliascntresetthe{lemma}
\theoremstyle{plain}
\crefname{theorem}{Theorem}{Theorems}
\crefname{proposition}{Proposition}{Propositions}
\crefname{lemma}{Lemma}{Lemmas}
\theoremstyle{definition}
\newtheorem{example}[lemma]{Example}
\newtheorem{definition}[lemma]{Definition}
\newtheorem{remark}[lemma]{Remark}
\DeclareMathOperator{\Tr}{Tr}
\DeclareMathOperator{\supp}{supp}
\DeclareMathOperator{\dist}{dist}
\newcommand{\HH}{\mathcal H}
\newcommand{\dd}{\mathrm d}
\newcommand{\one}{\mathbf1}
\newcommand{\dG}{\mathrm d\Gamma}
\newcommand{\norm}[1]{\lVert#1\rVert}

\date{November 8, 2026}

\definecolor{inkblue}{HTML}{147BB3}
\definecolor{brickred}{HTML}{BC4828}
\definecolor{darkink}{HTML}{192838}
\definecolor{mutedtext}{HTML}{496579}
\definecolor{lightgrid}{HTML}{D7E2E8}

\pgfmathdeclarefunction{smoothstep}{1}{%
  \pgfmathparse{(#1<=0) ? 0 : ((#1>=1) ? 1 :
    exp(-1/max(#1,0.0001)) /
    (exp(-1/max(#1,0.0001))+exp(-1/max(1-#1,0.0001))))}%
}

\pgfmathdeclarefunction{chiOne}{1}{%
  \pgfmathparse{
    smoothstep((#1-0.2)/0.55)*
    smoothstep((1.8-#1)/0.55)
  }%
}

\pgfmathdeclarefunction{chiTwo}{1}{%
  \pgfmathparse{
    smoothstep((#1-1.2)/0.55)*
    smoothstep((2.8-#1)/0.55)
  }%
}

\pgfmathdeclarefunction{bumpf}{1}{%
  \pgfmathparse{(abs((#1-1.5)/0.35)<1) ?
    1.28*exp(1-1/max(1-((#1-1.5)/0.35)^2,0.0001)) : 0}%
}

\begin{document}

\title{Lieb--Robinson bound for one-dimensional continuum fermions}

\author{Marius Lemm\thanks{marius.lemm@uni-tuebingen.de}}
\author{Hailun Lu\thanks{hailun.lu@uni-tuebingen.de}}
\author{Harman Preet Singh\thanks{harman-preet.singh@uni-tuebingen.de}}
\affil{\it\small Department of Mathematics, University of T\"ubingen, 72076 T\"ubingen, Germany}

\maketitle
\begin{abstract}
The central difficulty  in quantum many-body physics is that the Hilbert space dimension grows exponentially with the total degrees of freedom. A key feature to circumvent this problem is \textit{locality}, which  manifests dynamically through an  approximate causal structure in the form of a Lieb–Robinson bound (LRB). LRBs have been proved for quantum spin systems and certain lattice boson systems, but so far never for realistic \textit{continuum} systems.
We present the first LRB that applies to a standard system of one-dimensional continuum fermions. It applies to initial states with controlled moments of local particle number and energy and is thus thermodynamically stable. Our result breaks the main technical barrier that has existed for using the vast locality toolbox on continuum systems.

\end{abstract}


\section{Introduction and main results}\label{sec:main}

The core difficulty of the quantum many-body problem is what is known as the curse of dimensionality: the Hilbert space dimension grows exponentially with the number of constituents. A priori, this full complexity is needed to accurately describe the effects of interactions. If one assumes that the interactions are governed by a small parameter, then expansion methods (e.g., rigorous renormalization group or dilute limits) can be used to account for interactions perturbatively. In strongly interacting systems, the key insight to tame the curse of dimensionality is to use \textit{locality} to effectively break the system up into $\mathcal O(1)$-regions in which the exponential growth is effectively controllable. The mathematical manifestation of locality is in fact dynamical in nature: the \textit{Lieb--Robinson bound (LRB)}, named after its first appearance in \cite{lieb1972finite}, controls the spreading of a local observable under the Heisenberg dynamics. 

The ``unreasonable effectiveness of Lieb--Robinson bounds'' is well-documented and we only mention a few highlights here:
\begin{itemize}
    \item[(i)] In the 1970s, Robinson's first motivation to pursue LRBs was to use them to establish the existence and strong continuity of the thermodynamic limit of the dynamics in the sense of operator algebras \cite{lieb1972finite,brattelioperator}
    \item[(ii)]
    In the early 2000s Hastings famously discovered that LRBs combine productively with spectral gaps and time-frequency Fourier transforms  which led to a number of breakthroughs  on the  locality properties of gapped ground states\cite{HK2006,NS2006,Hastings2007area}. Consequently, LRBs have become central to the description of topological quantum phases via automorphic equivalence \cite{hastings2005quasiadiabatic,bachmann2012automorphic} and they are the decisive analytical tools to prove stability of the spectral gap \cite{bravyi2010topological,michalakis2013stability,NSY2017,NSY2022,del2025boundary,teufel2026bulk}.
    
    \item[(iii)] In condensed matter-physics, LRBs are the decisive tool for mathematically rigorous many-body adiabatic theory with applications to quantization of the Hall conductance \cite{bachmann2018adiabatic,HT2020bulk,HW2022,becker2026generalized,greenblatt2024adiabatic,henheik2026response} and other prethermal phenomena \cite{abanin2017rigorous,YL2023,mcdonough2025non}.

\item[(iv)] In quantum information theory,  LRBs control the transfer of quantum information and entanglement quite generally \cite{bravyi2006lieb}. On the one hand, this  implies upper bounds on the performance of quantum-enhanced messaging protocols \cite{epstein2017quantum}. On the other hand, LRBs underpin the main algorithms to efficiently simulate the dynamics by suitably reducing the global problem to a local one. This applies both to classical simulation of the Heisenberg dynamics of local observables \cite{chen2023speed} and to the digital quantum simulation of the global dynamics through the state-of-the-art Hastings-Haah-Kothari-Low algorithm \cite{haah2021quantum,kuwahara2024effective}.
\end{itemize}

For further background on LRBs, see the recent review \cite{chen2023speed}. 

The main problem we face is that all of these far-reaching and physically profound  applications of  LRBs (i)--(iv) are almost entirely limited to \textit{quantum spin systems} and \textit{lattice fermions}. The reason is that the standard proofs of LRBs require the interactions  to be bounded in operator norm sense.

In the last few years, the frontier has been pushed to an important new class of lattice bosons models: Bose--Hubbard Hamiltonians \cite{schuch2011information,yin2022finite,kuwahara2021lieb,FLS,LRSZ,kuwahara2024effective,kuwahara2024enhanced,lemm2025quantum,lemm2026lieb,elgart2026lieb,rubiliani2026lieb,lemm2026dissipation}; see also the earlier work \cite{nachtergaele2007lieb} on perturbations of harmonic oscillators. In these lattice boson systems, the local interactions are unbounded, but one is nonetheless able to derive \textit{LRBs for lattice bosons} which have two new features compared to their quantum spin system antecedent:
\begin{itemize}
    \item[(a)] The LRB depends on the initial boson density, but not otherwise on the initial state.
    \item[(b)] The effective velocity of information propagation is $v\sim t^{d-1}$ where $d$ is the spatial dimension. Thus, the light cone is ``bent'' and the velocity is no longer uniformly bounded in time. However, it is bounded \textit{uniformly in system size} (also called ``thermodynamically stable'') and this is sufficient for the applications (i)--(iv) described before. The bent light cone just changes some exponents in the applications.
\end{itemize}
Accordingly, the applications (i)-(iv) of the new bosonic LRBs are now being actively explored and it is presumably only a matter of time until we reach a quantum information theory of lattice bosons that --- subject to the changes implied by points (a) and (b) above --- reaches a similar level of generality as its quantum spin precursor.

The true frontier are then \textit{continuum systems}. These are both more fundamental and more realistic than their lattice counterparts and so it would be extremely desirable to unlock the far-reaching possibilities (i)--(iv) for them. LRBs have been shown to holds for certain systems of continuum fermions: ones with explicit ultraviolet cutoffs in the interaction \cite{narnhofer1990quantum,gebert2020lieb,hinrichs2025lieb}, ones acting on localized magnetic frames \cite{bachmann2025lieb} or ones with pseudo-relativistic kinetic operator $\sqrt{-\Delta_{x_i}+m^2}$ in  a semiclassical regime \cite{fresta2023effective,fresta2025effective}. 

However, the existing techniques do \textit{not} apply to a standard  many-body Hamiltonian of the form
\begin{equation}
  H_N= \sum_{i=1}^N (-\Delta_{x_i})+\sum_{1\leq i<j\leq N} W(x_i-x_j) 
\end{equation}

To explain this further, we note that for a Bose--Hubbard Hamiltonian the kinetic operator of individual bosons is hopping, i.e., the discrete Laplacian which of course has a bounded group velocity. In some sense, it was this fact which could be exploited to treat  Bose--Hubbard Hamiltonians in \cite{schuch2011information,yin2022finite,kuwahara2021lieb,FLS,LRSZ,kuwahara2024effective,kuwahara2024enhanced,lemm2025quantum,lemm2026lieb,elgart2026lieb,rubiliani2026lieb,lemm2026dissipation}. Roughly speaking, in a first step, one uses the bounded group velocity to control the accumulation of bosons locally. In a second step, this is used to approximate the bosons with a system of finite local dimension for which one can then derive an LRB using spin-system tools.

However, for the continuum Hamiltonian $H_N$, each particle's kinetic energy is described by the continuum Laplacian $-\Delta_{x_i}$ and therefore has unbounded group velocity. This is the reason why proving an LRB for any such Hamiltonian has remained completely open for a long time. To quote the last sentence of Bratteli-Robinson's 1997 edition \cite{brattelioperator}: \textit{``Finally, the description of continuous system is notably incomplete. But, this is a feature of all methods which have been applied to their analysis and it is doubtful that any complete understanding of these models will be obtained in the near future.''} LRBs for continuum systems have been even more sorely lacking since the advent of applications (ii)--(iv) in the meantime.

In the present paper, we break this longstanding barrier and prove the first Lieb--Robinson bounds for a continuum system. Specifically, we consider one-dimensional fermions in a torus geometry. Our LRB mirrors the two features (a) and (b) of the LRBs for lattice bosons recalled above:
\begin{itemize}
    \item[(a)'] The LRB depends on moments of the initial particle and energy density in the initial state.
    \item[(b)'] The effective velocity of information propagation is $v\sim e^{ct}$. Thus, the light cone is ``exponentially bent'' and the velocity is not uniformly bounded in time. However, it is bounded \textit{uniformly in system size}.
\end{itemize}
In the present paper, we only aim to derive the first LRB for a standard continuum system. We postpone all applications, in particular those in the direction of (i)--(iv) from above, to future work.

Next, we describe the setup and main result, which is stated as Theorem \ref{thm:main} below and extended to general polynomial observables in Theorem \ref{thm:polynomials}. Afterwards, in Section \ref{ssec:strategy}, we will describe our proof strategy. In a nutshell, we build on the blueprint provided by the lattice boson strategy, by rewriting the continuum system as a lattice Hamiltonian on ``pseudosites''. The local Hilbert space of a pseudosite is the Fock space over $L^2(I)$ for suitable intervals $I$. It is infinite-dimensional, but the occupation of high-energy modes can be controlled in terms of local energies, in some rough analogy with lattice bosons. This ``pseudolattice'' picture provides the first systematic conceptual framework for proving LRBs for continuum systems. 

Our new conceptual framework makes equal sense for higher-dimensional continuum fermions and bosons and we believe it will allow to cover all of these systems eventually. However, the technical implementation is analytically challenging and, for technical reasons, we currently require a strong form of the Pauli exclusion principle that only holds in 1D; see Remark \ref{remark:1Dfermions} for further details.

\subsection{Setup}

Write $\mathbb N=\{1,2,\ldots\}$ and $\mathbb N_0=\mathbb N\cup\{0\}$. Given an integer $L>0$, we identify $\mathbb T_L=\mathbb R/L\mathbb Z$ with $[0,L)$, equipped
with Lebesgue measure of mass $L$, and we set
\begin{equation}
 \norm{x}=\min\{x,L-x\},\qquad B_R(x)=\{y\in\mathbb T_L:\norm{x-y}<R\}.
\end{equation}
Because of the identification mentioned above, we can parameterize points on $\mathbb T_L$ as real points taken modulo $L$. Thus $\norm{x-y}$ denotes the periodic distance on the torus.

We shall describe many-body fermionic systems on $\mathbb T_{L}$ in the grand-canonical formalism, so that the associated space of states is the \emph{fermionic Fock space} $\mathcal F$, given by
\begin{equation}
    \mathcal F=\bigoplus_{N\ge0}\HH_N,
 \qquad \HH_N=L^2(\mathbb T_L)^{\wedge N},
\end{equation}
with $\HH_0=\mathbb C$.
For $f\in L^2(\mathbb T_L)$, we define the usual annihilation and creation operators $a(f),a^*(f)\in\mathcal B(\mathcal F)$ by
\begin{equation}
 (a(f)u_N)(x_1,\ldots,x_{N-1})
 =\sqrt N\int_{\mathbb T_L}\overline{f(y)}u_N(y,x_1,\ldots,x_{N-1})\,\dd y,
 \qquad a^*(f)=a(f)^*,
\end{equation}
for $N\ge1$, with zero annihilation on the vacuum. These operators satisfy the canonical anticommutation
relations (CAR)
\begin{equation}\label{eq:CAR}
 \{a(f),a(g)\}=0,\qquad \{a(f),a^*(g)\}=\langle f,g\rangle\one,
\end{equation}
which imply $\|a(f)\|=\|a^*(f)\|=\|f\|_2$. We use $a_x,a_x^*$ for the corresponding operator-valued distributions.

Furthermore, recall that any self-adjoint operator $A$ on a single-particle Hilbert space $\mathfrak g$  lifts to its second quantization $\mathrm d\Gamma(A)$ on the associated Fock space, whose sector-wise action is given by
\begin{equation}
 \dG(A)|_{\mathfrak g^{\wedge N}}\equiv A_N=\sum_{i=1}^N A^{(i)}.
\end{equation}
In particular, given a measurable set $U\subset\mathbb T_L$, the local particle number operator is given by $\mathcal N_U=\dG(\one_U)$, with $\mathcal N\equiv \mathcal N_{\mathbb T_L}$ being the global particle number operator.

We consider Hamiltonians of the form
\begin{equation}
    H= \mathrm d\Gamma(-\partial^2) + \frac{1}{2}\int_{\mathbb T^2_L} W(x-y) a^{*}_y a^{*}_x a_x a_y\, \mathrm dx\mathrm dy.
\end{equation}
The two-body interaction potential $W\in C^\infty(\mathbb T_L;\mathbb R)$ is an $L$-periodization of a function in $C^\infty(\mathbb R;\mathbb R)$ that is even, non-negative, and finite-range, meaning it satisfies
\begin{equation}\label{eq:W}
 W(s)=W(-s)\ge0,\qquad \supp W\subset B_{r_0}(0),
\end{equation}
for a fixed $r_0>0$. In particular, $r_0$  is independent of $L$ and $W$ and all its derivatives are bounded independently of $L$ as well.
On each $N$-particle sector $H$ takes the form
\begin{gather}
 H_N=-\sum_{i=1}^N\partial_i^2+\sum_{1\le i<j\le N}W(x_i-x_j),\label{eq:H}
\end{gather}
which, since $W$ is bounded and non-negative, is a self-adjoint and non-negative operator with domain $D(H_N)=H^2(\mathbb T_L^N)\cap\HH_N$.
On $\mathcal F$, $H=\bigoplus_{N\ge0}H_N$, with $H_0=0$, is then self-adjoint on the corresponding direct-sum domain.

In the following sections we denote Heisenberg time evolution of a bounded observable $A$ on $\mathcal F$ by
\[
\tau_t^H(A)=e^{itH}Ae^{-itH}.
\]

\subsection{Local energy moments}
Let $\eta\in C^\infty(\mathbb T_L;\mathbb R)$ be radial, with $0\leq\eta\leq1$ and $\sqrt{\eta}\in C^\infty(\mathbb T_L;\mathbb R)$, such that
\begin{equation}
\eta(x) = \begin{cases}
    1 & \|x\|\leq 1\\
    0 & \|x\|\geq 2.
\end{cases}
\end{equation}
For any $z\in\mathbb T_L$, we define
\begin{equation}\label{eq:eta}
 \eta_z(y):=\eta(y-z),\qquad
 B_z^W=B_{2+r_0}(z),
\end{equation}
where the superscript $W$ indicates the interaction neighborhood of $B_2(z)$: $B_z^W$ contains all points that can interact through $W$
with points of the interval $B_2(z)$ centered at $z$. 


\begin{definition}[Local energy]
    On $\mathcal F$, $\mathcal E_z$ is the nonnegative self-adjoint operator defined by
\begin{equation}
    \mathcal E_z := \mathrm d\Gamma(-\partial \eta_{z}\partial) + \frac14\int_{\mathbb T_L^2} (\eta_{z}(x)+\eta_{z}(y))W(x-y)a^*_{x}a^*_y a_y a_x\, \mathrm dx\mathrm dy,
\end{equation}
which is associated to the closure of the following energy form on
$C^\infty(\mathbb T_L^N)\cap\HH_N$:
\begin{equation}\label{eq:localform}
   \mathcal E_z[u]
   =\sum_{i=1}^N\int_{\mathbb T_L^N}\eta_z(x_i)|\partial_i u(\underline{x})|^2\,\dd \underline{x}
 +\frac12\sum_{1\leq i<j\leq N}\int_{\mathbb T_L^N}[\eta_z(x_i)+\eta_z(x_j)]
                  W(x_i-x_j)|u(\underline{x})|^2\,\dd \underline{x},
\end{equation}
with $\underline{x}=(x_1,\ldots,x_N)\in \mathbb T_L^N$.
\end{definition}

The form is closable because each weighted first derivative is
closable and the potential is bounded and nonnegative.
With a slight abuse of notation, we denote the form and its closure
by the same symbol $\mathcal E_z$.
On $\mathcal F$, $\mathcal E_z$ denotes the direct sum, with vacuum
value zero.

In what follows, we will only use the local energy and particle number operators centered at integer points $j=0,\dots,L-1$, which are the centers of our ``pseudosite intervals'' introduced later.

Write $\mathfrak S_1(\mathcal F)$ for the trace-class operators on $\mathcal F$. We call a $0\le\rho\in\mathfrak S_1(\mathcal F)$ with $\Tr\rho=1$ a state. For any nonnegative
selfadjoint $G$ on $\mathcal F$, we use the convention
\begin{equation}
 \Tr\rho G=\|G^{1/2}\rho^{1/2}\|_2^2
 =\sum_{\nu=1}^{\operatorname{rank}\rho}\lambda_\nu\|G^{1/2}u_\nu\|^2,\qquad
 \rho=\sum_{\nu=1}^{\operatorname{rank}\rho}\lambda_\nu|u_\nu\rangle\langle u_\nu|,
 \quad\lambda_\nu>0,
\end{equation}
where $\|\cdot\|_2$ is the Hilbert--Schmidt norm and a term outside the
form domain has value $+\infty$. The same convention applies to
nonnegative closed forms.
\begin{definition}
    For any $p>0$, and for any state $\rho$ on $\mathcal F$, we define the $p$-th local moment of $\rho$ to be
\begin{equation}\label{eq:MP}
 M_p(\rho)=1+\max_{0\le j<L}
       \Tr\rho(\mathcal E_j^p+\mathcal N_{B_j^W}^p).
\end{equation}
\end{definition}
We shall assume our initial state $\rho$ to satisfy
\begin{equation}\label{eq:momentbound}
    M_{p}(\rho) <\infty
\end{equation}
for all $p\in \mathbb N$. The following example provides physically relevant cases satisfying this local energy and particle-number moment bound.
\begin{example}[Quasi-free initial states]
For any quasi-free state $\rho_L$ on $\mathcal F$ with zero pairing,
let $\gamma_L(x,y)=\Tr(\rho_La_x^*a_y)$ denote the kernel of its
one-particle density matrix. A sufficient condition for
\eqref{eq:momentbound} to hold uniformly in $L$ is that $\gamma_L$
is smooth and
\begin{equation}\label{eq:quasifree-kernel-bounds}
 \sup_L\sup_{x,y\in\mathbb T_L}
 \bigl|\partial_x^r\partial_y^s\gamma_L(x,y)\bigr|<\infty
 \qquad\text{for every }r,s\in\mathbb N_0.
\end{equation}
Indeed, Wick's rule and the smoothness and finite range of $W$ then give
$\sup_L M_p(\rho_L)<\infty$ for every $p>0$. A concrete subclass of such states is that of Slater determinants built out of smooth functions.

In particular, for translation-invariant quasi-free states with
zero pairing, the normalized plane waves $L^{-1/2}e^{ikx}$,
$k\in(2\pi/L)\mathbb Z$, are independently occupied with
probabilities $\nu_L(k)\in[0,1]$, and
\begin{equation}
 \gamma_L(x,y)
 =\frac1L\sum_{k\in(2\pi/L)\mathbb Z}
       \nu_L(k)e^{ik(y-x)}.
\end{equation}
The kernel bounds \eqref{eq:quasifree-kernel-bounds} are then
equivalent to
\begin{equation}\label{eq:quasifree-momentum-moments}
 \sup_L\frac1L\sum_{k\in(2\pi/L)\mathbb Z}
       (1+k^2)^m\nu_L(k)<\infty
 \qquad\text{for every }m\in\mathbb N_0.
\end{equation}
A physically relevant example for which condition \eqref{eq:quasifree-momentum-moments} holds is provided by the free grand-canonical thermal states at inverse temperature $\beta>0$ and chemical potential
$\mu\in\mathbb R$,
\begin{equation}
 \rho_L=\frac{e^{-\beta\dG(-\partial^2-\mu)}}{\Tr e^{-\beta\dG(-\partial^2-\mu)}},\qquad
 \nu_L(k)=\frac1{1+e^{\beta(k^2-\mu)}},
\end{equation}
since their occupation probabilities decay exponentially in $k^2$.
It also holds whenever
$\nu_L(k)=0$ for $|k|>K$, with $K\ge0$ independent of $L$,
including the free Fermi ball
$\nu_L(k)=\one_{\{|k|\le K\}}$.
\end{example}

\subsection{Main result}
We bound the evolved graded commutator of two initially separated
observables in the Hilbert--Schmidt norm weighted by the initial state.

\begin{theorem}[Main result]\label{thm:main}
Let $L\geq 16(r_0+3)$, $k\in\mathbb N$; further, let $W$ satisfy \eqref{eq:W}.
For a given pair $x,y\in\mathbb T_L$, and let $f_x,f_y\in C^\infty(\mathbb T_L;\mathbb C)$ satisfy
\begin{equation}
 \supp f_x\subset B_1(x),\qquad \supp f_y\subset B_1(y).
\end{equation}
Then, there exist constants $c_k,C_k>0$ such that, for all states $\rho$ such that $M_{2k+6}(\rho)<\infty$, for all $\sigma,\nu\in\{-,+\}$, and for all $t\in\mathbb R$ we have
\begin{equation}\label{eq:LR-result}
 \|\{\tau_t^H(a^\sigma(f_x)),a^\nu(f_y)\}\rho^{1/2}\|_2
 \le C_k M_{2k+6}(\rho)^{1/2}\frac{e^{c_k|t|}}{\langle x-y\rangle^{k}},
\end{equation}
where $\langle x\rangle:=(1+\norm{x}^2)^{1/2}$.
\end{theorem}
The constants $c_k$ and $C_k$ also depend on $W,r_0$, the choice of cutoff function$ \sqrt{\eta}$ and the $H^{2k+4}$ norms
of $f_x,f_y$. Importantly, they are independent of $L,\rho,t$.

\subsection{Generalization to polynomial observables}
It is possible to generalize Theorem \ref{thm:main} to arbitrary polynomials in creation and annihilation operators.
For open $U\subset\mathbb T_L$ and
$\epsilon\in\{0,1\}$, let
\begin{equation}\label{eq:polynomials}
 \mathfrak P^\epsilon(U)=\operatorname{span}\left\{
 \prod_{i=1}^r a^{\sigma_i}(f_i):
 \begin{array}{l}
 r\in\mathbb N_0,\ r\equiv\epsilon\pmod2,\ \sigma_i\in\{-,+\},\\
 f_i\in C^\infty(\mathbb T_L;\mathbb C),\ \supp f_i\subset U
 \end{array}\right\}\subset\mathcal B(\mathcal F),
\end{equation}
where $a^-=a$, $a^+=a^*$ and an empty product is $1$.
An operator $A\in\mathcal B(\mathcal F)$ has parity $\epsilon$ if
\begin{equation}
 (-1)^{\mathcal N}A(-1)^{\mathcal N}=(-1)^\epsilon A.
\end{equation}
For $A,B\in\mathcal B(\mathcal F)$ of parities $\epsilon,\delta$,
respectively, we introduce the graded commutator
\begin{equation}
 [A,B]_{\rm gr}:=AB-(-1)^{\epsilon\delta}BA,
\end{equation}
which corresponds to the anticommutator in the case in which both $A,B$ are odd, and to the commutator in all other cases.

\begin{theorem}[Extension to polynomial observables]\label{thm:polynomials}
Under the same assumptions as in Theorem \ref{thm:main}, for any $\epsilon,\delta\in\{0,1\}$, for any
$A\in\mathfrak P^\epsilon(B_1(x))$ and
$B\in\mathfrak P^\delta(B_1(y))$, there exist constants $c_k,C_k>0$ such that
\begin{equation}\label{eq:LR-polynomials}
 \|[\tau_t^H(A),B]_{\rm gr}\rho^{1/2}\|_2
 \le C_k M_{2k+6}(\rho)^{1/2}\frac{e^{c_k|t|}}{\langle x-y\rangle^{k}}
\end{equation}
for all $t\in\mathbb R$.
\end{theorem}
The constants $c_k$ and $C_k$ now also depend on the finite polynomial expressions for $A,B$,  and
the $H^{2k+4}$ norms of their test functions.

The proof of \cref{thm:polynomials} is a simple adaptation of the proof of Theorem \ref{thm:main} which we relegate to Appendix~\ref{app:polynomials}.

\subsection{Proof strategy}\label{ssec:strategy}
Our main contribution is to develop a conceptual framework for deriving continuum LRBs through a lattice-based picture. Our goal with this lattice-based picture is to link to the blueprint for deriving LRBs for lattice bosons that has recently emerged. Of course, the continuum poses the fundamentally new challenge that the group velocity of individual electrons is unbounded, but nonetheless we have found extrapolating from the bosonic blueprint to be extremely fruitful in developing the proof. 

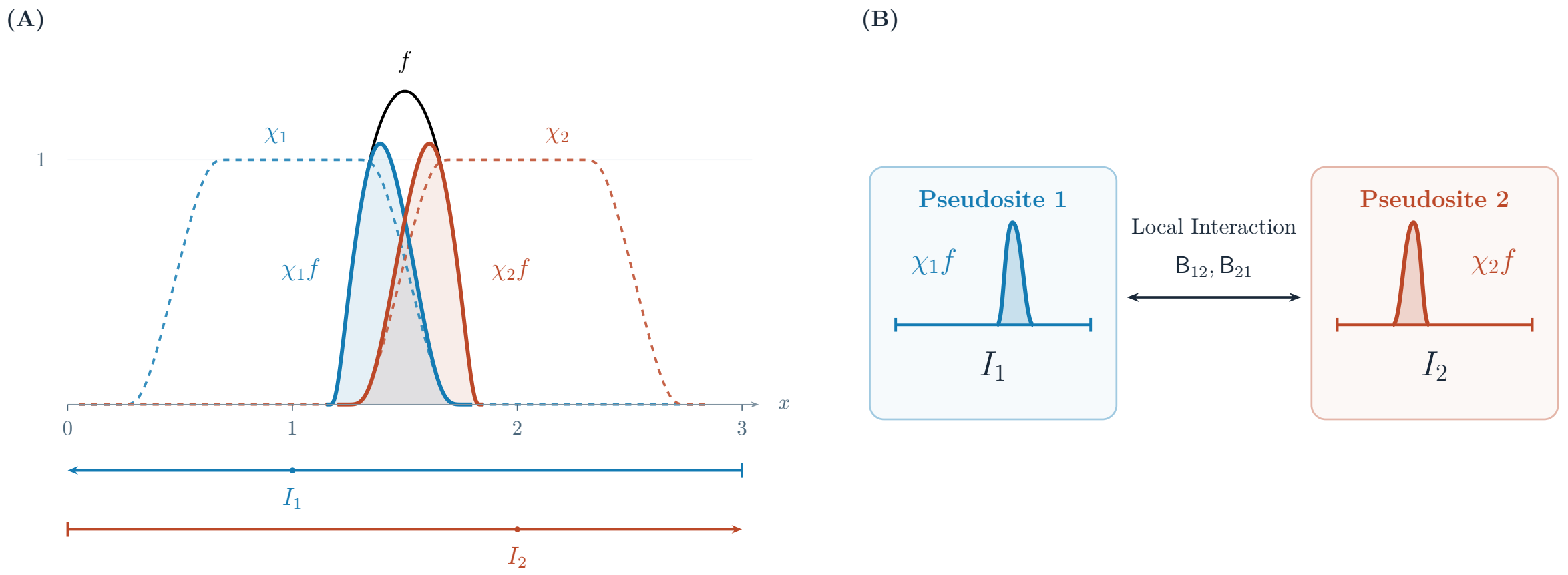
\begin{figure}[H]
\centering
\captionsetup{
  font=small,
  labelfont=bf,
  labelsep=period,
  justification=justified,
  singlelinecheck=false,
  skip=12pt
}

\resizebox{\linewidth}{!}{%
\begin{minipage}{26.7cm}

\begin{minipage}[t]{0.52\linewidth}
\vspace{0pt}
{\color{darkink}\bfseries (A)}\par\vspace{5pt}
\centering
\begin{tikzpicture}
\begin{axis}[
  width=\linewidth,height=11.2cm,
  xmin=0,xmax=3.1875,ymin=-0.74,ymax=1.55,
  domain=0.05:2.85,samples=281,
  axis lines=none,
  clip=false,
]
\draw[draw=lightgrid,line width=0.4pt]
  (axis cs:0,1) -- (axis cs:3.045,1);

\addplot[draw=none,fill=inkblue,fill opacity=0.11,forget plot]
  {chiOne(x)*bumpf(x)}\closedcycle;
\addplot[draw=none,fill=brickred,fill opacity=0.10,forget plot]
  {chiTwo(x)*bumpf(x)}\closedcycle;

\addplot[inkblue!85,dashed,line width=1.2pt,forget plot]
  {chiOne(x)};
\addplot[brickred!85,dashed,line width=1.2pt,forget plot]
  {chiTwo(x)};

\addplot[
  black,line width=1.35pt,
  domain=1.15:1.85,samples=141,forget plot
] {bumpf(x)};
\addplot[
  inkblue,line width=1.9pt,
  domain=1.15:1.80,samples=131,forget plot
] {chiOne(x)*bumpf(x)};
\addplot[
  brickred,line width=1.9pt,
  domain=1.20:1.85,samples=131,forget plot
] {chiTwo(x)*bumpf(x)};

\node[font=\large,text=black]
  at (axis cs:1.5,1.40) {$f$};
\node[font=\large,text=inkblue]
  at (axis cs:0.93,1.10) {$\chi_1$};
\node[font=\large,text=brickred]
  at (axis cs:2.18,1.10) {$\chi_2$};
\node[font=\large,text=inkblue,anchor=east]
  at (axis cs:1.16,0.55) {$\chi_1 f$};
\node[font=\large,text=brickred,anchor=west]
  at (axis cs:1.85,0.55) {$\chi_2 f$};

\draw[
  -{Stealth[length=4pt,width=3pt]},
  draw=mutedtext!70,line width=0.5pt
] (axis cs:0,0) -- (axis cs:3.075,0);

\pgfplotsinvokeforeach{0,1,2,3}{%
  \draw[mutedtext,line width=0.5pt]
    (axis cs:#1,0) -- (axis cs:#1,-0.025);
  \node[anchor=north,font=\small,text=mutedtext]
    at (axis cs:#1,-0.04) {$#1$};
}

\node[anchor=east,font=\small,text=mutedtext]
  at (axis cs:-0.06,1) {$1$};
\node[font=\small,text=mutedtext,anchor=west]
  at (axis cs:3.1275,0) {$x$};

\draw[
  {Stealth[length=5pt,width=4pt]}-,
  inkblue,line width=1.2pt
] (axis cs:0,-0.27) -- (axis cs:3,-0.27);
\draw[inkblue,line width=1.2pt]
  (axis cs:3,-0.24) -- (axis cs:3,-0.30);
\fill[inkblue]
  (axis cs:1,-0.27) circle[radius=1.4pt];
\node[anchor=north,font=\large,text=inkblue]
  at (axis cs:1,-0.31) {$I_1$};

\draw[
  -{Stealth[length=5pt,width=4pt]},
  brickred,line width=1.2pt
] (axis cs:0,-0.51) -- (axis cs:3,-0.51);
\draw[brickred,line width=1.2pt]
  (axis cs:0,-0.48) -- (axis cs:0,-0.54);
\fill[brickred]
  (axis cs:2,-0.51) circle[radius=1.4pt];
\node[anchor=north,font=\large,text=brickred]
  at (axis cs:2,-0.55) {$I_2$};

\end{axis}
\end{tikzpicture}
\end{minipage}\hfill
\begin{minipage}[t]{0.45\linewidth}
\vspace{0pt}
{\color{darkink}\bfseries (B)}\par\vspace{5pt}
\centering

\resizebox{\linewidth}{!}{%
\begin{tikzpicture}[x=1cm,y=1cm,font=\small]
\path[use as bounding box]
  (-0.1,-0.15) rectangle (12.1,8.1);

\draw[
  rounded corners=7pt,
  draw=inkblue!40,fill=inkblue!4,line width=0.9pt
] (0,1.55) rectangle (4.3,5.95);
\draw[
  rounded corners=7pt,
  draw=brickred!40,fill=brickred!4,line width=0.9pt
] (7.7,1.55) rectangle (12.0,5.95);

\node[font=\large\bfseries,text=inkblue]
  at (2.15,5.40) {Pseudosite 1};
\node[font=\large\bfseries,text=brickred]
  at (9.85,5.40) {Pseudosite 2};

\draw[inkblue,line width=1.3pt]
  (0.45,3.20) -- (3.85,3.20);
\draw[inkblue,line width=1.3pt]
  (0.45,3.08) -- (0.45,3.32);
\draw[inkblue,line width=1.3pt]
  (3.85,3.08) -- (3.85,3.32);

\draw[brickred,line width=1.3pt]
  (8.15,3.20) -- (11.55,3.20);
\draw[brickred,line width=1.3pt]
  (8.15,3.08) -- (8.15,3.32);
\draw[brickred,line width=1.3pt]
  (11.55,3.08) -- (11.55,3.32);

\path[fill=inkblue,fill opacity=0.20]
  (2.22,3.20)
  .. controls (2.34,3.20) and (2.35,4.98) .. (2.49,4.98)
  .. controls (2.65,4.98) and (2.70,3.20) .. (2.85,3.20)
  -- cycle;
\draw[inkblue,line width=2.0pt]
  (2.22,3.20)
  .. controls (2.34,3.20) and (2.35,4.98) .. (2.49,4.98)
  .. controls (2.65,4.98) and (2.70,3.20) .. (2.85,3.20);

\path[fill=brickred,fill opacity=0.20]
  (9.12,3.20)
  .. controls (9.27,3.20) and (9.32,4.98) .. (9.48,4.98)
  .. controls (9.62,4.98) and (9.63,3.20) .. (9.75,3.20)
  -- cycle;
\draw[brickred,line width=2.0pt]
  (9.12,3.20)
  .. controls (9.27,3.20) and (9.32,4.98) .. (9.48,4.98)
  .. controls (9.62,4.98) and (9.63,3.20) .. (9.75,3.20);

\node[text=inkblue,font=\Large]
  at (1.12,4.30) {$\chi_1 f$};
\node[text=brickred,font=\Large]
  at (10.88,4.30) {$\chi_2 f$};

\node[text=darkink,font=\LARGE]
  at (2.15,2.48) {$I_1$};
\node[text=darkink,font=\LARGE]
  at (9.85,2.48) {$I_2$};

\draw[
  {Stealth[length=6pt]}-{Stealth[length=6pt]},
  darkink,line width=1.2pt
] (4.48,3.68) -- (7.52,3.68);

\node[text=darkink,font=\normalsize]
  at (6.00,4.92) {Local Interaction};
\node[text=darkink,font=\large]
  at (6.00,4.20) {$\mathsf B_{12},\mathsf B_{21}$};

\end{tikzpicture}%
}
\end{minipage}

\end{minipage}%
}

\caption{\textbf{From the continuum to the pseudolattice.}
Smooth localization gives a lattice of pseudosite labels, with an infinite dimensional Hilbert space at each pseudosite.
\textbf{(A)} Localization on the original line. The black curve is $f$,
the dashed curves are the cutoffs $\chi_1$ and $\chi_2$, and the solid
colored curves are the localized pieces $\chi_1 f$ and $\chi_2 f$.
The intervals $I_1=B_2(1)$ and $I_2=B_2(2)$ are displayed on separate
rows below the axis, with their centers marked at $1$ and $2$.
Only $0\le x\le3$ is shown; the arrows indicate that $I_1$ continues to
the left and $I_2$ to the right.
\textbf{(B)} Each piece is placed on its own pseudosite, corresponding
to $\mathcal L_1 f=(\chi_1 f,\chi_2 f)$ on the two active pseudosites.
The operators $\mathsf B_{12}$ and $\mathsf B_{21}$ represent kinetic coupling between
neighboring pseudosites. The pair potential also gives couplings of
finite range.}
\label{fig:continuum-to-coupled-pseudosites}
\end{figure}

The lattice-based picture is constructed as follows; see Figure \ref{fig:continuum-to-coupled-pseudosites} for  a figure. We decompose space using a quadratic partition of unity. This yields an effective decomposition of space into overlapping integrals, which we think of as ``pseudosites''. Using Fock-space localization techniques, we can  isometrically  map the true continuum Fock space to a pseudolattice picture. Importantly, the effective pseudolattice Hamiltonian is \textit{local}. For the interaction this is clear, especially since we even assume it has finite range for simplicity: it couples a pseudotcite to all of those within the $W$-interaction range. Moreover, the boundary contributions to the kinetic energy are nearest-neighbor interactions between pseudosites and can be calculated by using the IMS localization formula \cite{CFKS} (which also saves a derivative).

Note that a pseudosite interval $I$ still hosts the antisymmetric Fock space over $L^2(I)$. This is of course infinite-dimensional, but the point is that through this Fock-space localization procedure \textit{we have arrived at an effective local lattice model with infinitely many local modes}. The occupancy of these local modes can be framed in terms of local particle number and kinetic energy, and therefore in terms of $M_p$-moments. In particular, initially, the occupancy of high modes is well-controlled by our assumed bound on $M_p$ and Markov's inequality.  \textit{If} we are able to control the dynamical growth of the  $M_p$-moments, then we can now ideas of truncation dynamics developed in the lattice boson context to approximate the effective dynamics in the pseudolattice picture by a kind of lattice fermion system. It turns out, though, that this control poses a major technical obstacle.

Physically, we certainly expect that when starting from an initial state of bounded local  energy moments as we do, then under the dynamics the local moment bounds may grow with time, but not with system size. Heuristically, any accumulation of energy in a local region must come from many and/or fast electrons reaching the region, but the creation of many fast electrons far away is in turn controlled by the growth of the same kind of local moment operator. This suggests that, as usual in dynamical proofs, one should aim to close a certain differential inequality within a suitable family of observables. Implementing this physically intuitive picture mathematically requires finding a family of observables that satisfies the following:
\begin{itemize}
    \item It controls all energy scales locally and high-energy scales far-away, with the energy scales controlled far away growing with the distance.
    \item Their dynamics satisfies a differential inequality that closes within the same family of observables.
\end{itemize}
However, it turns out that many natural candidates of observables, such as the $M_p's$ themselves or other variants like spatially smeared out local kinetic operators, do \textit{not} satisfy both of these requirements, either because they do not suitably link the scales, as required by the first bullet point, or because they do not close under taking commutators with the Hamiltonian, as required by the second bullet point.

Our technical contribution is then to introduce a new class of observables that fulfils both of these requirements. We call these new observables \textit{Momentum-adapted localization observables (MALOs)}; see Definition \ref{def:malo}. 
 First, we prove that MALOs satisfy a Gronwall-type propagation bound Proposition \ref{prop:propagation} and thus grow at most exponentially in time. Second, we show that they are controlled by the $M_p$'s (Proposition \ref{prop:comparison}) which means our assumption on the initial state also provides initial control on the MALOs.

At this stage, we can extend the truncation dynamics technique developed for lattice bosons to the pseudolattice picture to derive an LRB-type operator localization estimate in terms of MALOs (Proposition \ref{prop:truncation}), again in the pseudolattice picture. Combining this with the propagation bound for MALOs proved before and transferring back to the original Fock space picture, this rather easily implies prove Theorem \ref{thm:main}. 

\begin{remark}[Where we use that we are treating 1D fermions]\label{remark:1Dfermions}
    Our general idea of using pseudosites and lattice-boson techniques generalizes directly to continuum fermions and bosons in higher dimensions. However, there is one crucial step in controlling the dynamical growth of the MALO, where we use that we are treating 1D fermions: In Lemma \ref{sec:exclusion}, we    
    use a version of the Pauli exclusion principle to control the $\alpha$th moment of the number operator by the $\alpha$th moment of the local momentum energy. In general dimension $d$, the standard power counting (``filling the Fermi sphere'') shows that the Pauli exclusion principle yields $\alpha\times d$ for the power of momentum and this ``loss'' provides an obstacle to closing the differential inequalities in $d>1$.
\end{remark}

\subsection{Organization of the paper}
In Section \ref{sec:reduction}, we formally introduce our new momentum-adapted localization observables (MALOs) and record their key properties, Propositions \ref{prop:propagation}, \ref{prop:comparison}, and \ref{prop:truncation} mentioned above. They are then combined to prove Theorem \ref{thm:main}, alongside a flowchart that summarizes the proof strategy described above (Figure \ref{fig:continuum-proof-strategy}). 

Sections \ref{sec:kinetic} and  \ref{sec:interaction} are devoted to the proof of Proposition \ref{prop:propagation}. Since time derivatives produce a commutator with the Hamiltonian, we have to control both the commutator of the MALO with the kinetic energy and its commutator with the potential energy. This requires different arguments, which are separated into Sections \ref{sec:kinetic} and  \ref{sec:interaction}.

Section \ref{sec:reduction} contains the proof of Proposition \ref{prop:comparison}, which rests on expressing the kinetic energy as the full energy minus the potential enegy and controlling the potential energy by particle number.

Section \ref{sec:truncation} establishes the pseudolattice picture and, inspired by the lattice boson truncation schemens, prove an LRB provided one has sufficiently dynamical control on the MALO.

Appendix \ref{app:Kato} contains a short argument about the operator domain of the kinetic boundary contribution in the pseudolattice picture and Appendix \ref{app:polynomials} adapts the proof to polynomial observables in Theorem \ref{thm:polynomials}.

\section{Momentum-adapted localization observables (MALOs) and proof of the main result}\label{sec:reduction}

\subsection{Definition of momentum-adapted localization observables (MALOs)}
The quantity we propagate is a positive one-particle form localized near
an integer pseudosite center. We use it to control the error from removing
high pseudosite modes. It must satisfy two competing requirements:
initial local moments must control its sum over all pseudosites, and its
commutator with the interaction must be bounded by the form itself.
The weight defined below combines spatial decay at fixed momentum
with control of high momenta.

We first localize the one-particle space into bounded intervals.
Periodic Fourier modes on these intervals define the weighted form
and will later provide the finite-mode approximation used for locality.

\begin{figure}[!htbp]
\centering
\definecolor{tealgreen}{HTML}{008772}
\definecolor{violet}{HTML}{8053AA}

\captionsetup{
  font=small,
  labelfont=bf,
  labelsep=period,
  justification=justified,
  singlelinecheck=false,
  skip=12pt
}

\begin{tikzpicture}
\begin{axis}[
  width=\linewidth,
  height=0.40\linewidth,
  xmin=-3,xmax=3,
  ymin=-0.02,ymax=1.22,
  axis x line*=bottom,
  axis y line=none,
  axis line style={
    draw=mutedtext!70,
    line width=0.5pt
  },
  xtick={-3,-2,-1,0,1,2,3},
  xticklabels={
    $j-3$,$j-2$,$j-1$,$j$,$j+1$,$j+2$,$j+3$
  },
  ytick={0,1},
  yticklabels=\empty,
  tick align=outside,
  tick style={draw=mutedtext,line width=0.5pt},
  tick label style={font=\small,text=mutedtext},
  xlabel={position $x$},
  xlabel style={font=\small,text=mutedtext},
  clip=false,
]

\draw[draw=lightgrid,line width=0.4pt]
  (axis cs:-3,1) -- (axis cs:3,1);

\draw[violet,line width=1.25pt,dash pattern=on 4pt off 3pt]
  (axis cs:-2.82,0)
  .. controls (axis cs:-2.66,0) and (axis cs:-2.40,1)
  .. (axis cs:-2.24,1)
  -- (axis cs:-1.76,1)
  .. controls (axis cs:-1.60,1) and (axis cs:-1.34,0)
  .. (axis cs:-1.18,0);

\draw[inkblue,line width=1.25pt,dash pattern=on 4pt off 3pt]
  (axis cs:-1.82,0)
  .. controls (axis cs:-1.66,0) and (axis cs:-1.40,1)
  .. (axis cs:-1.24,1)
  -- (axis cs:-0.76,1)
  .. controls (axis cs:-0.60,1) and (axis cs:-0.34,0)
  .. (axis cs:-0.18,0);

\draw[tealgreen,line width=1.25pt,dash pattern=on 4pt off 3pt]
  (axis cs:-0.82,0)
  .. controls (axis cs:-0.66,0) and (axis cs:-0.40,1)
  .. (axis cs:-0.24,1)
  -- (axis cs:0.24,1)
  .. controls (axis cs:0.40,1) and (axis cs:0.66,0)
  .. (axis cs:0.82,0);

\draw[brickred,line width=1.25pt,dash pattern=on 4pt off 3pt]
  (axis cs:0.18,0)
  .. controls (axis cs:0.34,0) and (axis cs:0.60,1)
  .. (axis cs:0.76,1)
  -- (axis cs:1.24,1)
  .. controls (axis cs:1.40,1) and (axis cs:1.66,0)
  .. (axis cs:1.82,0);

\draw[violet,line width=1.25pt,dash pattern=on 4pt off 3pt]
  (axis cs:1.18,0)
  .. controls (axis cs:1.34,0) and (axis cs:1.60,1)
  .. (axis cs:1.76,1)
  -- (axis cs:2.24,1)
  .. controls (axis cs:2.40,1) and (axis cs:2.66,0)
  .. (axis cs:2.82,0);

\draw[
  black,line width=1.7pt,
  line cap=round,line join=round
]
  (axis cs:-3,0) -- (axis cs:-2,0)
  .. controls (axis cs:-1.65,0) and (axis cs:-1.35,1)
  .. (axis cs:-1,1)
  -- (axis cs:1,1)
  .. controls (axis cs:1.35,1) and (axis cs:1.65,0)
  .. (axis cs:2,0)
  -- (axis cs:3,0);

\node[font=\large,text=black]
  at (axis cs:0,1.12) {$\eta_j$};
\node[font=\large,text=violet]
  at (axis cs:-2,0.80) {$\chi_{j-2}$};
\node[font=\large,text=inkblue]
  at (axis cs:-1,0.80) {$\chi_{j-1}$};
\node[font=\large,text=tealgreen]
  at (axis cs:0,0.80) {$\chi_j$};
\node[font=\large,text=brickred]
  at (axis cs:1,0.80) {$\chi_{j+1}$};
\node[font=\large,text=violet]
  at (axis cs:2,0.80) {$\chi_{j+2}$};

\node[anchor=east,font=\small,text=mutedtext,xshift=-4pt]
  at (axis cs:-3,0) {$0$};
\node[anchor=east,font=\small,text=mutedtext,xshift=-4pt]
  at (axis cs:-3,1) {$1$};

\end{axis}
\end{tikzpicture}

\caption{\textbf{The cutoffs.}
The solid black curve is the energy cutoff $\eta_j$ from \eqref{eq:eta}, and the dashed
colored curves are those $\chi_j$ pseudosite cutoff functions \eqref{eq:partition} whose supports overlap it.
The cutoff $\eta_j$ has support $[j-2,j+2]$ and equals $1$ on
$[j-1,j+1]$. The support of $\chi_j$ lies inside this plateau;
the supports of $\chi_{j\pm1}$ lie inside the support of $\eta_j$;
and the supports of $\chi_{j\pm2}$ overlap the support of $\eta_j$
and extend beyond it.}
\label{fig:eta-chi-supports}
\end{figure}
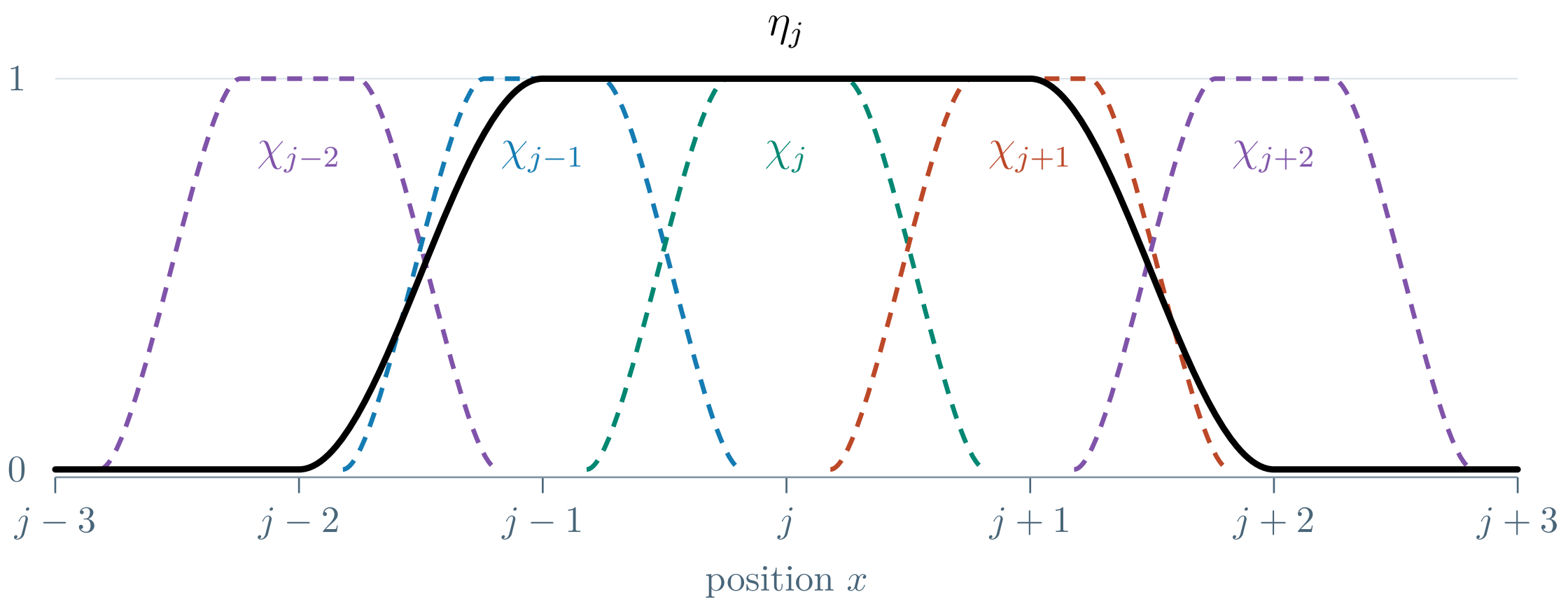

For centers $j\in\{0,\dots,L-1\}$, we define the \emph{$j$th pseudosite} centered at $j$ as the interval
\begin{equation}
I_j=B_2(j),
\end{equation}
and we fix a smooth non-negative quadratic partition of unity
$\{\chi_j\}_{j=0}^{L-1}$ on $\mathbb T_L$ with the following properties:
\begin{gather}
 \chi_j(y)=\chi_0(y-j),\qquad
 \sum_{j=0}^{L-1}\chi_j^2=1,\qquad
 \supp\chi_j\subset B_{7/8}(j).
 \label{eq:partition}
\end{gather}
For every $m\in\mathbb N_0$, we require $\|\partial^m\chi_j\|_\infty\le C_m$, uniformly in $L$ and independently of all model parameters. We refer to \ref{fig:eta-chi-supports} for a depiction of the functions $\chi_j$.

This partition of unity allows us to decompose a one-body state $\psi\in L^2(\mathbb T_L)$ into the collection $(\chi_j\psi)_{j=0}^{L-1}$ of its components in each pseudosite. This consideration motivates the introduction of the following definitions.
\begin{definition}[Pseudolattice picture]
    We define the \emph{$1$-body pseudolattice Hilbert space} $\widetilde{\mathfrak h}$ and the corresponding \emph{pseudolattice Fock space} $\widetilde{\mathcal F}$ as
    \begin{equation}
    \widetilde{\mathfrak h} := \bigoplus_{j=0}^{L-1}L^2(I_j),\qquad \widetilde{\mathcal F}:=\bigoplus_{N\ge0}\widetilde{\mathfrak{h}}^{\wedge N}.
    \end{equation}
    In the following sections, for any object $O$ related to $\mathcal F$, we shall denote its equivalent on $\widetilde{\mathcal F}$ with $\widetilde O$.
\end{definition}
\begin{definition}[Smooth localization isometry]
    The \emph{$1$-body smooth localization isometry} is the map
    \begin{gather}
\mathcal L_{1}: L^2(\mathbb T_L)\to\widetilde{\mathfrak h}\qquad \psi\longmapsto \mathcal L_{1}\psi = (\chi_j \psi)_{j=0,\dots,L-1},
 \end{gather}
 with adjoint
 \begin{equation}
\mathcal L_{1}^*: \widetilde{\mathfrak h}\to L^2(\mathbb T_L)\qquad \widetilde\psi\longmapsto \mathcal L_{1}^*\widetilde\psi = \sum_{j=0}^{L-1} \chi_j \widetilde\psi_j,
 \end{equation}
 where every product is extended by zero to $\mathbb T_L$.
 
  This in turn, by second quantization, induces a map between the respective Fock spaces, which we call the \emph{smooth localization isometry} $\mathcal L$:
 \begin{gather}
\mathcal L:
           \mathcal F\longrightarrow\widetilde{\mathcal F},\qquad \mathcal L:=\Gamma(\mathcal L_1)\equiv\bigoplus_{N\ge0}\mathcal L_{1}^{\otimes N}\label{eq:lift}
 \end{gather}
\end{definition} 
\begin{remark}
    Given $X\subset Y$, unlike the unitary decomposition $L^2(Y)\simeq L^2(X)\oplus L^2(X^c)$, the localization map $\mathcal L_1$ is an isometric embedding whose range is a proper subspace of $\widetilde{\mathfrak h}$:
    \begin{equation}
        \mathcal L_1^* \mathcal L_1 = \mathbf 1_{L^2(\mathbb T_L)}\qquad \mathcal L_1 \mathcal L_1^* = \boldsymbol{1}_{\mathrm{Ran} \mathcal L_1}.
    \end{equation}
    This is due to the overlap between consecutive elements of the partition $\{\chi_j\}_j$. The manifest advantage is that Sobolev regularity is preserved when passing to the pseudosite space.
\end{remark}

For $f\in L^2(\mathbb T_L)$ and $\sigma\in\{-,+\}$,
the localization isometries give
\begin{equation}\label{eq:field-lift}
 \mathcal L^*a^\sigma(\mathcal L_{1}f)\mathcal L= a^\sigma(f),\qquad
 \|a^\sigma(\mathcal L_{1}f)\|=\|f\|_2.
\end{equation}
The single-particle kinetic energy operator in each interval is then defined by using the negative Laplacian with periodic boundary conditions, namely
 \begin{gather}
 T_j=-\partial^2,\quad D(T_j)=H^2_{\rm per}(I_j),\qquad
 T=\bigoplus_{j=0}^{L-1}T_j.\label{eq:T}
\end{gather}
Therefore, the eigenvalues of $T_j$, counted with multiplicity, are of the form
$(\pi n/2)^2$, $n\in\mathbb Z$.
\begin{definition}[Momentum-adapted localization observable (MALO)]\label{def:malo}
    For $s\in\mathbb N$ and a reference pseudosite $j_0\in\{0,\ldots,L-1\}$, we set
    \begin{equation}
 g_{s,j}(\lambda):=\frac{(1+\lambda)^{s+1}}{\langle j\rangle^2+\lambda}
 \quad(j\in\mathbb T_L\cap \mathbb Z,\ \lambda\ge0)
 \end{equation}
 and we define the \emph{momentum-adapted localization observable} (MALO) of order $s$ to be
 \begin{equation}
 A_{s,j_0}:=\mathcal L^*\dG\Bigg(\bigoplus_{j=0}^{L-1} g_{s,j-j_0}(T_j)\Bigg)\mathcal L,
\end{equation}
in the form sense.
\end{definition}
We remark that using the periodic Laplacian is slightly more convenient  for us than the Dirichlet Laplacian, which would also be possible since we are implementing zero-b.c.\ truncations. The slight advantage of the periodic Laplacian is that some calculations that use the explicit eigenbasis are easier for exponentials than for sine functions.

We use the same symbol
$A_{s,j_0}$ also for its associated densely defined, closed, non-negative quadratic form on $\mathcal F$, namely
\begin{equation}\label{eq:Q}
 A_{s,j_0}[u]=\left\|\dG\Bigg(\bigoplus_{j=0}^{L-1} g_{s,j-j_0}(T_j)\Bigg)^{1/2}
                         \mathcal Lu\right\|^2,
\end{equation}
with form domain
\begin{equation}\label{eq:form-core}
 D(A_{s,j_0})=\{u\in\mathcal F:A_{s,j_0}[u]<\infty\} = \overline{\mathscr D}^{\,(\|\cdot\|^2+A_{s,j_0}[\cdot])^{1/2}}
\end{equation}
with $\mathscr D$ being a smooth core, namely
\begin{equation}\label{eq:smooth-domain}
 \mathscr D=\bigoplus_{N\ge0}^{\rm alg}
          (C^\infty(\mathbb T_L^N)\cap\HH_N).
\end{equation}
\begin{remark}
Some remark are in order.
\begin{itemize}
    \item[(i)] We note that the MALO achieves the first requirement described in the proof strategy, that it controls all energies locally (when $j\approx j_0$) and high-energies far away (when $|j-j_0|\gg 1$). The fact that it also satisfies the second requirement (the associated dynamical inequality closes) requires a proof and is the content of Proposition \ref{prop:propagation} below.
    \item[(ii)] We record the elementary inequalities
\begin{equation}\label{eq:scalar-basic}
 \langle j-j_0\rangle^{-2}(1+\lambda)^s\le g_{s,j-j_0}(\lambda)\le(1+\lambda)^s,
 \qquad g_{s,j-j_0}(\lambda)\le\langle j-j_0\rangle^{-2}(1+\lambda)^{s+1}
\end{equation}
 Near $j_0$, $\langle j-j_0\rangle$ is bounded,
so the lower bound recovers local derivatives of order $s$. The last
bound makes the sum over distant intervals convergent at the cost of one
additional derivative. Finally, the upper bound by $(1+\lambda)^s$
controls insertion of smooth field operators.
The first two bounds and the partition also imply that, on each fixed sector, the form norm is equivalent to the $H^s$ norm.
\item [(iii)]     For readers familiar with the ASTLO (adiabatic space time localization observables) method used in the lattice-boson setting in \cite{FLS,FLSa,LRSZ,LRZ,lemm2025quantum}, we remark that the MALOs are a completely different class of observables.
\end{itemize}
\end{remark}

Constants $C$ may change from line
to line. Unless stated otherwise, they depend only on the displayed
orders and on $W,r_0,\sqrt{\eta}$, with the universal localization cutoffs
fixed as above.

\paragraph{Creation and annihilation operators on the pseudosites.}
The local approximation will take values in the algebra of bounded observables over a
set of pseudosites. Therefore, we specify that algebra in terms of localized creation
and annihilation operators before stating the approximation result.
For pseudosite sets $Z\subset\{0,\ldots,L-1\}$, we define
\begin{equation}\label{eq:pseudositespace}    
 \widetilde{\mathfrak{h}}_Z:=\bigoplus_{j\in Z}L^2(I_j),\qquad
 \mathfrak A_Z:=\{a(\widetilde f),a^*(\widetilde f):\widetilde f\in\widetilde{\mathfrak{h}}_Z\}'' = \mathcal B(\mathcal F(\widetilde{\mathfrak{h}}_Z))\otimes1
       \subset\mathcal B(\widetilde{\mathcal F}).
\end{equation}
Here, the auxiliary creation and annihilation
operators are defined as above on the Fock space over $\widetilde{\mathfrak{h}}$. 
We write $\widetilde{\mathcal N}=\dG(1_{\widetilde{\mathfrak{h}}})$ for its total
particle-number operator. Notice that,
if $\supp f\subset B_1(x)$, then $\mathcal L_{1}f\in\widetilde{\mathfrak{h}}_X$, where
\begin{equation}\label{eq:field-support}
 X=\{j:\supp\chi_j\cap B_1(x)\ne\varnothing\}
       \subset\{j:\norm{j-x}<2\},\qquad |X|\le C.
\end{equation}

\subsection{Key facts about MALOs}
The first proposition propagates each form and controls its change under insertion of a smeared creation or annihilation operator. The second bounds the forms $A_{s,j_0}$ by the initial moments at
pseudosite centers. The third
uses these bounds to approximate its evolution by an operator supported near its
initial support. The approximation error yields the locality estimate.

\begin{proposition}[Propagation]\label{prop:propagation}
For any $s\in\mathbb N$, any $j_0\in\{0,\ldots,L-1\}$, and for all 
$t\in\mathbb R$, there exists $C_s$ such that
\begin{equation}\label{eq:propagation}
 e^{itH}A_{s,j_0}e^{-itH}\le e^{C_s|t|}A_{s,j_0},
\end{equation}
in the form sense on $D(A_{s,j_0})$.
\end{proposition}

\begin{proposition}[Initial comparison]\label{prop:comparison}
Let $s\in\mathbb N$, and let $p$ be the smallest even integer
with $p\ge s+1$. Then, there exists $C_s>0$ independent of $L,\rho$, such that
\begin{equation}\label{eq:initial-comparison}
 \max_{0\le j<L}\Tr\rho A_{s,j}\le C_s M_p(\rho).
\end{equation}
\end{proposition}

\begin{proposition}[Approximation by a local observable]\label{prop:truncation}
Let $k\in\mathbb N$, $\sigma\in\{-,+\}$, and
$f\in C^\infty(\mathbb T_L;\mathbb C)$ with $\supp f\subset B_1(x)$.
Let $1\le R\le L/4$.
For every $t\in\mathbb R$, there exists an odd operator
$a_R^\sigma(t,f)\in\mathfrak A_{x[R]}$, where
$x[R]=\{j\in\{0,\ldots,L-1\}:\norm{j-x}\le R\}$, such that
$\|a_R^\sigma(t,f)\|\le\|f\|_2$ and, for every state $\rho$,
\begin{equation}\label{eq:optimized-truncation}
 \|\bigl[\mathcal L\tau_t^H(a^\sigma(f))-a_R^\sigma(t,f)\mathcal L\bigr]\rho^{1/2}\|_2
 \quad\le\frac{C_k\|f\|_{H^{2k+4}}e^{c_k|t|}}{(1+R)^k}
 \left(1+\max_{0\le j<L}\Tr\rho A_{2k+4,j}\right)^{1/2},
\end{equation}
for positive constants $c_k, C_k$ independent of the chosen $\rho, f$.
\end{proposition}


\subsection{Proof of main result from key MALO properties}
\begin{figure}[H]
\centering
\definecolor{tealgreen}{HTML}{008772}

\captionsetup{
  font=small,
  labelfont=bf,
  labelsep=period,
  justification=justified,
  singlelinecheck=false,
  skip=12pt
}

\begin{tikzpicture}[
  proofbox/.style={
    rounded corners=3pt,
    line width=0.65pt,
    inner xsep=7pt,
    inner ysep=7pt,
    align=left,
    text=darkink,
    font=\small
  },
  proofarrow/.style={
    -{Stealth[length=5pt,width=4pt]},
    draw=darkink!85,
    line width=0.8pt
  },
  proofnote/.style={
    font=\footnotesize,
    align=center,
    text=mutedtext,
    inner sep=3pt
  }
]

\node[
  proofbox,
  draw=mutedtext!40,
  fill=darkink!3
] (moments) at (0,0) {
  \begin{tabular}{@{}c@{}}
  \multicolumn{1}{@{}c@{}}{\sffamily\bfseries Initial local moments}\\[5pt]
  $\displaystyle M_{2k+6}(\rho)<\infty$
  \end{tabular}
};

\node[
  proofbox,
  draw=inkblue!60,
  fill=inkblue!4
] (initial) at (8.25,0) {
  \begin{tabular}{@{}c@{}}
  \multicolumn{1}{@{}c@{}}{\sffamily\bfseries\color{inkblue} Initial MALO}\\[5pt]
  $\displaystyle
    \max_{0\le j<L}\operatorname{Tr}\rho A_{s,j}
    \le C_s M_{2k+6}(\rho)$
  \end{tabular}
};

\node[
  proofbox,
  draw=tealgreen!60,
  fill=tealgreen!4
] (propagated) at (8.25,-2.45) {
  \begin{tabular}{@{}c@{}}
  \multicolumn{1}{@{}c@{}}{\sffamily\bfseries\color{tealgreen} Propagated MALO}\\[5pt]
  $\displaystyle
    e^{itH}A_{s,j_0}e^{-itH}
    \le e^{C|t|}A_{s,j_0}$
  \end{tabular}
};

\node[
  proofbox,
  draw=brickred!60,
  fill=brickred!4
] (approximation) at (0,-2.45) {
  \begin{tabular}{@{}c@{}}
  \multicolumn{1}{@{}c@{}}{\sffamily\bfseries\color{brickred} LRB in pseudosite picture}\\[5pt]
  $\displaystyle
  \begin{gathered}
    \bigl\|(\mathcal L\tau_t^H(a^\sigma(f))
      -a_R^\sigma(t,f)\mathcal L)\rho^{1/2}\bigr\|_2\\[3pt]
    \le
      \frac{C\|f\|_{H^{2k+4}}e^{C|t|}}{(1+R)^k}
      M_{2k+6}(\rho)^{1/2}
  \end{gathered}$
  \end{tabular}
};

\draw[proofarrow]
  (moments.east) --
  node[proofnote,midway,above=2pt] {
    Proposition~\ref{prop:comparison}
  }
  (initial.west);

\draw[proofarrow]
  (initial.south) --
  node[proofnote,midway,right=4pt] {
    Proposition~\ref{prop:propagation}
  }
  (propagated.north);

\draw[proofarrow]
  (propagated.west) --
  node[proofnote,midway,above=2pt] {
    Proposition~\ref{prop:truncation}
  }
  (approximation.east);

\node[
  proofbox,
  draw=darkink!65,
  fill=darkink!3
] (conclusion) at (3.45,-5.6) {
  \begin{tabular}{@{}c@{}}
  \multicolumn{1}{@{}c@{}}{\sffamily\bfseries
    LRB in original picture 
  }\\[5pt]
  $\displaystyle
    \bigl\|
      \{\tau_t^H(a^\sigma(f_x)),a^\nu(f_y)\}
      \rho^{1/2}
    \bigr\|_2
    \le C_k M_{2k+6}(\rho)^{1/2}
       \frac{e^{c_k|t|}}{\langle x-y\rangle^k}$
  \end{tabular}
};

\draw[proofarrow]
  (approximation.south) --
  node[proofnote,midway,right=4pt,align=left] {
  Theorem \ref{thm:main}
  }
  (approximation.south |- conclusion.north);

\end{tikzpicture}

\caption[Proof strategy for the continuum locality bound]{%
\textbf{Proof strategy for the continuum locality bound.}
Under the hypotheses of Theorem~\ref{thm:main}, take $s=2k+4$
and $p=2k+6$. The arrows follow the order in which the estimates
are applied. Proposition~\ref{prop:comparison} bounds the initial
MALO expectation by the local moments,
Proposition~\ref{prop:propagation} controls its evolution, and
Proposition~\ref{prop:truncation}, combined with the initial comparison,
gives the displayed local approximation for $1\le R\le L/4$. Together with the operator-insertion 
estimate, these estimates yield the locality
bound of Theorem~\ref{thm:main},
as proved in Section~\ref{sec:reduction}.}
\label{fig:continuum-proof-strategy}
\end{figure}
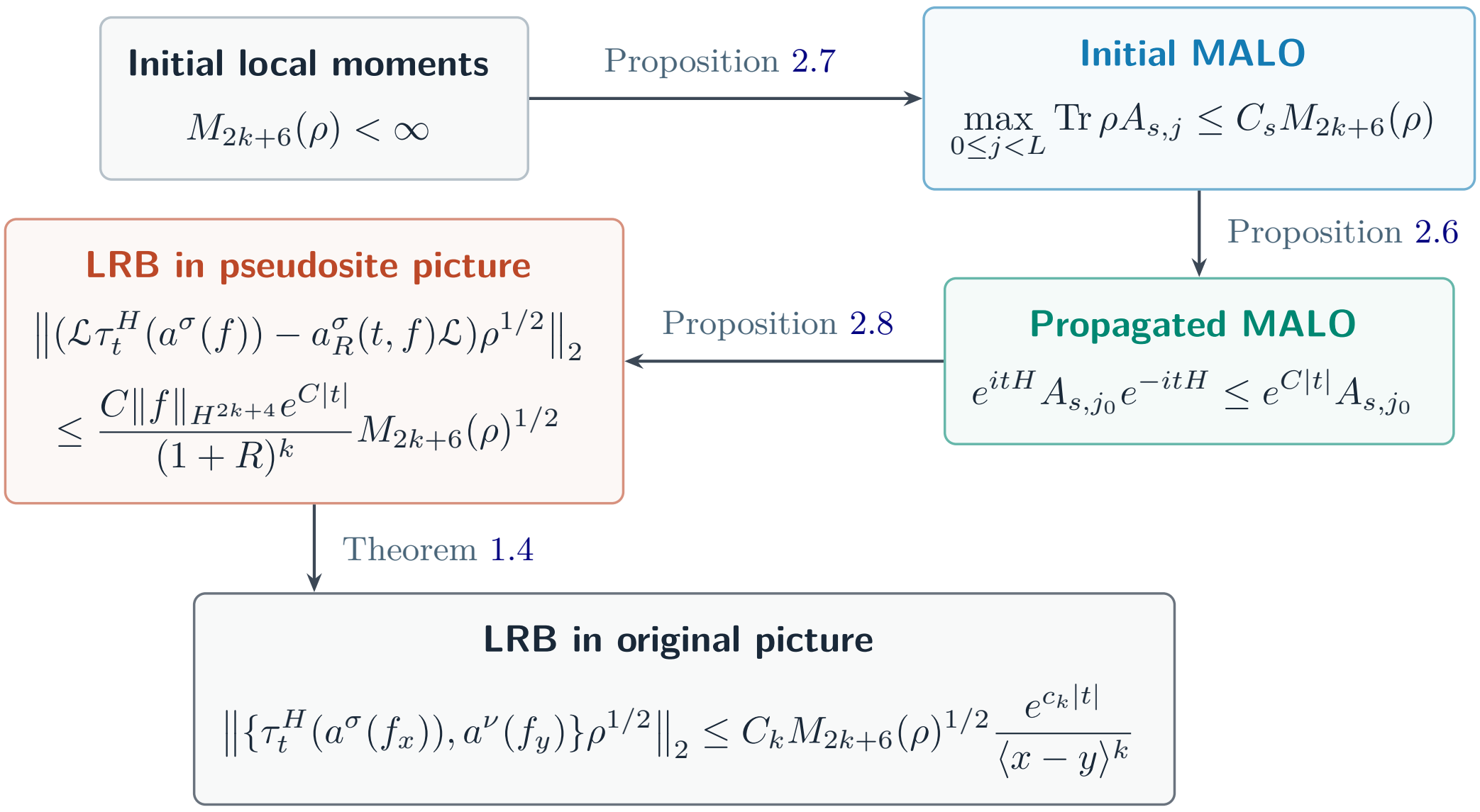

We begin by summarizing the proof strategy, which was already described in words in Subsection \ref{ssec:strategy} in the following flowchart, \ref{fig:continuum-proof-strategy}.

\begin{proof}[Proof of \cref{thm:main}]
We apply the local approximation at a radius proportional to the
separation of the supports. Set $s=2k+4$. The initial comparison
and the insertion estimate in \cref{prop:comparison,prop:propagation}, together with \eqref{eq:insertion}, give
\begin{equation}
 \begin{gathered}
 1+\max_{0\le j<L}\Tr\rho A_{s,j}\le C M_{2k+6}(\rho),\\
 \Tr a^\nu(f_y)\rho(a^\nu(f_y))^*
 +\max_{0\le j<L}\Tr a^\nu(f_y)\rho(a^\nu(f_y))^*A_{s,j}
 \le C\|f_y\|_{H^s}^2 M_{2k+6}(\rho).
 \end{gathered}
\end{equation}
For sufficiently large
$d=\norm{x-y}$, take $R=d/3\le L/6$. Since the supports of $a_R^\sigma(t,f_x)$
and $a^\nu(\mathcal L_{1}f_y)$ are disjoint by \eqref{eq:field-support}, their
anticommutator vanishes. Put
\begin{equation}
 D=\mathcal L\tau_t^H(a^\sigma(f_x))-a_R^\sigma(t,f_x)\mathcal L.
\end{equation}
The identities \eqref{eq:field-lift} give
\begin{equation}
 \mathcal L\{\tau_t^H(a^\sigma(f_x)),a^\nu(f_y)\}
 =Da^\nu(f_y)+a^\nu(\mathcal L_{1}f_y)D.
\end{equation}
Apply \eqref{eq:optimized-truncation} with $\rho$ and, when
$\beta:=\Tr a^\nu(f_y)\rho(a^\nu(f_y))^*>0$, with the state
$\rho_{f_y}:=\beta^{-1}a^\nu(f_y)\rho(a^\nu(f_y))^*$. The identity
\begin{equation}
 \|Da^\nu(f_y)\rho^{1/2}\|_2^2
 =\Tr a^\nu(f_y)\rho(a^\nu(f_y))^*D^*D
 =\beta\|D\rho_{f_y}^{1/2}\|_2^2
\end{equation}
and the preceding bounds yield
\begin{equation}
 \begin{aligned}
 \|\{\tau_t^H(a^\sigma(f_x)),a^\nu(f_y)\}\rho^{1/2}\|_2
 &\le\|Da^\nu(f_y)\rho^{1/2}\|_2+\|f_y\|_2\|D\rho^{1/2}\|_2\\
 &\le\frac{C\|f_x\|_{H^s}\|f_y\|_{H^s}e^{C|t|}}{(1+R)^k}
                   M_{2k+6}(\rho)^{1/2}.
 \end{aligned}
\end{equation}
Since $R=d/3$, this proves \eqref{eq:LR-result} for large $d$.
For small $d$, we use the trivial bound
\[
\|\{\tau_t^H(a^\sigma(f_x)),a^\nu(f_y)\}\|\le2\|f_x\|_2\|f_y\|_2
\]
to conclude the proof.
\end{proof}


Sections~\ref{sec:kinetic}--\ref{sec:truncation}
prove the three propositions in order. The propagation rests on the kinetic and
interaction commutator estimates, and the initial comparison rests on a product-rule comparison of kinetic and local-energy powers. For the local approximation, we
first remove high pseudosite modes and then remove boundary couplings.
Duhamel's formula places the unbounded truncation error on the exact
physical evolution, where the propagated form $A_{s,j_0}$ controls it.

\section{Propagation estimates: kinetic contribution}\label{sec:kinetic}
In this section we prove \cref{prop:propagation}. The propagated form
controls the mode-truncation error, and the insertion estimate allows
us to apply this control after a creation or annihilation operator acts. Fix $s\in\mathbb N$ and
a reference pseudosite $j_0\in\{0,\ldots,L-1\}$.

For $u\in\mathscr D$, we set $u_t:=e^{-itH}u$. Since $H$ preserves
particle number and $W$ is smooth and bounded, $u_t\in\mathscr D$.
On each of the finitely many sectors present in $u$, the map
$t\mapsto u_t$ is continuously differentiable in every Sobolev norm,
with derivative $-iHu_t$.

Since, on each fixed particle number sector, $A_{s,j_0}$ is bounded by a
(sector-dependent) multiple of the squared $H^s$ norm, $t\mapsto A_{s,j_0}[u_t]$ is differentiable and
\begin{equation}
 \frac{\dd}{\dd t}A_{s,j_0}[u_t]
 =\langle u_t,i[H,A_{s,j_0}]u_t\rangle.
\end{equation}
Writing $H=K+V$, with
\begin{equation}
 K=\dG(-\partial^2),\qquad
 V=\bigoplus_{n\ge0}\sum_{1\le i<j\le n}W(x_i-x_j),
\end{equation}
we obtain
\begin{equation}\label{eq:Q-time-derivative}
 \left|\frac{\dd}{\dd t}A_{s,j_0}[u_t]\right|
 \le |\langle u_t,[A_{s,j_0},K]u_t\rangle|
       +|\langle u_t,[A_{s,j_0},V]u_t\rangle|.
\end{equation}
We bound the first term in \eqref{eq:Q-time-derivative} by
$C_sA_{s,j_0}[u_t]$ in
\cref{lem:prop-kinetic}. We postpone an analogous bound on the second term, which depends on the two-body interaction, to \cref{sec:interaction}.
\begin{proposition}[Kinetic contribution]\label{lem:prop-kinetic}
For $u\in\mathscr D$,
$|\langle u,[A_{s,j_0},K]u\rangle|\le C_sA_{s,j_0}[u]$.
\end{proposition}

\subsection{Key estimates for the kinetic contribution}\label{app:ims}
This subsection proves two key estimates used in the proof of \cref{lem:prop-kinetic}.

Since all intervals in the pseudosite decomposition have length $4$, in this subsection we work on
$I=\mathbb R/(4\mathbb Z)$.

Let $\mathfrak g$ be a complex Hilbert space and let
$T=-\partial^2$ be the nonnegative selfadjoint operator on
$L^2(I;\mathfrak g)$ with $D(T)=H^2(I;\mathfrak g)$.

\begin{lemma}[Multiplication estimate]\label{lem:weighted-multiplication}
For any $j\in\mathbb R$, and for any $v\in C^\infty(I;\mathbb R)$ and
$w\in C^\infty(I;\mathfrak g)$,
\begin{equation}\label{eq:prop-multiplication}
 \begin{split}
 |\langle w,[g_{s,j}(T),v]w\rangle|
 \le C_s\|v\|_{C^{s+2}(I)}
       \|g_{s,j}(T)^{1/2}w\|
       \left\|\left(\frac{g_{s,j}(T)}{1+T}\right)^{1/2}w\right\|,
 \end{split}
\end{equation}
for some constant $C_s$ independent of $j,\mathfrak g$.
\end{lemma}

\begin{lemma}[Paired weighted estimate]\label{lem:weighted-first-order}
Given $a\in C^\infty(I;\mathbb R)$, define
$\mathsf B:H^1(I;\mathfrak g)\to L^2(I;\mathfrak g)$ by
\[
\mathsf Bf:=a\partial f+\frac{a'}{2}f.
\]
Let $j,k\in\mathbb R$ be such that $|\langle j\rangle-\langle k\rangle|\le1$; then, for all
$f,g\in C^\infty(I;\mathfrak g)$,
\begin{equation}\label{eq:weighted-first-order}
 \begin{split}
 |\langle f,(g_{s,j}(T)\mathsf B-\mathsf Bg_{s,k}(T))g\rangle|
 \le C\|a\|_{C^{s+3}(I)}
       \|g_{s,j}(T)^{1/2}f\|\,\|g_{s,k}(T)^{1/2}g\|.
 \end{split}
\end{equation}
for some constant $C_s$ independent of $j,k$, and $\mathfrak g$.
\end{lemma}

\begin{proof}[Proof of the two lemmas]
The main estimate is
\begin{equation}\label{eq:weighted-root-commutator}
 \|[g_{s,j}(T)^{1/2},v]w\|
 \le C_s\|v\|_{C^{s+2}(I)}
       \left\|\left(\frac{g_{s,j}(T)}{1+T}\right)^{1/2}w\right\|,
\end{equation}
for $v\in C^\infty(I;\mathbb C)$ and
$w\in C^\infty(I;\mathfrak g)$, uniformly in $b$ and $\mathfrak g$.
We prove it in the first two steps, and in the third
step we prove the two lemmas.
For simplicity of notation, we employ the shorthand $b:=\langle j\rangle$ and $c:=\langle k\rangle$.

\emph{Step 1: Differences of square roots of the weights.}
For $p,q\in\mathbb R$, put
$t=(1+p^2)^{1/2}$ and $u=(1+q^2)^{1/2}$.
Then $|t-u|\le|p-q|$ and $t,u\ge1$.
If $t\ge u$, then
\begin{equation}
 \left(\frac{t}{u}\right)^s
 \le\left(\frac{g_{s,j}(p^2)}{g_{s,j}(q^2)}\right)^{1/2}
 =\left(\frac{t}{u}\right)^{s+1}
   \left(\frac{b^2-1+u^2}{b^2-1+t^2}\right)^{1/2}
 \le\left(\frac{t}{u}\right)^{s+1}.
\end{equation}
For the lower bound, the last square root is at least $u/t$,
because $b^2-1\ge0$. For the upper bound, this square root is at most one.
In particular, the weight increases with $t$. The finite geometric sum
therefore gives, when $t\ge u$,
\begin{equation}
 \begin{aligned}
 \frac{|g_{s,j}(p^2)^{1/2}-g_{s,j}(q^2)^{1/2}|}
      {g_{s,j}(q^2)^{1/2}}
 &\le\left(\frac tu\right)^{s+1}-1\\
 &=\frac{t-u}{u}\sum_{\ell=0}^s\left(\frac tu\right)^\ell
 \le(s+1)\frac{|p-q|}{u}
                     (1+|p-q|)^s.
 \end{aligned}
\end{equation}
When $t\le u$, the same ratio bound with the momenta exchanged gives
\begin{equation}
 \left(\frac tu\right)^{s+1}
 \le\left(\frac{g_{s,j}(p^2)}{g_{s,j}(q^2)}\right)^{1/2}\le1.
\end{equation}
The normalized difference is then at most
$1-(t/u)^{s+1}\le(s+1)(u-t)/u$.
Thus, in both cases,
\begin{equation}\label{eq:fourier-root-difference}
 \begin{split}
 |g_{s,j}(p^2)^{1/2}-g_{s,j}(q^2)^{1/2}|
 &\le C|p-q|(1+|p-q|)^s
       \left(\frac{g_{s,j}(q^2)}{1+q^2}\right)^{1/2}.
 \end{split}
\end{equation}

\emph{Step 2: The square-root commutator.}
Let $p_n=\pi n/2$ and $e_n(x)=e^{ip_nx}/2$ for $n\in\mathbb Z$.
These are the orthonormal Fourier modes in $L^2(I)$.
For $v\in C^\infty(I;\mathbb C)$, define
\begin{equation}
 \widehat v_l=\frac14\int_Iv(x)e^{-ip_lx}\,\dd x,
 \qquad l\in\mathbb Z.
\end{equation}
Multiplication by $v$ has matrix $\widehat v_{n-m}$ in this basis.
For every integer $m\ge0$, Cauchy--Schwarz and Parseval give
\begin{equation}\label{eq:fourier-weighted-coefficients}
 \begin{aligned}
 \sum_l(1+|l|)^m|\widehat v_l|
 &\le\left(\sum_l(1+|l|)^{-2}\right)^{1/2}
       \left(\sum_l(1+|l|)^{2m+2}|\widehat v_l|^2\right)^{1/2}\\
 &\le C_m\left(\sum_{r=0}^{m+1}
                         \|\partial^rv\|_{L^2(I)}^2\right)^{1/2}
 \le C_m\|v\|_{C^{m+1}(I)}.
 \end{aligned}
\end{equation}
For the second line, we use
$(1+|l|)^{2m+2}\le C_m\sum_{r=0}^{m+1}|p_l|^{2r}$ and
$\sum_l|p_l|^{2r}|\widehat v_l|^2
=\|\partial^rv\|_{L^2(I)}^2/4$.

We first take $w$ to be a finite Fourier series, with coefficients
$w_n=\int_I\overline{e_n(x)}w(x)\,\dd x\in\mathfrak g$.
The Fourier coefficients of the commutator are
\begin{equation}
 ([g_{s,j}(T)^{1/2},v]w)_n
 =\sum_m\bigl(g_{s,j}(p_n^2)^{1/2}-g_{s,j}(p_m^2)^{1/2}\bigr)
       \widehat v_{n-m}w_m.
\end{equation}
Apply \eqref{eq:fourier-root-difference} and the triangle inequality in
$\ell^2(\mathbb Z;\mathfrak g)$, summing over the diagonals $n-m=l$.
Parseval's identity and \eqref{eq:fourier-weighted-coefficients}, with
$m=s+1$, give
\begin{equation}
 \begin{aligned}
 \|[g_{s,j}(T)^{1/2},v]w\|
 &\le C\sum_l(1+|l|)^{s+1}|\widehat v_l|
       \left(\sum_m\frac{g_{s,j}(p_m^2)}{1+p_m^2}
                         \|w_m\|_{\mathfrak g}^2\right)^{1/2}\\
 &\le C\|v\|_{C^{s+2}(I)}
       \left\|\left(\frac{g_{s,j}(T)}{1+T}\right)^{1/2}w\right\|.
 \end{aligned}
\end{equation}
Fourier truncations of a smooth vector converge in every Sobolev norm.
Since $g_{s,j}(p^2)^{1/2}\le(1+p^2)^{s/2}$ and multiplication by
$v$ preserves these spaces, passing to the limit proves
\eqref{eq:weighted-root-commutator} for all smooth $w$.

\emph{Step 3: Deduction of the two estimates.}
For real $v$, multiplication by $v$ is selfadjoint. Hence
\begin{equation}
 \begin{aligned}
 &|\langle w,[g_{s,j}(T),v]w\rangle|\\
 &\quad=\bigl|\langle g_{s,j}(T)^{1/2}w,
                        [g_{s,j}(T)^{1/2},v]w\rangle
       -\langle[g_{s,j}(T)^{1/2},v]w,
                        g_{s,j}(T)^{1/2}w\rangle\bigr|\\
 &\quad\le2\|g_{s,j}(T)^{1/2}w\|\,
                         \|[g_{s,j}(T)^{1/2},v]w\|.
 \end{aligned}
\end{equation}
Estimate \eqref{eq:weighted-root-commutator} proves
\cref{lem:weighted-multiplication}.

For the paired estimate, assume $|b-c|\le1$. Since $b,c\ge1$, we have
$b/2\le c\le2b$, and hence, for $p\in\mathbb R$,
\begin{equation}
 \begin{gathered}
 \frac12\le
 \left(\frac{g_{s,j}(p^2)}{g_{s,k}(p^2)}\right)^{1/2}\le2,\\
 \frac{|g_{s,j}(p^2)^{1/2}-g_{s,k}(p^2)^{1/2}|}
      {g_{s,k}(p^2)^{1/2}}
 =\frac{|\sqrt{c^2+p^2}-\sqrt{b^2+p^2}|}{\sqrt{b^2+p^2}}
 \le\frac1{\sqrt{1+p^2}}.
 \end{gathered}
\end{equation}
where in the last inequality we used that $b\mapsto\sqrt{b^2+p^2}$ is
$1$-Lipschitz. Therefore, for smooth $g$, we have
\begin{equation}\label{eq:root-parameter-comparison}
 \begin{gathered}
 \|g_{s,j}(T)^{1/2}g\|\le2\|g_{s,k}(T)^{1/2}g\|,\\
 \|(1+T)^{1/2}(g_{s,j}(T)^{1/2}-g_{s,k}(T)^{1/2})g\|
 \le\|g_{s,k}(T)^{1/2}g\|.
 \end{gathered}
\end{equation}
Since the Fourier multipliers commute with $\partial$, expansion gives
\begin{equation}
 \begin{aligned}
 g_{s,j}(T)^{1/2}\mathsf B-\mathsf Bg_{s,k}(T)^{1/2}
 &=[g_{s,j}(T)^{1/2},a]\partial
       +\tfrac12[g_{s,j}(T)^{1/2},a']\\
 &\quad+\mathsf B(g_{s,j}(T)^{1/2}-g_{s,k}(T)^{1/2}).
 \end{aligned}
\end{equation}
The first two terms are bounded by
\eqref{eq:weighted-root-commutator} and
\eqref{eq:root-parameter-comparison}, using
$\|(1+T)^{-1/2}\|\le1$ and $\|\partial(1+T)^{-1/2}\|\le1$.
For the last term, use
$\|\mathsf Bh\|\le C\|a\|_{C^1(I)}\|(1+T)^{1/2}h\|$ and the second bound
in \eqref{eq:root-parameter-comparison}. We obtain
\begin{equation}\label{eq:root-first-order-bound}
 \|(g_{s,j}(T)^{1/2}\mathsf B-\mathsf Bg_{s,k}(T)^{1/2})g\|
 \le C\|a\|_{C^{s+3}(I)}\|g_{s,k}(T)^{1/2}g\|.
\end{equation}
The same estimate holds with $j,k$ exchanged. All multipliers preserve
smoothness, so integration by parts is valid in the following identity:
\begin{equation}
 \begin{aligned}
 \langle f,(g_{s,j}(T)\mathsf B-\mathsf Bg_{s,k}(T))g\rangle &=\langle g_{s,j}(T)^{1/2}f,
       (g_{s,j}(T)^{1/2}\mathsf B-\mathsf Bg_{s,k}(T)^{1/2})g\rangle\\
 &\quad+\langle(g_{s,k}(T)^{1/2}\mathsf B-\mathsf Bg_{s,j}(T)^{1/2})f,
                         g_{s,k}(T)^{1/2}g\rangle.
 \end{aligned}
\end{equation}
Indeed, the two additional terms cancel because
$\langle \mathsf Bh_1,h_2\rangle=-\langle h_1,\mathsf Bh_2\rangle$ for smooth periodic
$h_1,h_2$, as $a$ is real-valued. Cauchy--Schwarz and
\eqref{eq:root-first-order-bound} now prove
\cref{lem:weighted-first-order}.
\end{proof}

\subsection{The kinetic contribution}
Define the real coefficients
\begin{equation}\label{eq:ims-coefficients}
 a_{jk}=\chi_j'\chi_k-\chi_j\chi_k',\qquad
 S=\sum_k|\chi_k'|^2 .
\end{equation}
Moreover, $a_{kj}=-a_{jk}$, and $a_{jk}=0$ unless the supports of
$\chi_j$ and $\chi_k$ overlap, which forces $\norm{j-k}\le1$.
Notice that $S$ is $1$-periodic on $\mathbb T_L$,
and its restriction to each interval is a smooth multiplication operator.

\begin{proof}[Proof of \cref{lem:prop-kinetic}]
Since the kinetic energy $K$ is translation invariant, we have
\[
[A_{s,j_0},K]=e^{-j_0\partial}[A_{s,0},K]e^{j_0\partial}
\]
Therefore, since $\mathscr D$ is an invariant set under translations,
we may, without loss of generality, set $j_0=0$ for the rest of the proof.

Secondly, the operator $A_{s,0}$ does not connect different particle number sectors, therefore it suffices to prove the claim on each sector separately.

We fix $N\ge1$ and consider $u\in C^\infty(\mathbb T_L^N)\cap\HH_N$.
The definition of $A_{s,0}$ gives
\begin{equation}\label{eq:Q-coordinates}
 \langle u,[A_{s,0},K]u\rangle
 =\sum_{i=1}^N\sum_{j=0}^{L-1}
 \langle u,[\chi_j(x_i)g_{s,j}(T_j)\chi_j(x_i),K_N]u\rangle,
\end{equation}
where $g_{s,j}(T_j)$ acts in coordinate $x_i$.
Since $K_N=-\sum_{m=1}^N\partial_m^2$, and operators in different
coordinates commute,
\begin{equation}
 [\chi_j(x_i)g_{s,j}(T_j)\chi_j(x_i),-\partial_m^2]=0
 \qquad(m\ne i).
\end{equation}
We are therefore left with
\begin{equation}
 \langle u,[A_{s,0},K]u\rangle
 =\sum_{i=1}^N\sum_j
 \langle u,[\chi_j(x_i)g_{s,j}(T_j)\chi_j(x_i),-\partial_i^2]u\rangle.
\end{equation}

\smallskip\noindent
\emph{Step 1.} Fix $i$ and regard $u$ as a smooth function of $x_i$ with values in
$\mathfrak g=L^2(\mathbb T_L^{N-1})$. We write primes for derivatives
in $x_i$ and set
$w_j=\chi_j u\in C^\infty_{\rm per}(I_j;\mathfrak g)$.

The product rule and the partition identities
$\sum_k\chi_k^2=1$, $\sum_k\chi_k\chi_k'=0$
underlying IMS \cite[Section~3.1]{CFKS} give
\begin{equation}\label{eq:direct-ims}
 \chi_j(-u'')=T_jw_j+Sw_j+
                \sum_k\mathsf B_{jk}w_k,
\end{equation}
with $\mathsf B_{jk}=2a_{jk}\partial+a_{jk}'$, as one may check by differentiating $\sum_k\chi_k^2=1$ twice.

\smallskip\noindent
\emph{Step 2.} We estimate the terms in \eqref{eq:direct-ims} using \cref{app:ims}.
The $T_j$ terms cancel because $T_j$ commutes with its Fourier multipliers.
For the $S$ contribution,
\eqref{eq:prop-multiplication} with $v=S$ and
$(1+T_j)^{-1}\le1$ yields
\begin{equation}\label{eq:ims-S-bound}
 |\langle w_j,[g_{s,j}(T_j),S]w_j\rangle|
 \le C_s\|g_{s,j}(T_j)^{1/2}w_j\|^2.
\end{equation}
It remains to estimate the first-order terms\footnote{For an overlapping pair choose integer representatives of its pseudosite
labels modulo $L$ with $k-j\in\{-1,0,1\}$.
The map
\begin{equation}
 f(r)\longmapsto f(r-(k-j)),\qquad
 f\in L^2(\mathbb R/(4\mathbb Z);\mathfrak g),
\end{equation}
expresses a pseudosite-$k$ function in the pseudosite-$j$ coordinate. It is unitary
and commutes with differentiation and every Fourier multiplier used. Both localized supports lie in the common interval interior.
All paired pseudosite terms below, including Fourier-multiplier outputs, are understood
under this identification.}.
Integration by parts gives
$\langle\mathsf B_{jk}w_k,w_j\rangle
=\langle w_k,\mathsf B_{kj}w_j\rangle$.
Using \eqref{eq:direct-ims} in both terms of the commutator and
reindexing the second first-order sum therefore gives
\begin{equation}\label{eq:ims-B}
 \begin{aligned}
 \sum_j\langle u,[\chi_j g_{s,j}(T_j)\chi_j,-\partial^2]u\rangle
=&\sum_j\langle w_j,[g_{s,j}(T_j),S]w_j\rangle\\&+\sum_{j,k}\langle w_j,
       (g_{s,j}(T_j)\mathsf B_{jk}
                  -\mathsf B_{jk}g_{s,k}(T_k))w_k\rangle.
 \end{aligned}
\end{equation}
Since $|\langle j\rangle-\langle k\rangle|\le1$ for overlapping intervals,
we may apply \cref{lem:weighted-first-order}, with $f=w_j$, $g=w_k$,
$a=2a_{jk}$, and $b=\langle j\rangle$, $c=\langle k\rangle$.
Together with the uniform bounds on
the derivatives of $a_{jk}$, it yields
\begin{equation}\label{eq:ims-fixed-pair-bound}
 \bigl|\langle w_j,(g_{s,j}(T_j)\mathsf B_{jk}
                 -\mathsf B_{jk}g_{s,k}(T_k))w_k\rangle\bigr|
                 \le C_s\|g_{s,j}(T_j)^{1/2}w_j\|
                     \|g_{s,k}(T_k)^{1/2}w_k\|.
\end{equation}
Summing \eqref{eq:ims-fixed-pair-bound} over the ordered pairs,
and using that only neighbouring intervals contribute, gives
\begin{equation}\label{eq:ims-pair-bound}
 \begin{aligned}
 \left|\sum_{j,k}\langle w_j,(g_{s,j}(T_j)\mathsf B_{jk}
                 -\mathsf B_{jk}g_{s,k}(T_k))w_k\rangle\right|
 &\le C_s\sum_j\|g_{s,j}(T_j)^{1/2}w_j\|
               \|g_{s,j+1}(T_{j+1})^{1/2}w_{j+1}\|\\
 &\le\frac{C_s}2\sum_j\bigl(
       \|g_{s,j}(T_j)^{1/2}w_j\|^2
       +\|g_{s,j+1}(T_{j+1})^{1/2}w_{j+1}\|^2\bigr)\\
 &=C_s\sum_j\|g_{s,j}(T_j)^{1/2}w_j\|^2,
 \end{aligned}
\end{equation}
with the understanding that the summations are taken modulo $L$.
Combining \eqref{eq:ims-pair-bound} with \eqref{eq:ims-S-bound} in
\eqref{eq:ims-B}, we obtain
\begin{equation}
 \left|\sum_j\langle u,
       [\chi_j g_{s,j}(T_j)\chi_j,-\partial^2]u\rangle\right|
 \le C_s\sum_j\|g_{s,j}(T_j)^{1/2}w_j\|^2.
\end{equation}

Finally, summing over $i$ produces
\begin{equation}
 \begin{aligned}
 |\langle u,[A_{s,0},K]u\rangle|
 \le C_s\sum_{i=1}^N\sum_j
       \|g_{s,j}(T_j)^{1/2}(\chi_j(x_i)u)\|^2=C_sA_{s,0}[u].
 \end{aligned}
\end{equation}
Orthogonal summation proves the
assertion for general $u\in\mathscr D$.
\end{proof}

\section{Propagation estimates: interaction contribution}\label{sec:interaction}
In this section we prove the following bound on the contribution to the time derivative of the MALO, coming from the two-body interaction. At the end of this section we also provide the proof of \cref{prop:propagation}.
\begin{proposition}[Interaction contribution]\label{lem:prop-interaction}
For $u\in\mathscr D$,
$|\langle u,[A_{s,j_0},V]u\rangle|\le C_sA_{s,j_0}[u]$.
\end{proposition}
\subsection{Exclusion estimate in the pseudosite modes}\label{sec:exclusion}
Before moving on to prove \cref{lem:prop-interaction}, we first prove the following lemma, which will play an important role in the proof of \cref{lem:prop-interaction}.
\begin{lemma}[Exclusion on pseudosite sets]\label{lem:exclusion}
Let $\varnothing\ne\mathcal C\subset\{0,\ldots,L-1\}$ be a set of pseudosites, and let 
$\widetilde{\mathcal N}_{\mathcal C}$ denote the number operator on $\widetilde{\mathfrak h}_\mathcal C$.

For any $\alpha\ge0$ and every nonnegative nondecreasing Borel function
$f:[0,\infty)\to[0,\infty)$,
\begin{equation}\label{eq:master-exclusion}
 \widetilde{\mathcal N}_{\mathcal C}^{\alpha}\dG(f(T_{\mathcal C}))
 \le C_{\alpha,|\mathcal C|}\dG((1+T_{\mathcal C})^{\alpha/2}f(T_{\mathcal C})).
\end{equation}
in the sense of spectral forms.
\end{lemma}

\begin{proof}
We first count the one-particle modes, then use their occupation numbers
to absorb the particle-number factor into the derivative weight.
The number of eigenmodes of $T_{\mathcal C}$ with eigenvalue at most
$\Lambda\ge0$, counted with multiplicity, is
\begin{equation}
 \mathscr N_{\mathcal C}(\Lambda)
 =|\mathcal C|\left(1+2\left\lfloor\frac{2\sqrt\Lambda}{\pi}\right\rfloor\right)
 \le C_{|\mathcal C|}(1+\Lambda)^{1/2}.
\end{equation}
In an occupation vector with $n\ge1$ particles in these intervals, list
their eigenvalues as $\lambda_1\le\cdots\le\lambda_n$.
No eigenmode is repeated, so
\begin{equation}
 \begin{gathered}
 i\le\mathscr N_{\mathcal C}(\lambda_i)
      \le C_{|\mathcal C|}(1+\lambda_i)^{1/2},\\
 \frac1n\sum_{i=1}^n(1+\lambda_i)^{\alpha/2}
 \ge\frac{c_{\alpha,|\mathcal C|}}n\sum_{i=1}^n i^\alpha
 \ge c'_{\alpha,|\mathcal C|}n^\alpha.
 \end{gathered}
\end{equation}
For nondecreasing sequences $(a_i)$ and $(b_i)$,
\begin{equation}
 n\sum_i a_ib_i-\Bigl(\sum_i a_i\Bigr)\Bigl(\sum_i b_i\Bigr)
 =\sum_{i<j}(a_j-a_i)(b_j-b_i)\ge0.
\end{equation}
Apply this identity with $a_i=(1+\lambda_i)^{\alpha/2}$ and
$b_i=f(\lambda_i)\ge0$. The preceding average bound gives
\begin{equation}
 n^\alpha\sum_i f(\lambda_i)
 \le C\sum_i(1+\lambda_i)^{\alpha/2}f(\lambda_i).
\end{equation}
When $n=0$, both sides of \eqref{eq:master-exclusion} vanish.
Expansion in the occupation basis therefore proves the inequality.
\end{proof}

\subsection{Proof of interaction contribution}
\begin{proof}[Proof of \cref{lem:prop-interaction}]
Since $V$ preserves particle number and $A_{s,j_0}$ does not connect
different sectors, it suffices to take
$u\in C^\infty(\mathbb T_L^N)\cap\HH_N$, for any given $N\ge1$.

\smallskip\noindent
\emph{Step 1.} We first localize the interaction involving a selected coordinate.
We set $\Phi=\mathcal Lu$ and we write
\begin{equation}
 \underline{j}=(j_1,\ldots,j_N)\in\{0,\ldots,L-1\}^N,\qquad
 I_{\underline{j}}=\prod_{k=1}^N I_{j_k}.
\end{equation}
The definition of $A_{s,j_0}$ gives
\begin{equation}\label{eq:interaction-coordinates}
 \begin{aligned}
 \langle u,[A_{s,j_0},V]u\rangle
 =\sum_{i=1}^N\sum_j
       \langle u,[\chi_j(x_i)g_{s,j-j_0}(T_j)\chi_j(x_i),V]u\rangle,
 \end{aligned}
\end{equation}
where $g_{s,j-j_0}(T_j)$ acts in coordinate $i$. Moreover,
\begin{equation}
 (\mathcal L(Vu))_{\underline{j}}
 =\left(\sum_{1\le m<n\le N}W(x_m-x_n)\right)
                      \Phi_{\underline{j}}.
\end{equation}
Given an interval label $j$, let
\begin{equation}
 \mathcal C_j=\{k:\supp\chi_k\cap B_{r_0+2}(j)\ne\varnothing\}
\end{equation}
be a neighbourhood containing all interval labels that can contribute to
the cutoff interaction. For $k\in\mathcal C_j$, \(\norm{j-k}\le r_0+3\); since the centers have unit spacing, this gives
\begin{equation}
 \sup_j|\mathcal C_j|\le C,\qquad
 \sup_k\#\{j:k\in\mathcal C_j\}\le C,
\end{equation}
with the constant $C$ being independent of $N,L,j_0$.

Fix $i\in\{1,\dots,N\}$ and $\underline{j}$, and $x_m\in\supp\chi_{j_m}$ for $m\ne i$.
For $x_i\in\supp\chi_{j_i}\subset B_{7/8}(j_i)$,
finite range gives $W(x_i-x_m)=0$ whenever $j_m\notin\mathcal C_{j_i}$.
With the remaining coordinates held fixed, we write
\begin{equation}
 \begin{gathered}
 w_{i,\underline{j}}(x_i)=\Phi_{\underline{j}}(x_1,\ldots,x_N),\qquad
 v_{i,\underline{j}}(x_i)=\sqrt{\eta_{j_i}(x_i)}\sum_{m\ne i}W(x_i-x_m),\\
 n_{i,\underline{j}}=\sum_{m\ne i}\one_{\mathcal C_{j_i}}(j_m).
 \end{gathered}
\end{equation}
Since the support of $w_{i,\underline{j}}$ is contained in the support of $\chi_{j_i}$, and since $\eta_{j_i}$ is constant on it, we have 
\begin{equation}\label{eq:supp}
 (1-\sqrt{\eta_{j_i}})w_{i,\underline{j}}=0.
\end{equation}
For the rest of the proof, we adopt the following notation:
\begin{equation}
 I_{\underline{j}\setminus\{i\}}=\prod_{m\ne i}I_{j_m},\qquad
 \dd \underline{x}_{\hat{\imath}}=\prod_{m\ne i}\dd x_m,
 \qquad x_m\in I_{j_m}.
\end{equation}
Terms in $V$ not involving coordinate $x_i$ commute with the MALO $g_{s,j_i-j_0}(T_{j_i})$. Combining this with \eqref{eq:supp} and the evenness
of $W$, we can therefore write \eqref{eq:interaction-coordinates} as
follows:
\begin{equation}\label{eq:interaction-localized}
 \langle u,[A_{s,j_0},V]u\rangle =\sum_{i=1}^N\sum_{\underline{j}}\int_{I_{\underline{j}\setminus\{i\}}}
       \langle w_{i,\underline{j}},
          [g_{s,j_i-j_0}(T_{j_i}),v_{i,\underline{j}}]w_{i,\underline{j}}\rangle\,
       \dd \underline{x}_{\hat{\imath}}.
\end{equation}

\smallskip\noindent
\emph{Step 2.} We apply the multiplication estimate.
We set $f_j(\lambda):=g_{s,j-j_0}(\lambda)/(1+\lambda)$.
Since $\eta_{j_i}$ vanishes outside $B_2(j_i)$,
at most $n_{i,\underline{j}}$ summands in $v_{i,\underline{j}}$ can be nonzero.
Furthermore, uniform bounds on the derivatives of $\sqrt{\eta_{j_i}}$ and $W$ give
\begin{equation}
 \|v_{i,\underline{j}}\|_{C^{s+2}(I_{j_i})}\le C_sn_{i,\underline{j}}.
\end{equation}
Therefore, at fixed $i,\underline{j}$ and remaining positions,
\eqref{eq:prop-multiplication} and the bound on $v_{i,\underline{j}}$ above give
\begin{equation}\label{eq:interaction-multiplication}
 |\langle w_{i,\underline{j}},[g_{s,j_i-j_0}(T_{j_i}),v_{i,\underline{j}}]w_{i,\underline{j}}\rangle|
 \le C_sn_{i,\underline{j}}\|g_{s,j_i-j_0}(T_{j_i})^{1/2}w_{i,\underline{j}}\|
                           \|f_{j_i}(T_{j_i})^{1/2}w_{i,\underline{j}}\|.
\end{equation}
By the definition of $A_{s,j_0}$,
\begin{equation}
 \sum_{i=1}^N\sum_{\underline{j}}\int_{I_{\underline{j}\setminus\{i\}}}
       \|g_{s,j_i-j_0}(T_{j_i})^{1/2}w_{i,\underline{j}}\|^2\,\dd \underline{x}_{\hat{\imath}}
 =A_{s,j_0}[u].
\end{equation}
Inserting \eqref{eq:interaction-multiplication} into
\eqref{eq:interaction-localized} and then applying Cauchy--Schwarz over
$i,\underline{j}$, and the remaining positions yields
\begin{equation}\label{eq:prop-transfer}
 \begin{aligned}
 |\langle u,[A_{s,j_0},V]u\rangle|
 &\le C_s\sum_{i=1}^N\sum_{\underline{j}}\int_{I_{\underline{j}\setminus\{i\}}}
       n_{i,\underline{j}}\|g_{s,j_i-j_0}(T_{j_i})^{1/2}w_{i,\underline{j}}\|\,
       \|f_{j_i}(T_{j_i})^{1/2}w_{i,\underline{j}}\|\,\dd \underline{x}_{\hat{\imath}}\\
 &\le C_sA_{s,j_0}[u]^{1/2}
       \left(\sum_{i=1}^N\sum_{\underline{j}}\int_{I_{\underline{j}\setminus\{i\}}}
       n_{i,\underline{j}}^2\|f_{j_i}(T_{j_i})^{1/2}w_{i,\underline{j}}\|^2\,\dd \underline{x}_{\hat{\imath}}\right)^{1/2}.
 \end{aligned}
\end{equation}
Since $j_i\in\mathcal C_{j_i}$, the number operator of \cref{lem:exclusion}
satisfies
\begin{equation}
 (\widetilde{\mathcal N}_{\mathcal C_{j_i}}\Phi)_{\underline{j}}
 =(n_{i,\underline{j}}+1)\Phi_{\underline{j}}.
\end{equation}
The Fourier multipliers preserve pseudosite labels, so their actions in each
coordinate commute with $\widetilde{\mathcal N}_{\mathcal C_{j_i}}$. Consequently,
\begin{equation}
 \begin{aligned}
 &\sum_{i=1}^N\sum_{\underline{j}}\int_{I_{\underline{j}\setminus\{i\}}}
       (n_{i,\underline{j}}+1)^2\|f_{j_i}(T_{j_i})^{1/2}w_{i,\underline{j}}\|^2\,\dd \underline{x}_{\hat{\imath}}\\
 &\qquad=\sum_j\sum_{i=1}^N\sum_{\underline{j}:\,j_i=j}\int_{I_{\underline{j}\setminus\{i\}}}
       (n_{i,\underline{j}}+1)^2\|f_j(T_j)^{1/2}w_{i,\underline{j}}\|^2\,\dd \underline{x}_{\hat{\imath}}\\
 &\qquad=\sum_j\langle\Phi,\widetilde{\mathcal N}_{\mathcal C_j}^2
       \dG(f_j(T_j))\Phi\rangle.
 \end{aligned}
\end{equation}
Since $n_{i,\underline{j}}^2\le(n_{i,\underline{j}}+1)^2$, inserting this identity into
\eqref{eq:prop-transfer} yields
\begin{equation}\label{eq:interaction-number-bound}
 |\langle u,[A_{s,j_0},V]u\rangle|
 \le C_sA_{s,j_0}[u]^{1/2}
       \left(\sum_j\langle\Phi,\widetilde{\mathcal N}_{\mathcal C_j}^2
       \dG(f_j(T_j))\Phi\rangle\right)^{1/2}.
\end{equation}

\smallskip\noindent
\emph{Step 3.} The exclusion estimate controls the last factor in
\eqref{eq:interaction-number-bound}.
In $f_j(T_{\mathcal C_j})$, the parameter $j$ is fixed
on every pseudosite of $\mathcal C_j$. The function $f_j$ is
nondecreasing, because
\begin{equation}
 \partial_\lambda\log f_j(\lambda)
 =\frac{s}{1+\lambda}-\frac1{\langle j-j_0\rangle^2+\lambda}
 \ge\frac{s-1}{1+\lambda}\ge0.
\end{equation}
We apply \cref{lem:exclusion}, with the specific choices $\alpha=2$ and $f=f_j$, which, together with
$(1+\lambda)f_j(\lambda)=g_{s,j-j_0}(\lambda)$, then gives the following form inequalities:
\begin{equation}\label{eq:exclusion-bound}
 \widetilde{\mathcal N}_{\mathcal C_j}^2\dG(f_j(T_j))
 \le\widetilde{\mathcal N}_{\mathcal C_j}^2\dG(f_j(T_{\mathcal C_j}))
 \le C_s\dG(g_{s,j-j_0}(T_{\mathcal C_j})).
\end{equation}
The weights on $\mathcal C_j$ are uniformly comparable: indeed, since $y\mapsto\langle y-j_0\rangle$ is $1$-Lipschitz,
\begin{equation}
 |\langle j-j_0\rangle-\langle k-j_0\rangle|\le \norm{j-k}
 \le r_0+3,
 \qquad k\in\mathcal C_j.
\end{equation}
Consequently, we have
\begin{equation}
 \frac{g_{s,j-j_0}(\lambda)}{g_{s,k-j_0}(\lambda)}
 =\frac{\langle k-j_0\rangle^2+\lambda}{\langle j-j_0\rangle^2+\lambda}
 \le(r_0+4)^2.
\end{equation}
Let us use the shorthand $h=\bigoplus_k g_{s,k-j_0}(T_k)$; since every pseudosite lies in finitely many sets $\mathcal C_j$,
\begin{equation}\label{eq:pseudosite-sum-bound}
 \sum_j g_{s,j-j_0}(T_{\mathcal C_j})
 =\bigoplus_k\sum_{j:\,k\in\mathcal C_j}g_{s,j-j_0}(T_k)
 \le C_s\bigoplus_k g_{s,k-j_0}(T_k)=C_sh.
\end{equation}
Summing both sides of \eqref{eq:exclusion-bound} over $j$, and using \eqref{eq:pseudosite-sum-bound} we obtain
\begin{equation}
 \sum_j\widetilde{\mathcal N}_{\mathcal C_j}^2\dG(f_j(T_j))
 \le C_s\dG(h).
\end{equation}
Since $\langle\Phi,\dG(h)\Phi\rangle=A_{s,j_0}[u]$, substitution into
\eqref{eq:interaction-number-bound} proves the bound on the $N$-particle sector.
For general $u\in\mathscr D$, it then suffices to decompose it into its sector components.
\end{proof}

\subsection{Proof of main propagation estimate}
\begin{proof}[Proof of \cref{prop:propagation}]
We prove \eqref{eq:propagation} by a standard density argument.
For $u\in\mathscr D$, combining \eqref{eq:Q-time-derivative} with
\cref{lem:prop-kinetic,lem:prop-interaction} and applying Gr\"onwall's
inequality gives
\begin{equation}
 \begin{aligned}
 \left|\frac{\dd}{\dd t}A_{s,j_0}[u_t]\right|
 &=|\langle u_t,i[H,A_{s,j_0}]u_t\rangle|
       \le C_sA_{s,j_0}[u_t],\\
 A_{s,j_0}[u_t]&\le e^{C_s|t|}A_{s,j_0}[u].
 \end{aligned}
\end{equation}
For $u\in D(A_{s,j_0})$, we pick a sequence $(u_n)_n\subset\mathscr D$ converging to
$u$ in the form norm by \eqref{eq:form-core}. Unitarity and lower
semicontinuity of the closed form give
\begin{equation}
 \begin{aligned}
 A_{s,j_0}[e^{-itH}u]
 &\le\liminf_{n\to\infty}A_{s,j_0}[e^{-itH}u_n]\\
 &\le e^{C_s|t|}\lim_{n\to\infty}A_{s,j_0}[u_n]=e^{C_s|t|}A_{s,j_0}[u].
 \end{aligned}
\end{equation}
This concludes the proof.
\end{proof}

\section{Initial comparison with local moments}\label{sec:static}
We recover the localized derivatives in $A_{s,j_0}$ from
powers of the local energy and the local particle number. The proof
uses the second-order energy estimate, the product rule for powers
of the one-coordinate kinetic operators, and their joint spectral
calculus with the nonnegative operator in the other coordinates.

\begin{lemma}[Localized derivative estimates]\label{lem:initial}
Let $j\in\{0,\ldots,L-1\}$.
For $N,a\in\mathbb N$, one has on $\HH_N$, in the form sense,
\begin{equation}\label{eq:initial-marked}
 \dG\bigl(\chi_j(1+T_j)^{2a}\chi_j\bigr)
 \le C_a\bigl(\mathcal E_j^{2a}
                   +\mathcal N_{B_j^W}^{2a}\bigr).
\end{equation}
For $u\in D(\mathcal E_j^a)\cap\HH_N$, each localized vector
$\chi_j(x_i)u$ belongs to $D((1+T_j)_i^a)$, viewing the function as taking values in the 
$L^2$ space of the remaining coordinates. The constant is independent
of $N,L,j$. On Fock space the right-hand form domain is
$D(\mathcal E_j^a)\cap D(\mathcal N_{B_j^W}^{a})$.
\end{lemma}

\begin{proof}[Proof of \cref{prop:comparison}]
We sum the localized derivative bounds with the spatial weights in $A_{s,j_0}$.
Fix $j_0\in\{0,\ldots,L-1\}$ and set $a=\lceil(s+1)/2\rceil=p/2$.
The scalar bound $g_{s,j-j_0}(\lambda)\le\langle j-j_0\rangle^{-2}(1+\lambda)^p$ and
\cref{lem:initial}, followed by orthogonal summation over particle
sectors, give the form inequalities
\begin{equation}
 A_{s,j_0}
 \le\sum_{j=0}^{L-1}\langle j-j_0\rangle^{-2}\dG\bigl(\chi_j(1+T_j)^p\chi_j\bigr)
 \le C\sum_{j=0}^{L-1}\langle j-j_0\rangle^{-2}
       (\mathcal E_j^p+\mathcal N_{B_j^W}^p).
\end{equation}
Taking traces therefore gives
\begin{equation}\label{eq:initial-static-sum}
 \begin{aligned}
 \Tr\rho A_{s,j_0}
 &\le C\sum_{j=0}^{L-1}\langle j-j_0\rangle^{-2}\Tr\rho
                 (\mathcal E_j^p+\mathcal N_{B_j^W}^p)\\
 &\le C_s\max_{0\le j<L}\Tr\rho
                 (\mathcal E_j^p+\mathcal N_{B_j^W}^p).
 \end{aligned}
\end{equation}
The last step uses unit spacing of the centers:
\begin{equation}
 \sum_{j=0}^{L-1}\langle j-j_0\rangle^{-2}
 \le1+2\sum_{n=1}^{\infty}(1+n^2)^{-1}<\infty.
\end{equation}
This bound is uniform in $L,j_0$. The form inequalities include the
required domain inclusion. Taking the maximum over $j_0$ proves
\eqref{eq:initial-comparison}.

\end{proof}

\subsection{Proof of the initial derivative estimate}
For the derivative estimate, we localize to the periodic interval, establish a second-order energy bound, and obtain higher-order Sobolev bounds by induction with constants independent of volume and particle number. The strong commutation used to reduce to a fixed number
of coordinates in $B_z^W$ is proved in
\cref{lem: strong commutation} below.

\begin{proof}[Proof of \cref{lem:initial}]

Fix $a,N\in\mathbb N$ and take $z=j$.
The projections in each product in
\eqref{eq:local-number-projections} reduce $\mathcal E_j$.
Work in one such subspace with $n\ge1$ coordinates in $B_j^W$,
relabelled $1,\ldots,n$, on
$L^2((B_j^W)^n\times((B_j^W)^c)^{N-n})$.
Coordinates outside $B_j^W$ contribute neither kinetic nor potential
terms. Indeed, $\eta_j(y')=0$ for $y'\notin B_j^W$, and finite range gives
\begin{equation}
 [\eta_j(y)+\eta_j(y')]W(y-y')=0
 \quad(y\in B_j^W,\ y'\notin B_j^W).
\end{equation}
Let $K_{j,i}\ge0$ be the operator associated with the closed kinetic
form in coordinate $i$, and put
\begin{equation}
 K_j=\sum_{i=1}^nK_{j,i},\qquad K_{j,i}=-\partial_i\eta_j(x_i)\partial_i,
 \qquad
 V_j=\frac12\sum_{1\le i<l\le n}
       [\eta_j(x_i)+\eta_j(x_l)]W(x_i-x_l).
\end{equation}
The potential is a bounded nonnegative multiplication operator on
this fixed sector, with $\|V_j\|\le Cn^2$. Hence
$\mathcal E_j=K_j+V_j$ and $D(\mathcal E_j)=D(K_j)$.
For each pair $1\le i<\ell\le n$, the summand
$\frac12[\eta_j(x_i)+\eta_j(x_\ell)]W(x_i-x_\ell)$
in $V_j$, together with its partial derivatives in $x_i$ and $x_\ell$
up to any fixed order, is bounded uniformly in $n,N,L,j$.

\emph{Step 1: Localization to the periodic interval.}
Let $w\in D(\mathcal E_j^a)$. The quadratic form associated with
$\chi_j(x_i)(1+T_j)_i^{2a}\chi_j(x_i)$, evaluated at $w$,
is the squared Sobolev norm of $\chi_j(x_i)w$, given by
\begin{equation}\label{eq:initial-pseudosite-sobolev}
 \|(1+T_j)_i^a(\chi_j(x_i)w)\|^2
 =\|\chi_j(x_i)w\|_{H_{\mathrm{per},x_i}^{2a}}^2.
\end{equation}
We establish the required Sobolev domain membership and bound below. Sobolev norms differentiate only coordinate $i$, viewing the function as taking values in the $L^2$ space of the remaining coordinates. It therefore suffices to prove
\begin{equation}\label{eq:initial-bootstrap-new}
 \sum_{i=1}^n
 \|\chi_j(x_i)w\|_{H_{\mathrm{per},x_i}^{2a}}^2
 \le C_a\|(\mathcal E_j+n)^aw\|^2,
 \qquad w\in D(\mathcal E_j^a).
\end{equation}
Step~2 proves the second-derivative bound, and Step~3 iterates it
up to order $2a$. The support of $\chi_j$ and all its derivatives
lies where $\eta_j=1$.

\emph{Step 2: Induction basis.}
For $w\in D(\mathcal E_j)=D(K_j)$, expansion gives
\begin{equation}\label{eq:local-energy-square-expansion}
 \|\mathcal E_j w\|^2
 =\|K_j w\|^2+\|V_j w\|^2+2\operatorname{Re}\langle K_j w,V_j w\rangle.
\end{equation}
Multiplication by $V_j$ preserves the kinetic form domain, so we can evaluate the mixed term using the kinetic form.
Since $\sqrt{\eta_j(x_i)}\,w$ has an $L^2$ weak derivative in $x_i$, we apply the product rule to $\eta_j(x_i)|w|^2$ and integrate by parts to obtain
\begin{equation}\label{eq:kinetic-potential-cross-term}
 2\operatorname{Re}\langle K_j w,V_j w\rangle
=2\sum_{i=1}^n\int\eta_j(x_i)V_j|\partial_iw|^2\,\dd\underline{x}-\int\sum_{i=1}^n\partial_i(\eta_j(x_i)\partial_iV_j)|w|^2\,\dd\underline{x}.
\end{equation}\label{eq:kinetic-potential-cross-term-bound}
There is no boundary term because $\eta_j=0$ near $\partial B_j^W$.
Since $V_j\ge0$, the first integral is nonnegative.
Each pair term in $V_j$ contributes derivatives in only its two coordinates, and hence
\begin{equation}\label{eq:local-pair-potential-derivative-bound}
 \left\|\sum_{i=1}^n\partial_i(\eta_j(x_i)\partial_iV_j)\right\|_\infty
 \le Cn^2.
\end{equation}
Since the nonnegative operators $K_{j,i}$ strongly commute, joint
spectral calculus gives the first inequality below. The second follows by substituting \eqref{eq:kinetic-potential-cross-term} into \eqref{eq:local-energy-square-expansion} and applying \eqref{eq:local-pair-potential-derivative-bound}.
\begin{equation}\label{eq:initial-kinetic-base}
 \sum_{i=1}^n\|K_{j,i}w\|^2\le\|K_j w\|^2
 \le\|\mathcal E_j w\|^2+Cn^2\|w\|^2.
\end{equation}
On a neighborhood of the support of every derivative of $\chi_j$,
we have $\eta_j=1$. The kinetic form controls $\partial_iw$ there, and the product rule gives
\begin{equation}\label{eq:initial-localized-second-derivative}
 -\partial_i^2(\chi_j(x_i)w)
 =\chi_j(x_i)K_{j,i}w
  -2\chi_j'(x_i)\partial_iw
  -\chi_j''(x_i)w.
\end{equation}
Since $\eta_j=1$ on $\operatorname{supp}\chi_j'$, the kinetic form gives
\begin{equation}
\label{eq:cutoff-first-derivative-sum}
\begin{aligned}
    \sum_{i=1}^n \|\chi_j'(x_i)\partial_i w\|^2
    \le \|\chi_j'\|_\infty^2
        \sum_{i=1}^n \int \eta_j(x_i)|\partial_i w|^2
        \,\mathrm{d}\underline{x}= \|\chi_j'\|_\infty^2 \langle w,K_jw\rangle.
\end{aligned}
\end{equation}
Thus the right-hand side of \eqref{eq:initial-localized-second-derivative} belongs to $L^2$.
Since $\chi_j(x_i)w$ is supported strictly inside $I_j$, it follows that $\chi_j(x_i)w\in H_{\mathrm{per},x_i}^2$. Using \eqref{eq:initial-localized-second-derivative} and \eqref{eq:cutoff-first-derivative-sum}, we obtain
\begin{equation}
\begin{aligned}
    \sum_{i=1}^n \|\chi_j(x_i)w\|_{H_{\mathrm{per},x_i}^2}^2
    &\le C\left(
        \sum_{i=1}^n \|K_{j,i}w\|^2+ \langle w,K_jw\rangle
        + n\|w\|^2
    \right)\\
    &\le C\|(\mathcal E_j+n)w\|^2.
\end{aligned}
\end{equation}
The second inequality follows from \eqref{eq:initial-kinetic-base}, the form inequality
$0\le K_j\le\mathcal E_j$, and $n\ge1$. This proves \eqref{eq:initial-bootstrap-new} for $a=1$. The same argument applies when $\chi_j$ is replaced by any of its derivatives needed in Step~3, since their supports lie in the region where $\eta_j=1$.

\emph{Step 3: Induction steps and summation.}
For the fixed coordinate $y=x_i$, put
$\mathfrak g_i=L^2((B_j^W)^{n-1}\times((B_j^W)^c)^{N-n})$ and let
\begin{equation}\label{eq:positive-spectator}
 H_{j,i}=\sum_{k\ne i}K_{j,k}+
   \frac12\sum_{\substack{k<l\\k,l\ne i}}
       [\eta_j(x_k)+\eta_j(x_l)]W(x_k-x_l).
\end{equation}
This is a nonnegative selfadjoint operator on $\mathfrak g_i$,
independent of $y$, and strongly commuting with $K_{j,i}$.
The remaining interaction $v_{j,i}(y)$ consists of the $n-1$ pair terms involving $y=x_i$. Each summand and its $y$-derivatives of any fixed order are uniformly bounded as multiplication operators on $\mathfrak g_i$:
\begin{equation}\label{eq:one-coordinate-interaction-bounds}
 v_{j,i}(y)=\frac12\sum_{k\ne i}
       [\eta_j(y)+\eta_j(x_k)]W(y-x_k),\qquad
 \sup_y\|\partial_y^\ell v_{j,i}(y)\|_{\mathcal B(\mathfrak g_i)}
 \le C_\ell n\quad(\ell\in\mathbb N_0).
\end{equation}
Bounded perturbation gives
\begin{equation}\label{eq:marked-equation-new}
 K_{j,i}+H_{j,i}=\mathcal E_j-v_{j,i},\qquad D(K_{j,i}+H_{j,i})=D(\mathcal E_j).
\end{equation}
The induction only requires the localized vectors
$\chi_j^{(\ell)}(x_i)w$, $0\le\ell\le2a$.
Since $\eta_j=1$ on a neighborhood of $\supp\chi_j$,
$K_{j,i}$ acts as $(T_j)_i=-\partial_i^2$ on each of these
vectors. We first write the product-rule identity for
$\chi_j(x_i)w$; the same identity holds with $\chi_j$ replaced
by any of these derivatives. Thus, in $L^2(I_j;\mathfrak g_i)$,
\begin{equation}\label{eq:initial-localized-equation}
 \begin{aligned}
 (1+(T_j)_i+H_{j,i})(\chi_j(x_i)w)=\chi_j(x_i)\mathcal E_jw
       +(1-v_{j,i})\chi_j(x_i)w-2\partial_i(\chi_j'(x_i)w)
       +\chi_j''(x_i)w.
 \end{aligned}
\end{equation}
Denote the right-hand side of \eqref{eq:initial-localized-equation} by $r$. Taking Fourier coefficients in $x_i$ gives
$(1+\lambda_m+H_{j,i})\widehat {\chi_j(x_i)w}_m=\widehat r_m$, where $\lambda_m$ is the eigenvalue of $T_j$ for the $m$-th Fourier mode. Since $H_{j,i}\ge0$, each Fourier mode satisfies
$(1+\lambda_m)\|\widehat{\chi_j(x_i)w}_m\|_{\mathfrak g_i}
\le\|\widehat r_m\|_{\mathfrak g_i}$. For $s\in\mathbb N_0$, we obtain
\begin{equation}
\begin{aligned}
    \|\chi_j(x_i)w\|_{H^{s+2}_{\mathrm{per},x_i}}^2=\sum_m(1+\lambda_m)^{s+2}\|\widehat {\chi_j(x_i)w}_m\|_{\mathfrak g_i}^2
    \le \sum_m (1+\lambda_m)^s
        \|\widehat r_m\|_{\mathfrak g_i}^2
    = \|r\|_{H^s_{\mathrm{per},x_i}}^2.
\end{aligned}
\end{equation}
In particular, $\chi_j(x_i)w\in H^{s+2}_{\mathrm{per},x_i}$ whenever $r\in H^s_{\mathrm{per},x_i}$. Applying the product rule and using \eqref{eq:one-coordinate-interaction-bounds} and \eqref{eq:initial-localized-equation}, we obtain
\begin{equation}\label{eq:localized-sobolev-gain}
\begin{aligned}
    \|\chi_j(x_i)w\|_{H^{s+2}_{\mathrm{per},x_i}}
    \le C_s\bigl(
        &\|\chi_j(x_i)\mathcal E_jw\|_{H^s_{\mathrm{per},x_i}}
        + n\|\chi_j(x_i)w\|_{H^s_{\mathrm{per},x_i}}\\
        &+ \|\chi_j'(x_i)w\|_{H^{s+1}_{\mathrm{per},x_i}}
        + \|\chi_j''(x_i)w\|_{H^s_{\mathrm{per},x_i}}
    \bigr).
\end{aligned}
\end{equation}
$C_s$ is independent of $H_{j,i},n,N,L, j$. 

The preceding estimate controls the $H^{s+2}$-norm of $\chi_jw$
by the $H^s$-norms of $\chi_j\mathcal E_jw$ and $n\chi_jw$,
together with lower-order Sobolev norms involving derivatives
of $\chi_j$. We prove \eqref{eq:initial-bootstrap-new} by induction on the power $b$, including the estimates for the derivatives of $\chi_j$ needed at each order. Each step uses lower-order estimates involving at most three additional derivatives of $\chi_j$. Thus, for fixed $a$, only finitely many derivatives are needed, and the corresponding constants can be absorbed into $C_a$. Step~2 gives the base case $b=1$.

Suppose \eqref{eq:initial-bootstrap-new} holds at order $b-1$. Let $w\in D(\mathcal E_j^b)$, where $2\le b\le a$.
Then $w,\mathcal E_jw\in D(\mathcal E_j^{b-1})$.
Applying \eqref{eq:localized-sobolev-gain} with $s=2b-3$ and
cutoff $\chi_j'$, then using the embedding
$H_{\mathrm{per},x_i}^{2b-2}\subset H_{\mathrm{per},x_i}^{2b-3}$
and the induction hypothesis at order $b-1$, gives
\begin{equation}
\label{eq:localized-cutoff-derivative}
\begin{aligned}
\sum_{i=1}^n \|\chi_j'(x_i)w\|_{H_{\mathrm{per},x_i}^{2b-1}}^2
&\le C_a\sum_{i=1}^n\Big(
  \|\chi_j'(x_i)\mathcal E_jw\|_{H_{\mathrm{per},x_i}^{2b-3}}^2
  +n^2\|\chi_j'(x_i)w\|_{H_{\mathrm{per},x_i}^{2b-3}}^2\\
&\qquad\quad
  +\|\chi_j''(x_i)w\|_{H_{\mathrm{per},x_i}^{2b-2}}^2
  +\|\chi_j'''(x_i)w\|_{H_{\mathrm{per},x_i}^{2b-3}}^2
  \Big)\\
&\le C_a\Big(
  \|(\mathcal E_j+n)^{b-1}\mathcal E_jw\|^2
  +n^2\|(\mathcal E_j+n)^{b-1}w\|^2
  \Big).
\end{aligned}
\end{equation}
Taking $s=2b-2$ in \eqref{eq:localized-sobolev-gain}
and summing the squared norms over $i$, we obtain the following
estimate using \eqref{eq:localized-cutoff-derivative} and
the induction hypothesis:
\begin{equation}
\begin{aligned}
\sum_{i=1}^n \|\chi_j(x_i)w\|_{H_{\mathrm{per},x_i}^{2b}}^2
&\le C_a\Big(
  \|(\mathcal E_j+n)^{b-1}\mathcal E_jw\|^2
  +n^2\|(\mathcal E_j+n)^{b-1}w\|^2
  \Big)\\
&\le C_a\|(\mathcal E_j+n)^b w\|^2.
\end{aligned}
\end{equation}
The last step follows from spectral calculus, using
$t^2+n^2\le(t+n)^2$ for $t\ge0$. Taking $b=a$ proves
\eqref{eq:initial-bootstrap-new} on $D(\mathcal E_j^a)$.

Combining \eqref{eq:initial-bootstrap-new} with
\eqref{eq:initial-pseudosite-sobolev} gives
\begin{equation}\label{eq:initial-coordinate-sector}
 \begin{aligned}
 \sum_{i=1}^n\|(1+T_j)_i^a(\chi_j(x_i)w)\|^2&=\sum_{i=1}^n \|\chi_j(x_i)w\|_{H_{\mathrm{per},x_i}^{2a}}^2
 \\&\le C_a\|(\mathcal E_j+n)^aw\|^2\le C_a\bigl(\|\mathcal E_j^aw\|^2+n^{2a}\|w\|^2\bigr).
 \end{aligned}
\end{equation}
The last inequality uses $(t+n)^{2a}\le2^{2a-1}(t^{2a}+n^{2a})$.

On $L^2(\mathbb T_L^N)$, define the coordinate projections, for $1\le i\le N$
and $0\le n\le N$,
\begin{equation}\label{eq:local-number-projections}
 P_i=\one_{B_j^W}(x_i),
 \qquad
 \Pi_n=
 \sum_{\substack{S\subset\{1,\ldots,N\}\\|S|=n}}
 \prod_{i\in S}P_i\prod_{k\notin S}(1-P_k).
\end{equation}
Since $\mathcal N_{B_j^W}=\sum_{i=1}^NP_i$, the $\Pi_n$
are its spectral projections. Each $\Pi_n$ is permutation
invariant and therefore preserves $\HH_N$.

Return to $u\in D(\mathcal E_j^a)\cap\HH_N$ and decompose
$u=\sum_{n=0}^N\Pi_nu$.
On $\Pi_0u$, the local energy and all localized vectors vanish.
For $n\ge1$, apply \eqref{eq:initial-coordinate-sector} to each
product component of $\Pi_nu$ in
\eqref{eq:local-number-projections}.
For fixed $i$, localization annihilates the components with
$i\notin S$, while the remaining components stay orthogonal
in the other coordinates. Summing squared norms gives
\begin{equation}
 \begin{aligned}
 \sum_{i=1}^N
 \|(1+T_j)_i^a(\chi_j(x_i)u)\|^2
 &\le C_a\sum_{n=0}^N
 \bigl(
   \|\mathcal E_j^a\Pi_nu\|^2
   +n^{2a}\|\Pi_nu\|^2
 \bigr)\\
 &=C_a\bigl(
   \|\mathcal E_j^au\|^2
   +\|\mathcal N_{B_j^W}^au\|^2
 \bigr).
 \end{aligned}
\end{equation}
Here we use the strong commutation property proved in
\cref{lem: strong commutation} below and
$\mathcal N_{B_j^W}\Pi_n=n\Pi_n$.
This proves \eqref{eq:initial-marked} and its domain assertion
on $\HH_N$, with the same constant.
Orthogonal summation over $N$, including the vacuum, proves
the Fock-space statement, with right-hand form domain
$D(\mathcal E_j^a)\cap D(\mathcal N_{B_j^W}^a)$.

\end{proof}
\begin{lemma}[Strong commutation]\label{lem: strong commutation}
The operators $\mathcal E_j$ and $\mathcal N_{B_j^W}$ strongly commute,
meaning that their spectral projections commute. 
\end{lemma}

\begin{proof}[Proof of \cref{lem: strong commutation}]
Fix $N\in\mathbb N$. Realize the closure of the
form~\eqref{eq:localform} on $L^2(\mathbb T_L^N)$, before
restricting to its antisymmetric subspace, and use the
coordinate projections $P_i$ introduced above.

Since $\supp\eta_j\subset B_j^W$, choose real smooth cutoffs
$0\le\varphi_\varepsilon\le1$ converging almost everywhere
to $\one_{B_j^W}$, with derivatives supported where
$\eta_j=0$. The product rule gives
\[
 \eta_j(x_k)^{1/2}
 \partial_k(\varphi_\varepsilon(x_i)u)
 =
 \varphi_\varepsilon(x_i)
 \eta_j(x_k)^{1/2}\partial_ku,
 \qquad 1\le k\le N.
\]
This identity extends to the form domain by approximation.
Dominated convergence then gives convergence to $P_iu$
in the form norm, so $P_i$ preserves that domain.
For $u\in D(\mathcal E_j)$ and every $v$ in the form domain,
\[
 \mathcal E_j(v,P_iu)
 =\mathcal E_j(P_iv,u)
 =\langle P_iv,\mathcal E_ju\rangle
 =\langle v,P_i\mathcal E_ju\rangle.
\]
The representation theorem therefore yields
\begin{equation}\label{eq:reducing-coordinates}
 P_iD(\mathcal E_j)\subset D(\mathcal E_j),
 \qquad
 \mathcal E_jP_iu=P_i\mathcal E_ju
 \quad(u\in D(\mathcal E_j)).
\end{equation}
Thus each $P_i$, and hence each $\Pi_n$, commutes with
the spectral projections of $\mathcal E_j$.
Since $\Pi_n$ and $\mathcal E_j$ preserve $\HH_N$,
their restrictions commute.
The $\Pi_n$ are the spectral projections of
$\mathcal N_{B_j^W}$, proving strong commutation on each
sector. Orthogonal direct summation proves the
Fock-space assertion.
\end{proof}


\section{LRB in the pseudolattice picture}\label{sec:truncation}

In this section we prove \cref{prop:truncation}.

The auxiliary Hamiltonian
$\widetilde H$ constructed below reproduces the physical evolution on
$\operatorname{Ran}\mathcal L$. Restriction of its local couplings to
finitely many modes makes them bounded (\cref{lem:modecut}); the error
is controlled by the weighted forms of \cref{prop:propagation}.
A commutator estimate for the restricted dynamics
(\cref{lem:trunc-commutator}) then bounds the error from removing
the boundary couplings (\cref{lem:spacecut}). The mode and boundary
errors are combined in the proof of the proposition at the end of
the section.

For later convenience, we introduce the following notation for the pseudosite picture.

For a set of pseudosites $Z\subset\{0,\ldots,L-1\}$, we write $\mathcal F_Z=\bigoplus_{N\ge0}\widetilde{\mathfrak{h}}_Z^{\wedge N}$ for the Fock space over the pseudosites in $Z$. Furthermore, we denote by
$P_Z\in\mathcal B(\widetilde{\mathfrak{h}})$
\begin{equation}
      (P_Z\widetilde f)_j=\one_Z(j)\widetilde f_j
\end{equation}
the one-body projection onto the pseudosites in $Z$. Also, given an operator $A_Z$ on $\widetilde{\mathfrak{h}}_Z$, we will often identify it with its extension $A_Z\oplus0$ to $\widetilde{\mathfrak{h}}$.

\subsection{Auxiliary dynamics and local couplings}
On the pseudosite Fock space $\widetilde{\mathcal F}$, we introduce the auxiliary Hamiltonian $\widetilde H=\bigoplus_{N\ge0}\widetilde H_N$, defined by
\begin{equation}\label{eq:aux-H}
    \widetilde H_N=\sum_{i=1}^N(T+\mathsf C)^{(i)}+\widetilde V_N,
\end{equation}
where $T=\oplus_{j=0}^{L-1} T_j$, and the auxiliary kinetic operator $\mathsf C:D(T)\to\widetilde{\mathfrak{h}}$ is constructed from \eqref{eq:direct-ims}, namely
\begin{equation}\label{eq:aux-kinetic}
 (\mathsf Cf)_j=Sf_j+\sum_k \mathsf B_{jk}f_k
 =Sf_j+\sum_k(2a_{jk}\partial+a_{jk}')f_k.
\end{equation}
and where the two-body interaction $\widetilde V_N$ is given by the following bounded operator:
\begin{equation}
 (\widetilde V_N\Phi)_{\underline{j}}(\underline{x})
 :=\sum_{i<m} W_{j_ij_m}(x_i,x_m)\Phi_{\underline{j}}(\underline{x})\qquad W_{jk}(x,y):=\sqrt{\eta_{j}(x)\eta_{k}(y)}W(x-y).
\end{equation}
\begin{remark}\label{rmk:lattice}
    Notice that the single-pseudosite kinetic energy $T_j$ can be interpreted as an (unbounded) on-pseudosite term acting on  $\widetilde{\mathfrak h}_{\{j\}}\equiv L^2(I_j)$; on the other hand $\mathsf C$ (respectively, $\widetilde V_N$) can be seen as a sum of nearest-neighbor terms, since $\mathsf B_{jk}=0$ if $\|j-k\|\geq 2$ (resp. finite-range two-pseudosite terms, since $W_{jk}$ vanishes if $\|j-k\|\geq 4+r_0$).
\end{remark}
\begin{lemma}\label{lem:aux-kinetic}
The auxiliary kinetic operator $\mathsf C$ defined in \eqref{eq:aux-kinetic} is infinitesimally $T$-bounded. In particular, \(
 T+\mathsf C\) is self-adjoint on \(D(T+\mathsf C)=D(T)\).
\end{lemma}
We defer the proof to Appendix \ref{app:Kato}. As a standard consequence of \cref{lem:aux-kinetic} we then have that $\widetilde H_N$ is self-adjoint on $D(\sum_{i=1}^{N} T^{(i)})\cap\widetilde{\mathfrak h}^{\wedge N}$, therefore making $\widetilde H$ self-adjoint on its natural domain.

We now write the decomposition described in \cref{rmk:lattice} explicitly. We set $\kappa=\lceil2(r_0+5)\rceil$, and we define the family of pseudosite sets
\begin{equation}
 \mathscr Z=\{\{j\}:0\le j<L\}
       \cup\{\{j,l\}:1\le \norm{j-l}\le\kappa\}.
\end{equation}
For $Z\in\mathscr Z$, we define the following first-order operator on
$D(T_Z)\subset\widetilde{\mathfrak{h}}_Z$
\begin{equation}\label{eq:hZ}
 \mathsf C_Z:=\sum_{j,l:\{j,l\}=Z}P_{\{j\}}\mathsf C P_{\{l\}},
\end{equation}
which then vanishes for pairs
$Z=\{j,l\}$ with $\norm{j-l}\ge2$.
Analogously, define the bounded two-particle operator
$\mathsf W_Z\in\mathcal B(\widetilde{\mathfrak{h}}_Z^{\wedge 2})$ by
\begin{equation}
 (\mathsf W_Z\Phi)_{jj'}(y,y'):=
 \one_{\{Z=\{j,j'\}\}}
 W_{j,j'}(y,y')\Phi_{jj'}(y,y').
\end{equation}
For $N\in\mathbb N$, we define the local interaction
\begin{equation}
 \Psi_Z\big|_{\widetilde{\mathfrak{h}}_Z^{\wedge N}}
 =\sum_{i=1}^N(\mathsf C_Z)_i+\sum_{1\le i<m\le N}(\mathsf W_Z)_{im},
\end{equation}
with $\Psi_Z=0$ on the vacuum sector. Then we have that
\begin{equation}\label{eq:local-decomposition}
 \widetilde H=\dG(T)+\sum_{Z\in\mathscr Z}\Psi_Z.
\end{equation}
\begin{proposition}[Auxiliary Hamiltonian]\label{prop:aux-H}
$\widetilde H$ is the Hamiltonian describing the dynamics in the pseudolattice picture. That is, for every $t\in\mathbb R$, we have
\begin{equation}\label{eq:unitary-lift}
 e^{-itH}=\mathcal L^*e^{-it\widetilde H}\mathcal L.
\end{equation}
\end{proposition}
\begin{proof}
The direct IMS identity \eqref{eq:direct-ims}, together with the fact that $\chi_j$ is smooth and
supported in the interior of $I_j$, yields
\begin{equation}\label{eq:intertwine}
 \mathcal L_{1}D(K_1)\subset D(T),\qquad
 K_1u=\mathcal L_{1}^*(T+\mathsf C)\mathcal L_{1}u
\end{equation}
for any $u\in C^{\infty}(\mathbb T_L)$, and by density they extend to $D(K_1)=H^{2}(\mathbb T_L)$.

Moreover, $(\bigoplus_N\widetilde V_N)\mathcal L=\mathcal LV$ because
$\sqrt{\eta_j}\chi_j=\chi_j$, therefore $\mathcal L_{1}^{\otimes N}$ takes $D(H_N)$ into $D(\widetilde H_N)$,
and by \eqref{eq:intertwine} we have $\mathcal LH_N=\widetilde H_N\mathcal L$.

Applying to $(H_N-z)^{-1}u$, with $z\in\mathbb C\setminus\mathbb R$, this gives
$(\widetilde H_N-z)^{-1}\mathcal L=\mathcal L(H_N-z)^{-1}$. The identities
for $z$ and $\overline z$ show that $\operatorname{Ran}\mathcal L$
reduces $\widetilde H_N$, and the spectral theorem then yields \eqref{eq:unitary-lift}.
\end{proof}

\subsection{Restriction to finitely many modes}
For any energy $E\ge1$, and for any set of pseudosites $Z$, we introduce the energy cut-off projector
\begin{equation}\label{eq:mode-cutoff}
 \pi_Z(E)=\one_{[1,E]}(1+T_Z)\in\mathcal B(\widetilde{\mathfrak{h}}_Z).
\end{equation}
and, on $\widetilde{\mathfrak{h}}_Z^{\wedge N}$, we then define the truncated couplings
\begin{equation}\label{eq:mode-cutoff-coupling}
 \Psi_Z^{(E)}=
 \sum_{i=1}^N\pi_i(\mathsf C_Z)_i\pi_i
 +\sum_{i<m}^N\pi_i\pi_m(\mathsf W_Z)_{im}\pi_i\pi_m,
 \qquad \pi_i:=\pi_Z(E)_i.
\end{equation}

\begin{lemma}[Restriction to finitely many modes]\label{lem:modecut}
Let $Z\in\mathscr Z$, and $E\ge1$. The operator
$\Psi_Z^{(E)}$ is bounded, even, and self-adjoint; also, there exists $C>0$ independent of $E$ such that
\begin{equation}\label{eq:modecut-norm}
 \|\Psi_Z^{(E)}\|\le CE.
\end{equation}
Furthermore, for any $p,N\in \mathbb N$, and for any vector smooth, periodic vector $u$
\begin{equation}\label{eq:modecut-error}
 \|(\Psi_Z-\Psi_Z^{(E)})u\|^2
 \le C_p\frac{\|\dG((1+T_Z)^{p+1})u\|^2}{E^{2p}}.
\end{equation}
for a constant $C_p$ independent of $E$ and $N$.
\end{lemma}

\begin{proof}
\textit{Step 1: Boundedness of the truncated coupling.}
Denoting by $d_E$ the rank of $\pi_Z(E)$, eigenvalue counting in one dimension gives
\begin{equation}
 d_E\le C\sqrt E,\qquad
 \|\pi_Z(E)\mathsf C_Z\pi_Z(E)\|\le C\sqrt E,\qquad
 \|\pi_Z(E)^{\otimes2}\mathsf W_Z\pi_Z(E)^{\otimes2}\|\le C.
\end{equation}
For the middle bound, we have used that
$\|(1+T_Z)^{1/2}\pi_Z(E)\|\le\sqrt E$.

On the Fock
space generated by these modes, the Pauli exclusion principle implies that particle number is bounded by
$d_E\le C\sqrt E$.
Thus the one-body sum in \eqref{eq:mode-cutoff-coupling} has norm at
most $d_EC\sqrt E\le CE$, and the pair sum has norm at most
$\binom{d_E}2\|\mathsf W_Z\|\le CE$. This proves
\eqref{eq:modecut-norm}.

\smallskip\noindent\textit{Step 2: Estimates for the spectral tails.}
Since the operators in the claimed estimate preserve the number of particles in \(Z\), by orthogonality it is sufficient to consider \(u\) with exactly \(n\) particles in the pseudosites contained in \(Z\), with \(n\le N\). Let us write $G_i=(1+T_Z)^{(i)}$, with $i=1,\dots,n$.

By definition, $\Psi_Z=\sum_{i}(\mathsf C_Z)_i + \sum_{i<m}(\mathsf W_Z)_{im}$, therefore we may estimate each term separately.
First, we decompose the one-body difference as
\begin{equation}
 (\mathsf C_Z)_i-\pi_i(\mathsf C_Z)_i\pi_i
 =(1-\pi_i)(\mathsf C_Z)_i+\pi_i(\mathsf C_Z)_i(1-\pi_i).
\end{equation}
By spectral calculus, we have the Markov inequalities
\begin{equation}
    \|(1-\pi_i)u\|\le E^{-p}\|G_i^p u\|\qquad \|G_i^{1/2}(1-\pi_i)u\|\le E^{-p}\|G_i^{p+1/2}u\|,
\end{equation}
which, together with the product rule, then imply
$\|G_i^p(\mathsf C_Z)_iu\|\le C_p\|G_i^{p+1/2}u\|$. Hence
\begin{equation}
 \begin{aligned}
 \|((\mathsf C_Z)_i-\pi_i(\mathsf C_Z)_i\pi_i)u\|
 &\le E^{-p}\|G_i^p(\mathsf C_Z)_iu\|
       +C\|G_i^{1/2}(1-\pi_i)u\|\\
 &\le C_pE^{-p}\|G_i^{p+1/2}u\|.
 \end{aligned}
\end{equation}
For the pair term put $\Pi=\pi_i\pi_m$ and use
\begin{equation}
 (\mathsf W_Z)_{im}-\Pi(\mathsf W_Z)_{im}\Pi
 =(1-\Pi)(\mathsf W_Z)_{im}+\Pi(\mathsf W_Z)_{im}(1-\Pi),\qquad
 1-\Pi=(1-\pi_i)+\pi_i(1-\pi_m).
\end{equation}
Since the multiplication operator $\mathsf W_Z$ is bounded and smooth in the $i$-th variable, we have 
$\|G_i^p(\mathsf W_Z)_{im}u\|\le C_p\|G_i^pu\|$, and similarly for $m$. The
spectral-tail inequalities give
\begin{equation}
 \|((\mathsf W_Z)_{im}-\pi_i\pi_m(\mathsf W_Z)_{im}\pi_i\pi_m)u\|^2
 \le C_pE^{-2p}
       (\|G_i^pu\|^2+\|G_m^pu\|^2).
\end{equation}

\smallskip\noindent\textit{Step 3: Summation and exclusion principle.}
Applying Cauchy--Schwarz over the one-body and pair sums gives factors $n$ and
$\binom n2$, respectively. In particular, for the pair contribution,
\begin{equation}
 \begin{aligned}
 \binom n2\sum_{i<m}\bigl(\|G_i^pu\|^2+\|G_m^pu\|^2\bigr)
 =\binom n2(n-1)\sum_i\|G_i^pu\|^2
 \le n^3\sum_i\|G_i^pu\|^2.
 \end{aligned}
\end{equation}
Consequently
\begin{equation}
 \begin{aligned}
 \|(\Psi_Z-\Psi_Z^{(E)})u\|^2
 &\le C_pE^{-2p}\left(
       n\sum_i\|G_i^{p+1/2}u\|^2
           +n^3\sum_i\|G_i^pu\|^2\right)\\
 &\le C_pE^{-2p}
       \left\langle u,\sum_iG_i^{2p+3/2}u\right\rangle
 \le C_pE^{-2p}
       \left\langle u,\sum_iG_i^{2p+2}u\right\rangle.
 \end{aligned}
\end{equation}
where in the second passage we applied \cref{lem:exclusion} with the following choices: $\mathcal C=Z$, with
$\alpha=1$ and $f(\lambda)=(1+\lambda)^{2p+1}$ for the one-body term,
and with $\alpha=3$ and $f(\lambda)=(1+\lambda)^{2p}$ for the pair term. The last inequality follows from $G_i\ge1$.
\end{proof}

\subsection{Removal of boundary couplings}
Given an observable localized near $x\in\mathbb T_L$, we want to
approximate its full evolution by dynamics acting only on a pseudosite
ball $x[R]=\{j\in\{0,\ldots,L-1\}:\norm{j-x}\le R\}$. We pass from the full
$\widetilde H$ to such a Hamiltonian in two stages.

First, we replace
every interaction $\Psi_Z$ meeting $x[R]$ by its truncated version $\Psi_Z^{(E)}$, producing the intermediate Hamiltonian $\widehat H_{R,E}$, which therefore differs from $\widetilde H$ in $O(R)$ interaction terms, uniformly in $L$.

Next, removing the truncated couplings crossing the boundary we are left with the desired Hamiltonian $H_{R,E}$, which contains
the kinetic energy and truncated interactions inside $x[R]$, up to an exterior
Hamiltonian $H_{\mathrm{out}}$ whose dynamics does not affect observables supported inside. In particular, $H_{R,E}$ generates the approximating evolution of the observable.

We define
\begin{align}
 \widehat H_{R,E}
 &:=\dG(T)+\sum_{Z\cap x[R]\ne\varnothing}\Psi_Z^{(E)}
                 +\sum_{Z\cap x[R]=\varnothing}\Psi_Z,
                                                        \label{eq:Hhat}\\
 H_{R,E}&:=\dG(T_{x[R]})+
                        \sum_{Z\subset x[R]}\Psi_Z^{(E)}.
                                                        \label{eq:HR}
\end{align}
The following lemma shows that the dynamics generated by $\widehat H_{R,E}$ and $H_{R,E}$ are locally close, with an error that decays exponentially in the distance of the observable from the boundary of $x[R]$.

\begin{lemma}[Removal of boundary couplings]\label{lem:spacecut}
Let $1\le R\le L/4$ and $E\ge1$. For nonempty $X\subset x[R]$,
$\widetilde u\in\widetilde{\mathfrak{h}}_X$, $\sigma\in\{-,+\}$, and
$r_X=\max_{j\in X}\norm{j-x}$, one has, for every $t\in\mathbb R$,
\begin{equation}\label{eq:spacecut}
 \|\tau_t^{\widehat H_{R,E}}(a^\sigma(\widetilde u))
               -\tau_t^{H_{R,E}}(a^\sigma(\widetilde u))\|
 \le C_0\|\widetilde u\|\,|X|e^{-c(R-r_X)}(e^{c_1E|t|}-1).
\end{equation}
for constants $C_0,c,c_1>0$ dependent on $\kappa$ and independent of $E,R,L,$ and $t$.
\end{lemma}

\begin{proof}
The exterior operator
\begin{equation}
 H_{\rm out}=\dG(T_{(x[R])^c})+
                          \sum_{Z\subset(x[R])^c}\Psi_Z
\end{equation}
Since it is supported on $x[R]^c$, it strongly commutes with both $H_{R,E}$ and $a^\sigma(\widetilde u)$. Moreover,
\begin{equation}
 \widehat H_{R,E}=H_{R,E}+H_{\rm out}
      +\sum_{\substack{Z\cap x[R]\ne\varnothing\\
                        Z\cap(x[R])^c\ne\varnothing}}\Psi_Z^{(E)}.
\end{equation}
Since $R\le L/4$, $x[R]$ is an arc of consecutive pseudosites, so at
most $4\kappa^2$ pairs cross its boundary. For such a pair
$Z=\{j,l\}$ with $l\notin x[R]$, the triangle inequality and
$\norm{j-l}\le\kappa$ give
\begin{equation}
 \dist(X,Z)\ge R-r_X-\kappa.
\end{equation}
The boundary sum is bounded, and
\begin{equation}
 \tau_t^{H_{R,E}+H_{\rm out}}(a^\sigma(\widetilde u))
       =\tau_t^{H_{R,E}}(a^\sigma(\widetilde u)),
\end{equation}
Duhamel's formula and \cref{lem:trunc-commutator} therefore give, for $t\ge0$,
\begin{equation}
 \begin{aligned}
 \|\tau_t^{\widehat H_{R,E}}(a^\sigma(\widetilde u))
                    -\tau_t^{H_{R,E}}(a^\sigma(\widetilde u))\|
 &\le
 \sum_{\substack{Z\cap x[R]\ne\varnothing\\
                 Z\cap(x[R])^c\ne\varnothing}}
 \int_0^t
 \|[\Psi_Z^{(E)},\tau_r^{H_{R,E}}(a^\sigma(\widetilde u))]\|\,\dd r\\
 &\le C\|\widetilde u\||X|Ee^{-(R-r_X)/\kappa}
                     \int_0^t e^{c_1E r}\,\dd r\\
 &\le C_0\|\widetilde u\||X|e^{-c(R-r_X)}(e^{c_1Et}-1),
 \end{aligned}
\end{equation}
with $c=1/\kappa$ and $C_0=C/c_1$. Reversing the signs of both Hamiltonians gives the
bound for negative times.
\end{proof}
For nonempty pseudosite sets $X,Y$,let us indicate with
\( \dist(X,Y):=\min_{j\in X,l\in Y}\norm{j-l}\) their distance.
\begin{lemma}[LRB bound]\label{lem:trunc-commutator}
There exist constants $C_0,c_1>0$ dependent on $\kappa$ such that the following holds. Let $1\le R\le L/4$ and $E\ge1$. For any nonempty
$X\subset x[R]$, $\widetilde u\in\widetilde{\mathfrak{h}}_X$, $\sigma\in\{-,+\}$,
and for any $Y\in\mathscr Z$, and for every $t\in\mathbb R$, we have
\begin{equation}\label{eq:trunc-pathbound}
 \begin{gathered}
 \|[\tau_t^{H_{R,E}}(a^\sigma(\widetilde u)),\Psi_Y^{(E)}]\|
 \le C_0\|\widetilde u\|\|\Psi_Y^{(E)}\||X|
          e^{c_1E|t|-\dist(X,Y)/\kappa}.
 \end{gathered}
\end{equation}
for constants $C_0,c_1>0$ dependent on $\kappa$ and independent of $E,R,L,$ and $t$.
\end{lemma}

\begin{proof}
Without loss of generality, we may assume\footnote{Otherwise, we use the trivial bound $\|[\tau_t(A),B]\|\leq 2\|A\|\|B\|$ for all $t\in\mathbb R$.} $X\cap Y=\varnothing$. We may also bring the time evolution on $\Psi_Y^{(E)}$.

\smallskip\noindent\textit{Step 1.}
Following \cite[Section~3]{NS}, we devise an interaction picture to remove the unbounded on-(pseudo)site contributions to $H_{R,E}$. We set $\mathsf K=\dG(T_{x[R]})$ and define, for $r\in\mathbb R$,
\begin{equation}
 \Psi_Z(r)=e^{ir\mathsf K}\Psi_Z^{(E)}e^{-ir\mathsf K},\qquad
 U(r)=e^{ir\mathsf K}e^{-irH_{R,E}}.
\end{equation}
Due to $\mathsf K$ being a sum of on-pseudosite terms,
$\Psi_Z(r)$ is even, supported in $Z$, and norm-continuous, with
$\|\Psi_Z(r)\|=\|\Psi_Z^{(E)}\|$. Moreover,
\begin{equation}
 iU'(r)=\sum_{Z\subset x[R]}\Psi_Z(r)U(r),\qquad 
 \tau_r^{H_{R,E}}(\Psi_Y^{(E)})=U(r)^*\Psi_Y(r)U(r).
\end{equation}

Therefore, by Duhamel interpolation we have
\begin{equation}\label{eq:duh}
 [\tau_t^{H_{R,E}}(\Psi_Y^{(E)}),a^\sigma(\widetilde u)]=i\sum_{Z\subset x[R]}\int_0^t
    [U(r)^*[\Psi_Z(r),\Psi_Y(t)]U(r),a^\sigma(\widetilde u)]\,\dd r.
\end{equation}

\smallskip\noindent\textit{Step 2.}
With the unbounded $\mathsf K$ handled, we are left with the time-dependent interactions
\(\Psi_Z(r), Z\in\mathscr Z,\ Z\subset x[R]\). For this, we apply a standard Lieb-Robinson bound on the finite pseudosite lattice CAR algebra
\(\operatorname{CAR}\!\left(\oplus_{j=0}^{L-1}
 \bigl(\operatorname{Ran}\pi_{\{j\}}(E)+\operatorname{span}\{\widetilde u_j\}\bigr)\right)\).

Following \cite{NSY2017}, we choose the positive symmetric kernel
\begin{equation}
 G(j,l)=\frac{e^{-\norm{j-l}/\kappa}}{32(1+\norm{j-l})^2},
\end{equation}
which satisfies, uniformly in $L$,
\begin{equation}
 \sup_j\sum_lG(j,l)\le\frac3{32},\qquad
 \sum_mG(j,m)G(m,l)\le\frac38G(j,l).
\end{equation}
By \cref{lem:modecut} and the finite range of $\Psi_Z$, we have
\begin{equation}
 \sum_{\substack{Z\subset x[R]\\j,l\in Z}}\|\Psi_Z(r)\|
 \le CE\,\one_{\{\norm{j-l}\le\kappa\}}
 \le C_\kappa E\,G(j,l),
\end{equation}
uniformly in $r$. Thus the hypotheses of \cite{NSY2017} are verified.

Applying then \cite[Theorem~3.1(i)]{NSY2017} to the integrand in \eqref{eq:duh} with respect to $r$, and using $\|\Psi_Y(t)\|=\|\Psi_Y^{(E)}\|$ and $\|a^\sigma(\widetilde u)\|=\|\widetilde u\|$,
we obtain
\begin{equation}
 \|[\tau_t^{H_{R,E}}(\Psi_Y^{(E)}),a^\sigma(\widetilde u)]\|
 \le C_0\|\Psi_Y^{(E)}\|\|\widetilde u\||X|e^{c_1Et-\dist(X,Y)/\kappa}.
\end{equation}
Reversing $\mathsf K$ and
all couplings gives the bound for $t<0$.
\end{proof}
\begin{remark}
By \cref{lem:modecut}, increasing $E$ improves the mode-cutoff error
bound needed to compare $\widetilde H$ with $\widehat H_{R,E}$, but at the same time it increases the norm bound $CE$ for the truncated interactions, which was employed in these previous two lemmas; in the the proof of
\cref{prop:truncation} we will balance the two errors by choosing $E$ proportional
to $R/(1+|t|)$.
\end{remark}

\subsection{Proof of Proposition~\ref{prop:truncation}}
The following lemma gives a form bound for creation and annihilation operators with respect to the MALOs.
\begin{lemma}\label{lem:insertion}
    For any $\sigma\in\{-,+\}$ and any
$f\in C^\infty(\mathbb T_L;\mathbb C)$, there exists a constant $C_s>0$ such that
\begin{equation}\label{eq:insertion}
 a^{-\sigma}(f)(1+A_{s,j_0})a^\sigma(f)
 \le C_s\|f\|_{H^s(\mathbb T_L)}^2(1+A_{s,j_0})
\end{equation}
in the form sense.
\end{lemma}
\begin{proof}
Let us adopt the shorthands $h:=\bigoplus_{j} g_{s,j-j_0}(T_j)$ and $\widetilde f:=\mathcal L_1 f$. For $u\in D(A_{s,j_0})$, denote $\mathcal Lu=:\widetilde u\in D(\mathrm d\Gamma(h)^{1/2})$. We use the localization identity  $\mathcal La^{\sigma}(f)= a^\sigma(\widetilde f)\mathcal L$, to get
\begin{equation}
    A_{s,j_0}[a^\sigma(f)u]=\|\mathrm d\Gamma(h)^{1/2}a^{\sigma}(\widetilde f)\widetilde u\|^2.
\end{equation}
Since the pseudosite Fourier modes form an orthonormal eigenbasis $(e_\nu)$ of $h$,
with eigenvalues $\lambda_\nu>0$, we obtain the spectral identity
\begin{equation}\label{eq:spectra-id}
 A_{s,j_0}[a^\sigma(f)u]=\|\dG(h)^{1/2} a^{\sigma}(\widetilde f)\widetilde u\|^2
 =\sum_\nu\lambda_\nu\|a(e_\nu)a^{\sigma}(\widetilde f)\widetilde u\|^2.
\end{equation}
We now verify its well-posedness. The CAR give
\begin{equation}\label{eq:car}
 a(e_\nu)a(\widetilde f)=-a(\widetilde f)a(e_\nu),\qquad
 a(e_\nu)a^*(\widetilde f)=\langle e_\nu,\widetilde f\rangle-a^*(\widetilde f)a(e_\nu).
\end{equation}
Then, the second
identity in \eqref{eq:car} and $\|a^*(\widetilde f)\|=\|\widetilde f\|$ give
\begin{equation}\label{eq:car2}
 \begin{aligned}
 \sum_{\nu=1}^m\lambda_\nu\|a(e_\nu)a^*(\widetilde f)\widetilde u\|^2
 &\le2\|\widetilde f\|^2\sum_{\nu=1}^m\lambda_\nu\|a(e_\nu)\widetilde u\|^2+2\sum_{\nu=1}^m\lambda_\nu
       |\langle e_\nu,\widetilde f\rangle|^2\|\widetilde u\|^2\\
 &\le2\|\widetilde f\|^2\|\dG(h)^{1/2}\widetilde u\|^2
       +2\|h^{1/2}\widetilde f\|^2\|\widetilde u\|^2\\
       &= 2\|f\|^2 A_{s,j_0}[u] + 2 \|h^{1/2}\widetilde f\|^2 \|u\|^2
 \end{aligned}
\end{equation}
Taking $m\to\infty$, and employing \eqref{eq:spectra-id}, we prove that $a^*(f)u\in D(A_{s,j_0})$.
Analogously, the first CAR identity yields
\begin{equation}
 A_{s,j_0}[a(f)u]=\|\dG(h)^{1/2}a(\widetilde f)\widetilde u\|^2
 =\sum_\nu\lambda_\nu\|a(\widetilde f)a(e_\nu)\widetilde u\|^2
 \le\|f\|^2 A_{s,j_0}[u],
\end{equation}
so $a(f)u$ also belongs to $D(A_{s,j_0})$.

On the other hand, each component of $\widetilde f=(\chi_j f)_j$ is smooth, so we may apply \eqref{eq:scalar-basic} and the product rule to get
\begin{equation}\label{eq:kin-erg}
 \|h^{1/2}\widetilde f\|^2
 \le\sum_j\|(1+T_j)^{s/2}(\chi_jf)\|^2
 \le C_s\|f\|_{H^s(\mathbb T_L)}^2.
\end{equation}
Finally it is enough to insert \eqref{eq:kin-erg} into \eqref{eq:car2} to obtain
\begin{equation}
    A_{s,j_0}[a^\sigma(f)u] \leq C_{s}\|f\|_{H^s(\mathbb T_L)}^2(\|u\|^2 + A_{s,j_0}[u]), 
\end{equation}
hence proving the claim.
\end{proof}
We are now ready to prove \cref{prop:truncation}.
\begin{proof}[Proof of \cref{prop:truncation}]
Here, we shall use the shorthand $\mathsf a=a^\sigma(\mathcal L_1f)$, and we shall set $p=k+1$ and $s=2p+2=2k+4$, and let $\gamma$ be the constant in
\eqref{eq:propagation} at order $s$.

First of all, for the case of large $R$, we make the ansatz  
\[
a_R^\sigma(t,f)=\tau_t^{H_{R,E}}(\mathsf a)= \tau_t^{H_{R,E}}(a^\sigma(\mathcal L_1f)),
\]
with cutoff $E$ to be suitably chosen in terms of $R$ and $t$ (see Step~3). Indeed, it satisfies all of the requirements of the claim: its support lies in $\mathfrak A_{x[R]}$ because both $\mathsf a$ and $H_{R,E}$
are supported on the pseudosite set $x[R]$; it is odd and has norm $\|f\|_2$;
it is independent of the initial state $\rho$.

\eqref{eq:unitary-lift}
\begin{equation}
 \|[\mathcal L\tau_t^H(a^\sigma(f))
       -a_R^\sigma(t,f)\mathcal L]\rho^{1/2}\|_2 = \|[\tau_t^{\widetilde H}(\mathsf a)
       -\tau^{H_{R,E}}_t(\mathsf a)]\mathcal L\rho^{1/2}\|_2.
 \end{equation}
Telescoping $\tau^{\widehat H_{R,E}}_t$ and using $\|\mathcal L\rho^{1/2}\|_2=1$, we then obtain
 \begin{equation}\label{eq:est}
 \|[\mathcal L\tau_t^H(a^\sigma(f))
       -a_R^\sigma(t,f)\mathcal L]\rho^{1/2}\|_2\le\|[\tau_t^{\widetilde H}(\mathsf a)
       -\tau_t^{\widehat H_{R,E}}(\mathsf a)]\mathcal L\rho^{1/2}\|_2
       +\|\tau_t^{\widehat H_{R,E}}(\mathsf a)-\tau_t^{H_{R,E}}(\mathsf a)\|.
\end{equation}
We will dedicate Step~1 (respectively~2) to estimate the first (resp. second) term, and Step~3 to appropriately choose the cutoff.

\smallskip\noindent\textit{Step 1: bound for the first term.}
 We put $\Delta:=\widetilde H-\widehat H_{R,E}$, and for each $Z\cap x[R]\neq \varnothing$, we choose a center $j_Z\in Z$.
Since $\norm{j-j_Z}\le\kappa$ for any $j\in Z$, \eqref{eq:scalar-basic}
gives the form bound
\begin{equation}
 \mathcal L^*\dG((1+T_Z)^s)\mathcal L\le CA_{s,j_Z},
\end{equation}
which, together with the stronger bound established in Step~3 of the proof of \cref{lem:modecut}, yields the norm estimate
\begin{equation}\label{eq:trunc-totalerror}
  \|\Delta\mathcal Lu\|\le C E^{-p}\sum_{Z\cap x[R]\neq\varnothing} A_{s,j_Z}[u]^{1/2}
\end{equation}
for any $u\in\mathscr D$.
Writing $\widetilde U(t)=e^{-it\widetilde H}$, $\widehat U(t)=e^{-it\widehat H_{R,E}}$, by a Duhamel iteration we obtain
\begin{equation}\label{eq:trunc-Duhamel}
 [\tau_t^{\widetilde H}(\mathsf a)-\tau_t^{\widehat H_{R,E}}(\mathsf a)]\mathcal Lu=i\int_0^t\Bigl[
       \widehat U(r-t)\Delta \widetilde U(-r)\mathsf a\widetilde U(t)
       -\widehat U(-t)\mathsf a\widehat U(r)\Delta \widetilde U(t-r)
                       \Bigr]\mathcal Lu\,\dd r.
\end{equation}
We estimate the norm of the integrand in \eqref{eq:trunc-Duhamel} by
\begin{equation}
 \begin{aligned}
  \|\Delta &\widetilde U(-r)\mathsf a\widetilde U(t)\mathcal Lu\|
       +\|\mathsf a\|\,\|\Delta \widetilde U(t-r)\mathcal Lu\|\\
 &\le CE^{-p}e^{\gamma |r|/2}
          \sum_{Z} A_{s,j_Z}[a^\sigma(f)e^{-itH}u]^{1/2}
       +C\|f\|_2E^{-p}e^{\gamma|t-r|/2}
          \sum_Z A_{s,j_Z}[u]^{1/2}\\
 &\le C\|f\|_{H^s}E^{-p}e^{\gamma|t|}
          \sum_Z\bigl(\|u\|^2+A_{s,j_Z}[u]\bigr)^{1/2},
 \end{aligned}
\end{equation}
where in the first passage we used \eqref{eq:trunc-totalerror} and then \cref{prop:propagation} regarding time propagation of the MALO; in the last passage we employed \cref{lem:insertion} followed by \cref{prop:propagation} again. Integrating over $r$ we obtain the same bound multiplied by $|t|$:
\begin{equation}\label{eq:trunc-energy}
 \|(\tau_t^{\widetilde H}(\mathsf a)-\tau_t^{\widehat H_{R,E}}(\mathsf a))
                                  \mathcal Lu\|
 \le C\|f\|_{H^s}|t|E^{-p}e^{\gamma|t|}
          \sum_Z\bigl(\|u\|^2+A_{s,j_Z}[u]\bigr)^{1/2}.
\end{equation}

By denseness, the bound \eqref{eq:trunc-energy} extends to all $u$ in the joint form domain of the $A_{s,j}$. We may assume $\max_j\Tr\rho A_{s,j}<\infty$, since otherwise the claimed estimate is trivial. Then every eigenvector of $\rho$ with positive eigenvalue belongs to this domain, and hence we get
\begin{equation}\label{eq:trunc-energycomparison}
 \begin{aligned}
 \|(\tau_t^{\widetilde H}(\mathsf a)-\tau_t^{\widehat H_{R,E}}(\mathsf a))
                                  \mathcal L\rho^{1/2}\|_2
 &\le C\|f\|_{H^s}|t|E^{-p}e^{\gamma|t|}
          \sum_Z(1+\Tr\rho A_{s,j_Z})^{1/2}\\
 &\le C\|f\|_{H^s} |t|E^{-p}e^{\gamma|t|} R Q_s(\rho),
 \end{aligned}
\end{equation}
where in the last passage we used that the number of terms in the sum is bounded by $C R$, and where we introduced $Q_s(\rho)$ as
\begin{equation}
  Q_s(\rho):=\sup_{j\in\{0,\ldots,L-1\}}(1+\Tr\rho A_{s,j})^{1/2}.
\end{equation}

\smallskip\noindent\textit{Step 2: bound for the second term.}
Let $R\ge R_0:=4\kappa+8$. Then the support $X$ of $\mathsf a$, given
in \eqref{eq:field-support}, lies in $x[R]$, and it has $r_X<2$. Applying
\cref{lem:spacecut} with $\widetilde u=\mathcal L_{1}f$ yields
\begin{equation}
 \begin{aligned}
 \|\tau_t^{\widehat H_{R,E}}(\mathsf a)
           -\tau_t^{H_{R,E}}(\mathsf a)\|
 \le C\|f\|_{2}e^{-cR}(e^{c_1E|t|}-1).
 \end{aligned}
\end{equation}

\smallskip\noindent\textit{Step 3: choice of cutoff.}
Inserting the bounds from Steps~1 and~2 into \eqref{eq:est} and using $Q_s(\rho)\ge1$ finally yields
\begin{equation}
 \begin{aligned}
 \|[\mathcal L\tau_t^H(a^\sigma(f))
       -a_R^\sigma(t,f)\mathcal L]\rho^{1/2}\|_2
 &\le C\|f\|_{H^s}Q_{s}(\rho)\left[R|t|E^{-p}e^{\gamma|t|}+e^{-cR}(e^{c_1E|t|}-1)\right].
 \end{aligned}
\end{equation}
We now pick
$0<\varepsilon\le1$ such that $c_1\varepsilon<c/2$; notice that it is independent of $k$.

For
$R\ge\max\{R_0,(1+|t|)/\varepsilon\}$, choosing the energy cutoff as
$E:=\varepsilon R/(1+|t|)\ge1$ gives
\begin{equation}
 \begin{aligned}
 R|t|E^{-p}e^{\gamma|t|}
 &=\varepsilon^{-p}R^{-k}|t|(1+|t|)^pe^{\gamma|t|}
 \le C_k(1+R)^{-k}e^{(\gamma+k+2)|t|},\\
 e^{-cR}(e^{c_1E|t|}-1)
 &\le e^{-cR/2}
 \le C_k(1+R)^{-k},
 \end{aligned}
\end{equation}
therefore proving the claim for large $R$.

For $R<\max\{R_0, (1+|t|)/\varepsilon\}$, our pick is $a_R^\sigma(t,f)=0$, which yields an error of at most
$\|f\|_2$ for all $t\in\mathbb R$, while
$1+R\le(1+R_0+\varepsilon^{-1})(1+|t|)$ and $(1+|t|)^k\le e^{k|t|}$, so
the same bound holds after possibly enlarging the prefactor.
\end{proof}

\appendix
\section{Relative boundedness of the kinetic coupling}\label{app:Kato}
\begin{proof}[Proof of \cref{lem:aux-kinetic}]
For \emph{smooth} $f,g\in D(T)$, integration by parts yields
\begin{equation}
 \langle f_j,\mathsf B_{jk}g_k\rangle_{L^2(I_j)}
 =\langle \mathsf B_{kj}f_j, g_k\rangle_{L^2(I_k)},
\end{equation}
because $a_{kj}=-a_{jk}$ and the overlap coefficients vanish near
both pseudosite boundaries. Summing over $j,k$ and adding $S$ then proves symmetry of $\mathsf C$ on smooth vectors.

Next, we obtain the relative bound for smooth $f$: using the fact that $a_{jk}=0$ if $\|j-k\|\geq 2$, for any $\varepsilon >0$ we have
\begin{equation}
 \begin{aligned}
 \|\mathsf Cf\|^2
 &\le C\sum_j\Bigg(\|f_j\|^2+
             \sum_{k:\norm{j-k}\le1}(\|f_k'\|^2+\|f_k\|^2)\Bigg)\\
 &\le C\sum_k(\|f_k'\|^2+\|f_k\|^2)=C\|(1+T)^{1/2}f\|^2 \\&\le\varepsilon^2\|Tf\|^2+C_\varepsilon^2\|f\|^2,
 \end{aligned}
\end{equation}
where in the last inequality we used
$C(1+\lambda)\le\varepsilon^2\lambda^2+C_\varepsilon^2$ for
$\lambda\ge0$. This bound establishes both that $\mathsf C$ is symmetric on $D(T)$ and that it is infinitesimally $T$-bounded.
By the Kato-Rellich theorem, we then conclude that $T+\mathsf C$ is self-adjoint. 
\end{proof}

\section{Extension to polynomial observables}\label{app:polynomials}
In this appendix, we prove Theorem \ref{thm:polynomials}.

The polynomial result follows from the initial comparison and the
insertion and approximation estimates below. Insertion is controlled by
iteration in the form norm; approximation also requires a product
expansion for the boundary error.

Replacing each $a^\sigma(f)$ in a physical polynomial
$A$ by $a^\sigma(\mathcal L_{1}f)$ defines $\widetilde A$ with
\begin{equation}\label{eq:A-lift}
 \widetilde A\mathcal L=\mathcal LA,\qquad
 \widetilde A^*\mathcal L=\mathcal LA^*,\qquad
 \|\widetilde A\|=\|A\|.
\end{equation}
The intertwining identities follow from \eqref{eq:field-lift} by
multiplication and linearity. Fock factorization over
$\operatorname{Ran}\mathcal L_{1}$ and its orthogonal complement proves
the norm identity and shows that the lift is independent of the chosen
polynomial expression. If $A\in\mathfrak P^\epsilon(B_1(x))$, then its corresponding $\widetilde A$
has also parity $\epsilon$ and its support is contained 
\begin{equation}\label{eq:A-support}
 X=\{j:\supp\chi_j\cap B_1(x)\ne\varnothing\}
       \subset\{j:\norm{j-x}<2\},\qquad |X|\le C.
\end{equation}

\begin{lemma}[Action of polynomial observables]\label{lem:polynomial-insertion}
For $s\in\mathbb N$, a reference pseudosite $j_0$, and every finite polynomial
$A$ in smooth creation and annihilation operators, the following holds in the form sense on
$D(A_{s,j_0})$:
\begin{equation}\label{eq:polynomial-insertion}
 A^*(1+A_{s,j_0})A\le C_A(1+A_{s,j_0}).
\end{equation}
for a constant $C_A$
dependent on the chosen expression for $A$, in particular the $H^s$-norms of its test functions, uniformly in $L,j_0$.
\end{lemma}

\begin{lemma}[Approximation of polynomial observables]\label{lem:polynomial-truncation}
Let $k\in\mathbb N$, $A\in\mathfrak P^\epsilon(B_1(x))$, and
$1\le R\le L/4$. For every $t\in\mathbb R$, there is
$A_R(t)\in\mathfrak A_{x[R]}$, of parity $\epsilon$, such that
$\|A_R(t)\|\le\|A\|$ and, for every state $\rho$,
\begin{equation}\label{eq:polynomial-truncation}
 \|\bigl(\mathcal L\tau_t^H(A)-A_R(t)\mathcal L\bigr)\rho^{1/2}\|_2\le\frac{C_Ae^{(\gamma+k+2)|t|}}{(1+R)^k}
 \left(1+\max_{0\le j<L}\Tr\rho A_{2k+4,j}\right)^{1/2}.
\end{equation}
\end{lemma}
\begin{remark}
   Here, $\gamma$ is the constant in \eqref{eq:propagation} at order $2k+4$. Moreover, the constant $C_A$ depends on $k$ and on the
chosen finite expression for $A$; in particular, it depends on the $H^{2k+4}$-norms of its test functions.
\end{remark}

\begin{proof}[Proof of \cref{thm:polynomials}]
Set $s=2k+4$. The initial comparison and
\eqref{eq:polynomial-insertion} yield
\begin{equation}
 \begin{gathered}
 1+\max_j\Tr\rho A_{s,j}\le C M_{2k+6}(\rho),\\
 \Tr B\rho B^*+\max_j\Tr B\rho B^*A_{s,j}
 \le C_B\left(1+\max_j\Tr\rho A_{s,j}\right)
 \le C_B M_{2k+6}(\rho).
 \end{gathered}
\end{equation}
For sufficiently large $d=\norm{x-y}$, take $R=d/3\le L/6$. The supports of
$A_R(t)$ and $\widetilde B$ are disjoint, so
$[A_R(t),\widetilde B]_{\rm gr}=0$. Set
\begin{equation}
 D=\mathcal L\tau_t^H(A)-A_R(t)\mathcal L.
\end{equation}
The intertwining identities give
\begin{equation}
 \begin{aligned}
 \mathcal L[\tau_t^H(A),B]_{\rm gr}
 &=DB-(-1)^{\epsilon\delta}\widetilde B D,\\
 \|[\tau_t^H(A),B]_{\rm gr}\rho^{1/2}\|_2
 &\le\|DB\rho^{1/2}\|_2+\|B\|\|D\rho^{1/2}\|_2.
 \end{aligned}
\end{equation}
Apply \eqref{eq:polynomial-truncation} with $\rho$ and, if
$\beta=\Tr B\rho B^*>0$, with $\beta^{-1}B\rho B^*$, multiplying the
latter estimate by $\beta^{1/2}$. Since
\begin{equation}
 \|DB\rho^{1/2}\|_2^2=\Tr B\rho B^*D^*D
 =\|D(B\rho B^*)^{1/2}\|_2^2,
\end{equation}
the preceding weighted-form bounds give \eqref{eq:LR-polynomials}.
If $\beta=0$, then $B\rho^{1/2}=0$. For bounded $d$, the estimate follows from
$\|[\tau_t^H(A),B]_{\rm gr}\|\le2\|A\|\|B\|$.
\end{proof}

\subsection{Adaptation of the insertion estimate}
\begin{proof}[Proof of \cref{lem:polynomial-insertion}]
For $s\in\mathbb N$ and a reference pseudosite $j_0$, the single-operator
estimate \eqref{eq:insertion} gives boundedness of every smooth
creation or annihilation operator on $D(A_{s,j_0})$ in the norm
\begin{equation}
 u\longmapsto\bigl(\|u\|^2+A_{s,j_0}[u]\bigr)^{1/2}.
\end{equation}
Iterating this, we obtain the bound for every monomial. The triangle inequality
in this norm gives \eqref{eq:polynomial-insertion} for finite linear
combinations.
\end{proof}

\subsection{Adaptation of the local approximation}
\begin{proof}[Proof of \cref{lem:polynomial-truncation}]
The mode comparison follows from \eqref{eq:trunc-Duhamel} and the
polynomial insertion estimate. The boundary comparison then follows
by expanding the difference of two evolved products.

Set $s=2k+4$ and $p=k+1$. Sums over $Z$ run over the at most $CR$
members of $\mathscr Z$ meeting $x[R]$, with $j_Z\in Z$ chosen
as in the proof of \cref{prop:truncation}. For $u\in\mathscr D$,
$t\ge0$, and $0\le r\le t$, \eqref{eq:polynomial-insertion} and
\eqref{eq:propagation} give
\begin{equation}
 \begin{aligned}
 A_{s,j_Z}[e^{irH}Ae^{-itH}u]
 &\le C_Ae^{\gamma(r+t)}
          \bigl(\|u\|^2+A_{s,j_Z}[u]\bigr),\\
 A_{s,j_Z}[e^{-i(t-r)H}u]
 &\le e^{\gamma(t-r)}A_{s,j_Z}[u].
 \end{aligned}
\end{equation}
For the observable $\widetilde A$, the intertwining identities
\eqref{eq:A-lift} and \eqref{eq:unitary-lift} place the two mode errors
in \eqref{eq:trunc-Duhamel} on the vectors above. The estimate
\eqref{eq:trunc-totalerror} takes square roots of these form bounds,
so both exponential factors are at most $e^{\gamma t}$.
Reversing both generators covers
negative times. The same extension from $\mathscr D$ to the common
form domain and triangle inequality over eigenvectors as in the
proof of \cref{prop:truncation} therefore give
\begin{equation}
 \begin{aligned}
 \|(\tau_t^{\widetilde H}(\widetilde A)
       -\tau_t^{\widehat H_{R,E}}(\widetilde A))\mathcal L\rho^{1/2}\|_2 &\le C_A|t|E^{-(k+1)}e^{\gamma|t|}
               \sum_Z\left(1+\Tr\rho A_{s,j_Z}\right)^{1/2}\\
 &\le C_AR|t|E^{-(k+1)}e^{\gamma|t|}
               \left(1+\max_j\Tr\rho A_{s,j}\right)^{1/2}.
 \end{aligned}
\end{equation}
If the right side is finite, all positive-weight eigenvectors lie in
the required common form domain; otherwise the estimate is automatic.

For the boundary comparison, first let
$P=\prod_{i=1}^m a^{\sigma_i}(f_i)$ be a monomial with $m\ge1$ and
$\supp f_i\subset B_1(x)$. Put $c_i=a^{\sigma_i}(\mathcal L_{1}f_i)$, and
abbreviate $\alpha=\tau_t^{\widehat H_{R,E}}$ and
$\beta=\tau_t^{H_{R,E}}$ in the following calculation. Multiplicativity
of these evolutions gives the ordered product identity
\begin{equation}
 \alpha(\widetilde P)-\beta(\widetilde P)
 =\sum_{i=1}^m\left(\prod_{\ell<i}\alpha(c_\ell)\right)
       \bigl(\alpha(c_i)-\beta(c_i)\bigr)
       \left(\prod_{\ell>i}\beta(c_\ell)\right).
\end{equation}
All products are in increasing index order, with empty products equal
to $1$. For $R\ge R_0=4\kappa+8$, every field has interval support in
$x[R]$ with $r_X<2$. Since $\|c_i\|=\|f_i\|_2$, the single-operator
estimate \eqref{eq:spacecut} yields
\begin{equation}
 \begin{aligned}
 \|\alpha(\widetilde P)-\beta(\widetilde P)\|
 &\le\sum_{i=1}^m\left(\prod_{\ell\ne i}\|f_\ell\|_2\right)
                     \|\alpha(c_i)-\beta(c_i)\|\\
 &\le Cm\left(\prod_{i=1}^m\|f_i\|_2\right)
                   e^{-cR}(e^{c_1E|t|}-1).
 \end{aligned}
\end{equation}
Summing with the
coefficients in the chosen expression for $A$ therefore gives
\begin{equation}
 \|\tau_t^{\widehat H_{R,E}}(\widetilde A)
       -\tau_t^{H_{R,E}}(\widetilde A)\|
 \le C_Ae^{-cR}(e^{c_1E|t|}-1).
\end{equation}
Combining the two comparisons and using \eqref{eq:unitary-lift} gives
\begin{equation}
 \begin{aligned}
 &\|\bigl(\mathcal L\tau_t^H(A)
       -\tau_t^{H_{R,E}}(\widetilde A)\mathcal L\bigr)\rho^{1/2}\|_2\\
 &\quad\le C_A\left[R|t|E^{-(k+1)}e^{\gamma|t|}
                    +e^{-cR}(e^{c_1E|t|}-1)\right]
          \left(1+\max_j\Tr\rho A_{s,j}\right)^{1/2}.
 \end{aligned}
\end{equation}
Choose $0<\varepsilon\le1$ with $c_1\varepsilon<c/2$. For
$R\ge\max\{R_0,(1+|t|)/\varepsilon\}$, set
$E=\varepsilon R/(1+|t|)$ and
$A_R(t)=\tau_t^{H_{R,E}}(\widetilde A)$. The cutoff calculation in the
proof of \cref{prop:truncation} bounds the two terms in brackets by
$C_k(1+R)^{-k}e^{(\gamma+k+2)|t|}$. Since $H_{R,E}$ acts only in
$x[R]$ and is even, $A_R(t)$ has the required support and parity;
unitary conjugation and \eqref{eq:A-lift} give $\|A_R(t)\|=\|A\|$.

For the remaining radii, set $A_R(t)=0$. The error is at most $\|A\|$,
and $1+R\le(1+R_0+\varepsilon^{-1})(1+|t|)$, so
$(1+|t|)^k\le e^{k|t|}$ gives the same bound after enlarging $C_A$.
\end{proof}

\section*{Acknowledgments}
This is supported by the Deutsche Forschungsgemeinschaft (DFG, German Research Foundation) through the grants TRR 352 – Project-ID 470903074 and FOR 5413 – Project-ID 465199066, as well as by the European Union through ERC Starting Grant MathQuantProp, Grant Agreement 101163620.\footnote{Views and opinions expressed are however those of the authors only and do not necessarily
	reflect those of the European Union or the European Research Council Executive Agency. Neither
	the European Union nor the granting authority can be held responsible for them.}

\section*{AI declaration}
The authors acknowledge help from ChatGPT (6 Astra) in
finding and checking mathematical arguments and for assistance with parts of the writing.
The authors take full responsibility for the complete manuscript. 

\begin{footnotesize}
\bibliographystyle{alpha}
\bibliography{references}

@article{nachtergaele2007lieb,
	title={{Lieb-{R}obinson bounds for harmonic and anharmonic lattice systems}},
	author={Nachtergaele, Bruno and Raz, Hillel and Schlein, Benjamin and Sims, Robert},
	journal={arXiv preprint, arXiv:0712.3820},
	year={2007}
}

@article{kuwahara2024enhanced,
	title={{Enhanced {L}ieb-{R}obinson bounds for a class of {B}ose-{H}ubbard type Hamiltonians}},
	author={Kuwahara, Tomotaka and Lemm, Marius},
	journal={arXiv preprint, arXiv:2405.04672},
	year={2024}
}

@article{lieb1972finite,
	title={{The finite group velocity of quantum spin systems}},
	author={Lieb, Elliott H and Robinson, Derek W},
	journal={Communications in mathematical physics},
	volume={28},
	number={3},
	pages={251--257},
	year={1972},
	publisher={Springer}
}

@Article{FLS,
	author     = {Faupin, J\'{e}r\'{e}my and Lemm, Marius and Sigal, Israel Michael},
	journal    = {Commun. Math. Phys.},
	title      = {{On {L}ieb-{R}obinson bounds for the {B}ose-{H}ubbard model}},
	year       = {2022},
	issn       = {0010-3616,1432-0916},
	number     = {3},
	pages      = {1011--1037},
	volume     = {394},
	doi        = {10.1007/s00220-022-04416-8},
	fjournal   = {Communications in Mathematical Physics},
	mrclass    = {82C20 (81P45 81V10 81V73)},
	mrreviewer = {H.\ Hogreve},
	url        = {https://doi.org/10.1007/s00220-022-04416-8},
}

@article{lemm2025quantum,
  title={{On the quantum dynamics of long-ranged Bose-Hubbard Hamiltonians}},
  author={Lemm, Marius and Rubiliani, Carla and Zhang, Jingxuan},
  journal={arXiv preprint arXiv:2505.01786},
  year={2025}
}

@article{elgart2026lieb,
  title={{Lieb-Robinson bounds for Bose-Hubbard Hamiltonians via path expansions}},
  author={Elgart, Alexander and Lemm, Marius},
  journal={arXiv preprint arXiv:2611.02263},
  year={2026}
}

@article{rubiliani2026lieb,
  title={{Lieb-Robinson Light Cones in Long-Range Bosonic Systems}},
  author={Rubiliani, Carla and Shang, Cheng and Kuwahara, Tomotaka},
  journal={arXiv preprint arXiv:2611.00776},
  year={2026}
}

@Article{LRZ,
  author  = {Lemm, Marius and Rubiliani, Carla and Zhang, Jingxuan},
  journal = {To appear in J. Math. Pures Appl.},
  title   = {{On the microscopic propagation speed of long-range quantum many-body systems}},
  year    = {2026},
}

@Article{FLSa,
	author   = {Faupin, J\'{e}r\'{e}my and Lemm, Marius and Sigal, Israel Michael},
	journal  = {Phys. Rev. Lett.},
	title    = {{Maximal speed for macroscopic particle transport in the {B}ose-{H}ubbard model}},
	year     = {2022},
	issn     = {0031-9007,1079-7114},
	number   = {15},
	pages    = {Paper No. 150602, 6},
	volume   = {128},
	doi      = {10.1103/physrevlett.128.150602},
	fjournal = {Physical Review Letters},
	mrclass  = {82C70},
}

@Article{LRSZ,
	author    = {Lemm, Marius and Rubiliani, Carla and Sigal, Israel Michael and Zhang, Jingxuan},
	journal   = {Phys. Rev. A},
	title     = {{Information propagation in long-range quantum many-body systems}},
	year      = {2023},
	month     = {Dec},
	pages     = {L060401},
	volume    = {108},
	doi       = {10.1103/PhysRevA.108.L060401},
	issue     = {6},
	numpages  = {6},
	publisher = {American Physical Society},
}

@Article{NS,
	author        = {Bruno Nachtergaele and Robert Sims},
	title         = {{On the dynamics of lattice systems with unbounded on-site terms in the Hamiltonian}},
	year          = {2014},
	month         = oct,
	archiveprefix = {arXiv},
	eprint        = {1410.8174},
	primaryclass  = {math-ph},
}

@article{kuwahara2021lieb,
	title={{Lieb-{R}obinson bound and almost-linear light cone in interacting boson systems}},
	author={Kuwahara, Tomotaka and Saito, Keiji},
	journal={Physical Review Letters},
	volume={127},
	number={7},
	pages={070403},
	year={2021},
	publisher={APS}
}

@article{kuwahara2024effective,
	title={{Effective light cone and digital quantum simulation of interacting bosons}},
	author={Kuwahara, Tomotaka and Vu, Tan Van and Saito, Keiji},
	journal={Nature Communications},
	volume={15},
	number={1},
	pages={2520},
	year={2024},
	publisher={Nature Publishing Group UK London}
}

@article{yin2022finite,
	title={{Finite speed of quantum information in models of interacting bosons at finite density}},
	author={Yin, Chao and Lucas, Andrew},
	journal={Physical Review X},
	volume={12},
	number={2},
	pages={021039},
	year={2022},
	publisher={APS}
}

@article{schuch2011information,
	title={{Information propagation for interacting-particle systems}},
	author={Schuch, Norbert and Harrison, Sarah K and Osborne, Tobias J and Eisert, Jens},
	journal={Physical Review A},
	volume={84},
	number={3},
	pages={032309},
	year={2011},
	publisher={APS}
}

@article{fresta2023effective,
	title={{Effective dynamics of extended fermi gases in the high-density regime}},
	author={Fresta, Luca and Porta, Marcello and Schlein, Benjamin},
	journal={Communications in Mathematical Physics},
	volume={401},
	number={2},
	pages={1701--1751},
	year={2023},
	publisher={Springer}
}

@inproceedings{hinrichs2025lieb,
  title={{On Lieb--Robinson bounds for a class of continuum fermions}},
  author={Hinrichs, Benjamin and Lemm, Marius and Siebert, Oliver},
  booktitle={Annales Henri Poincar{\'e}},
  volume={26},
  number={1},
  pages={41--80},
  year={2025},
  organization={Springer}
}

@inproceedings{gebert2020lieb,
  title={{Lieb--Robinson Bounds and Strongly Continuous Dynamics for a Class of Many-Body Fermion Systems in R d}},
  author={Gebert, Martin and Nachtergaele, Bruno and Reschke, Jake and Sims, Robert},
  booktitle={Annales Henri Poincar{\'e}},
  volume={21},
  number={11},
  pages={3609--3637},
  year={2020},
  organization={Springer}
}

@article{lemm2026lieb,
  title={{Lieb-Robinson bounds for Bose-Hubbard Hamiltonians: A review with a simplified proof}},
  author={Lemm, Marius and Rubiliani, Carla},
  journal={arXiv preprint arXiv:2603.26209},
  year={2026}
}

@article{narnhofer1990quantum,
  title={{Quantum field theories with Galilei-invariant interactions}},
  author={Narnhofer, H and Thirring, W},
  journal={Physical review letters},
  volume={64},
  number={16},
  pages={1863},
  year={1990},
  publisher={APS}
}

@article{brattelioperator,
  title={{Operator Algebras and Quantum Statistical Mechanics II Equilibrium States Models in Quantum Statistical Mechanics}},
  author={Bratteli, Ola and Robinson, Derek W},
  year={1997},
  publisher={Springer}
}

@article{fresta2025effective,
  title={{Effective Dynamics of Local Observables for Extended Fermi Gases in the High-Density Regime}},
  author={Fresta, Luca and Porta, Marcello and Schlein, Benjamin},
  journal={Communications in Mathematical Physics},
  volume={406},
  number={10},
  pages={247},
  year={2025},
  publisher={Springer}
}

@article{lemm2026dissipation,
  title={{Dissipation-enabled operator-norm locality bounds for bose-hubbard hamiltonians}},
  author={Lemm, Marius and M{\"o}bus, Tim},
  journal={arXiv preprint arXiv:2610.01669},
  year={2026}
}

@article{bachmann2018adiabatic,
  title={{The adiabatic theorem and linear response theory for extended quantum systems}},
  author={Bachmann, Sven and De Roeck, Wojciech and Fraas, Martin},
  journal={Communications in Mathematical Physics},
  volume={361},
  number={3},
  pages={997--1027},
  year={2018},
  publisher={Springer}
}

@article{teufel2026bulk,
  title={{Bulk gap stability and spectral flow for fermions with weak long-range interactions}},
  author={Teufel, Stefan},
  journal={arXiv preprint arXiv:2610.40299},
  year={2026}
}

@article{chen2023speed,
  title={{Speed limits and locality in many-body quantum dynamics}},
  author={Chen, Chi-Fang and Lucas, Andrew and Yin, Chao},
  journal={Reports on Progress in Physics},
  volume={86},
  number={11},
  pages={116001},
  year={2023},
  publisher={IOP Publishing}
}

@article{epstein2017quantum,
  title={{Quantum speed limits for quantum-information-processing tasks}},
  author={Epstein, Jeffrey M and Whaley, K Birgitta},
  journal={Physical Review A},
  volume={95},
  number={4},
  pages={042314},
  year={2017},
  publisher={APS}
}

@article{bravyi2006lieb,
  title={{Lieb-Robinson bounds and the generation of correlations and topological quantum order}},
  author={Bravyi, Sergey and Hastings, Matthew B and Verstraete, Frank},
  journal={Physical review letters},
  volume={97},
  number={5},
  pages={050401},
  year={2006},
  publisher={APS}
}

@inproceedings{becker2026generalized,
  title={{The generalized adiabatic theorem for extended lattice systems}},
  author={Becker, Lennart and Teufel, Stefan and Wesle, Marius},
  booktitle={Annales Henri Poincar{\'e}},
  pages={1--42},
  year={2026},
  organization={Springer}
}

@article{hastings2005quasiadiabatic,
  title={{Quasiadiabatic continuation of quantum states: The stability of topological ground-state degeneracy and emergent gauge invariance}},
  author={Hastings, Matthew B and Wen, Xiao-Gang},
  journal={Physical Review B—Condensed Matter and Materials Physics},
  volume={72},
  number={4},
  pages={045141},
  year={2005},
  publisher={APS}
}

@article{bachmann2012automorphic,
  title={{Automorphic equivalence within gapped phases of quantum lattice systems}},
  author={Bachmann, Sven and Michalakis, Spyridon and Nachtergaele, Bruno and Sims, Robert},
  journal={Communications in Mathematical Physics},
  volume={309},
  number={3},
  pages={835--871},
  year={2012},
  publisher={Springer}
}

@article{bravyi2010topological,
  title={{Topological quantum order: stability under local perturbations}},
  author={Bravyi, Sergey and Hastings, Matthew B and Michalakis, Spyridon},
  journal={Journal of mathematical physics},
  volume={51},
  number={9},
  year={2010},
  publisher={AIP Publishing}
}

@article{michalakis2013stability,
  title={{Stability of frustration-free Hamiltonians}},
  author={Michalakis, Spyridon and Zwolak, Justyna P},
  journal={Communications in Mathematical Physics},
  volume={322},
  number={2},
  pages={277--302},
  year={2013},
  publisher={Springer}
}

@article{del2025boundary,
  title={{Boundary effects and the stability of the low energy spectrum of the AKLT model}},
  author={Del Vecchio, Simone and Fr{\"o}hlich, J{\"u}rg and Pizzo, Alessandro and Ranallo, Alessio},
  journal={Journal of Functional Analysis},
  pages={111269},
  year={2025},
  publisher={Elsevier}
}

@article{haah2021quantum,
  title={{Quantum algorithm for simulating real time evolution of lattice Hamiltonians}},
  author={Haah, Jeongwan and Hastings, Matthew B and Kothari, Robin and Low, Guang Hao},
  journal={SIAM Journal on Computing},
  volume={52},
  number={6},
  pages={FOCS18--250},
  year={2021},
  publisher={SIAM}
}

@article{Hastings2007area,
    author = {Hastings, Matthew B.},
    title = {{An area law for one-dimensional quantum systems}},
    journal = {Journal of Statistical Mechanics: Theory and Experiment},
   year = {2007-08-20},
    volume = {2007},
    number = {08},
    pages = {P08024},
    doi = {10.1088/1742-5468/2007/08/P08024},
    ids = {Hastings2007},
}

@article{HK2006,
    author = {Hastings, Matthew B. and Koma, Tohru},
    title = {{Spectral Gap and Exponential Decay of Correlations}},
    journal = {Communications in Mathematical Physics},
   year = {2006-04},
    volume = {265},
    number = {3},
    pages = {781--804},
    doi = {10.1007/s00220-006-0030-4},
    eprint = {math-ph/0507008},
    eprinttype = {arXiv},
    ids = {hastings2006spectral},
}

@book{CFKS,
  title={{Schr{\"o}dinger operators: with applications to quantum mechanics and global geometry}},
  author={Cycon, Hans L and Froese, Richard G and Kirsch, Werner and Simon, Barry and Cycon, Hans L},
  year={1987},
  publisher={Springer}
}

@article{HT2020bulk,
    author = {Henheik, Joscha and Teufel, Stefan},
    title = {{Adiabatic theorem in the thermodynamic limit: Systems with a gap in the bulk}},
    journal = {Forum of Mathematics, Sigma},
   year = {2022},
    volume = {10},
    doi = {10.1017/fms.2021.80},
    eprint = {2012.15239},
    eprintclass = {math-ph},
    eprinttype = {arXiv},
    ids = {HT2020b},
}

@article{HW2022,
    author = {Henheik, Joscha and Wessel, Tom},
    title = {{On adiabatic theory for extended fermionic lattice systems}},
    journal = {Journal of Mathematical Physics},
   year = {2022-12-01},
    volume = {63},
    number = {12},
    pages = {121101},
    doi = {10.1063/5.0123441},
    eprint = {2208.12220},
    eprintclass = {math-ph},
    eprinttype = {arXiv},
}

@article{NS2006,
    author = {Nachtergaele, Bruno and Sims, Robert},
    title = {{{Lieb-Robinson} Bounds and the Exponential Clustering Theorem}},
    journal = {Communications in Mathematical Physics},
   year = {2006-03},
    volume = {265},
    number = {1},
    pages = {119--130},
    doi = {10.1007/s00220-006-1556-1},
    eprint = {math-ph/0506030},
    eprinttype = {arXiv},
    ids = {nachtergaele2006lieb},
}

@inproceedings{NSY2017,
    author = {Nachtergaele, Bruno and Sims, Robert and Young, Amanda},
    editor = {Bonetto, Federico and Borthwick, David and Harrell, Evans and Loss, Michael},
    title = {{{Lieb}-{Robinson} bounds, the spectral flow, and stability of the spectral gap for lattice fermion systems}},
    booktitle = {Mathematical Results in Quantum Physics},
   year = {2018},
    series = {Contemporary Mathematics},
    number = {717},
    publisher = {American Mathematical Society},
    pages = {93--115},
    doi = {10.1090/conm/717},
    eprint = {1705.08553},
    eprintclass = {math-ph},
    eprinttype = {arXiv},
    ids = {nachtergaele2017liebrobinson},
}

@article{NSY2022,
    author = {Nachtergaele, Bruno and Sims, Robert and Young, Amanda},
    title = {{Quasi-Locality Bounds for Quantum Lattice Systems. {Part} {II}. {Perturbations} of Frustration-Free Spin Models with Gapped Ground States}},
    journal = {Annales Henri Poincaré},
   year = {2022-02},
    volume = {23},
    number = {2},
    pages = {393--511},
    doi = {10.1007/s00023-021-01086-5},
    eprint = {2010.15337},
    eprinttype = {arXiv},
    ids = {NSY2020,nachtergaele2022quasi},
}

@article{YL2023,
    author = {Yin, Chao and Lucas, Andrew},
    title = {{Prethermalization and the Local Robustness of Gapped Systems}},
    journal = {Physical Review Letters},
   year = {2023-08},
    volume = {131},
    number = {5},
    pages = {050402},
    doi = {10.1103/PhysRevLett.131.050402},
    eprint = {2209.11242},
    eprintclass = {cond-mat.str-el},
    eprinttype = {arXiv},
    ids = {YL2022,YL23},
}

@article{mcdonough2025non,
  title={{Non-perturbatively slow spread of quantum correlations in non-resonant systems}},
  author={McDonough, Ben T and Lemm, Marius and Lucas, Andrew},
  journal={arXiv preprint arXiv:2510.11831},
  year={2025}
}

@inproceedings{bachmann2025lieb,
  title={{Lieb--Robinson Bounds in the Continuum Via Localized Frames}},
  author={Bachmann, Sven and Nittis, Giuseppe De},
  booktitle={Annales Henri Poincar{\'e}},
  volume={26},
  number={1},
  pages={1--40},
  year={2025},
  organization={Springer}
}

@article{greenblatt2024adiabatic,
  title={{Adiabatic evolution of low-temperature many-body systems}},
  author={Greenblatt, Rafael L and Lange, Markus and Marcelli, Giovanna and Porta, Marcello},
  journal={Communications in Mathematical Physics},
  volume={405},
  number={3},
  pages={75},
  year={2024},
  publisher={Springer}
}

@article{henheik2026response,
  title={{Response theory at low temperature for weakly interacting quantum spin systems}},
  author={Henheik, Joscha and Lemm, Marius and Wessel, Tom},
  journal={arXiv preprint arXiv:2610.25249},
  year={2026}
}

@article{abanin2017rigorous,
  title={{A rigorous theory of many-body prethermalization for periodically driven and closed quantum systems}},
  author={Abanin, Dmitry and De Roeck, Wojciech and Ho, Wen Wei and Huveneers, Fran{\c{c}}ois},
  journal={Communications in Mathematical Physics},
  volume={354},
  number={3},
  pages={809--827},
  year={2017},
  publisher={Springer}
}
\end{footnotesize}
\end{document}